\documentclass[12pt]{article}
\usepackage{epsfig}
\usepackage{psfrag}
\usepackage{enumitem}
\usepackage{latexsym}
\usepackage{indentfirst}
\usepackage{fancyhdr}
\usepackage{dsfont}
\usepackage{amssymb}
\usepackage{amsmath}
\usepackage{amsfonts}
\usepackage{pifont}
\usepackage{cite}
\usepackage{multirow}
\usepackage{bbold}
\usepackage{color}
\usepackage[footnotesize]{caption2}
\usepackage{graphicx}
\usepackage[center,footnotesize,hang]{subfigure}
\usepackage{url}
\usepackage{array}

\begin{document}
\begin{titlepage}
\vspace*{-1cm}


\vskip 2.5cm
\begin{center}
{\Large\bf TFH Mixing Patterns, Large $\theta_{13}$ and $\Delta(96)$ Flavor Symmetry}
\end{center}
\vskip 0.2  cm
\vskip 0.5  cm
\begin{center}
{ Gui-Jun Ding}~\footnote{Email: dinggj@ustc.edu.cn}
\\
\vskip .2cm
{\it Department of Modern Physics,}
\\
{\it University of Science and Technology of China, Hefei, Anhui
230026, China}

\end{center}
\vskip 0.7cm
\begin{abstract}
\noindent

We perform a comprehensive analysis of the Toorop-Feruglio-Hagedorn (TFH) mixing patterns within the family symmetry $\Delta(96)$. The general neutrino mass matrix for the TFH mixing and its symmetry properties are investigated. The possible realizations of the TFH mixing in $\Delta(96)$ are analyzed in the minimalist approach. We propose two dynamical models which produce the TFH mixing patterns at leading order. The full flavor symmetries are $\Delta(96)\times Z_3\times Z_3$ and $\Delta(96)\times Z_5 \times Z_2$ respectively. The next to leading order terms introduce corrections of order $\lambda^2_c$ to the three mixing angles in both models. The allowed mixing patterns are studied under the condition that the Klein four subgroups and the cyclic $Z_N$ subgroups with $N\geq3$ are preserved in the neutrino and the charged lepton sector respectively. We suggest that the deformed tri-bimaximal mixing is a good leading order approximation to understanding a largish reactor angle.

\end{abstract}
\end{titlepage}
\setcounter{footnote}{1}
\vskip2truecm

\section{Introduction}
In the past years, neutrino physics made great progress, neutrino
oscillation and tiny neutrino masses have been established. The mass
squared differences and the mixing angles are known with good
accuracy. Recently the T2K \cite{Abe:2011sj} and MINOS \cite{minos}
Collaborations reported the evidence for a relatively large
$\theta_{13}$ at the level of $2.5\sigma$ and $1.7\sigma$
respectively. Combining the data from T2K, MINOS and other
experiments, the global fit to the neutrino oscillation data
indicates $\theta_{13}>0$ at $3\sigma$ confidence level
\cite{Schwetz:2011zk,Fogli:2011qn}, although the magnitude of
$\theta_{13}$ still suffers from large uncertainties, and it yields
\begin{equation}
\label{eq1}\sin^2\theta_{13}= 0.013^{+0.007}_{-0.005} (0.016^{+0.008}_{-0.006}),~~~~~~~\mathrm{at}~~ 1\sigma\;,
\end{equation}
where the result shown in parentheses corresponds to the inverted neutrino mass hierarchy. A non-zero $\theta_{13}$ is also verified by the first results from Double CHOOZ \cite{double_chooz}.

So far considerable efforts have been devoted to the well-known
tri-bimaximal (TB) mixing \cite{TBmix}, it is found that the TB
mixing can be naturally produced via the spontaneous breaking of
certain discrete flavor symmetries, and many beautiful models were
constructed, please see
Refs.\cite{Altarelli:2010gt,Ishimori:2010au,Grimus:2011fk} for
review. The deviations from the TB mixing can arise due to the
corrections from the higher-dimensional operators, and the three
mixing angles generally receive corrections of the same order of
magnitude except some special dynamics are introduced
\cite{Lin:2009bw}. Therefore the reactor angle is expected to be at
most of order $\lambda^2_c$, where $\lambda_c$ is the Cabibbo angle,
since the experimentally allowed departure of $\theta_{12}$ from its
TB value $\sin^2\theta_{12}=1/3$ is of order $\lambda_c^2$. If it is
confirmed that $\theta_{13}$ is close to its present central value
in the near future, the TB mixing may not be a good leading order
(LO) approximation any more. Many proposals have been put forward to
accommodate a non-zero $\theta_{13}$
\cite{Toorop:2011jn,deAdelhartToorop:2011re,theta_13}. In fact,
there were already some discussions \cite{Xing:2008ie,King:2009qt}
about non-zero reactor mixing angle even before the T2K data. In
general, there are two approaches to understanding a largish
$\theta_{13}$. The first one is starting from some textures which
could be outside the current $3\sigma$ allowed range, and the
sizable next to leading order (NLO) corrections would pull the three
mixing angles into the experimentally preferred range. Ref.
\cite{Altarelli:2009gn} is a typical realization of this approach,
where the bimaximal mixing is derived as the LO mixing pattern, the
solar and the reactor angles get corrections of order $\lambda_c$
while the atmospheric angle keeps intact at NLO. The second one is
to directly produce some LO approximation textures, which are in
good agreement with the present data, then the NLO corrections
should be mild in this case. By analyzing the symmetry breaking of
the finite modular group $\Gamma_{N}$, Feruglio et al. proposed that
two attractive mixing textures with
$\sin^2\theta_{13}=(2-\sqrt{3})/6$,
$\sin^2\theta_{12}=(8-2\sqrt{3})/13$,
$\sin^2\theta_{23}=(5+2\sqrt{3})/13$, $\delta_{CP}=\pi$ and
$\sin^2\theta_{13}=(2-\sqrt{3})/6$,
$\sin^2\theta_{12}=\sin^2\theta_{23}=(8-2\sqrt{3})/13$,
$\delta_{CP}=0$ can be generated if we choose $\Delta(96)$ as the
flavor symmetry and further break it into the Klein four ($K_4$) and
$Z_3$ subgroups in the neutrino and charged lepton sectors
respectively \cite{Toorop:2011jn,deAdelhartToorop:2011re}.
Henceforth, we shall call these two mixing patterns as TFH1 and TFH2
textures respectively, where THF is the abbreviation of the author
names Toorop-Feruglio-Hagedorn. They are denoted as {\tt M1} and
{\tt M2} in Refs.\cite{Toorop:2011jn,deAdelhartToorop:2011re}.
Obviously TFH1 and TFH2 predict the same solar mixing angle
$\theta_{12}$ and the reactor mixing angle $\theta_{13}$, they only
differ in the suggested value for the atmospheric mixing angle
$\theta_{23}$. Note that $\theta_{12}$ for TFH mixing is slightly
below the $3\sigma$ upper limits of the two global fits in
Refs.\cite{Schwetz:2011zk,Fogli:2011qn}, $\theta_{13}$ lies in the
$3\sigma$ range of Ref. \cite{Fogli:2011qn} nevertheless it is
larger than the $3\sigma$ upper bound of Ref.\cite{Schwetz:2011zk}.
For the TFH1 mixing, $\theta_{23}$ is marginally above the $3\sigma$
upper limit of Refs.\cite{Schwetz:2011zk,Fogli:2011qn}, while it is
a bit larger (smaller) than the $3\sigma$ lower bound of Ref.
\cite{Fogli:2011qn} (Ref.\cite{Schwetz:2011zk}) in the case of TFH2
mixing.

Given the recent experimental evidences for a relatively large $\theta_{13}$ \cite{Abe:2011sj,minos,double_chooz}, it is interesting to investigate whether we can and how to consistently derive the TFH textures within the framework of $\Delta(96)$ family symmetry, this is the motivation of the present work. In this paper, we perform a comprehensive exploration of the THF textures based on $\Delta(96)$, two flavor models giving rise to TFH1 and TFH2 textures are built. In both models, the neutrino sector is invariant under a $K_4$ subgroup at LO while $\Delta(96)$ is broken completely in the charged lepton sector, and the hierarchies among the charged lepton masses are produced. Furthermore, we study all the possible mixing patterns and the generated groups if the flavor symmetry group $\Delta(96)$ is broken into the $K_4$ and $Z_N$ ($N\geq3$) subgroups in the neutrino and charged lepton sectors respectively.

This article is organized as follows. In section 2, we discuss the symmetry properties of the general neutrino mass matrix which leads to the TFH mixing patterns in the flavor basis. Some basic features of $\Delta(96)$ group are recalled in section 3. We proceed to the model building in the so-called minimalist approach in section 4, where the different assignments of the involved fields are presented. The structures of the models for the TFH1 and TFH2 mixing are described in section 5 and section 6 respectively, where both the vacuum alignment and the next to leading order corrections are discussed in detail. In section 7, we explore the possible mixing patterns by requiring that the $K_4$ and $Z_N(N\geq3)$ subgroup are preserved in the neutrino and charged lepton sectors respectively. Finally section 8 is devoted to our conclusions. We report the representation matrices for the group generators and the connection with the Escobar-Luhn basis in Appendix A. The collection of the Clebsch-Gordan coefficients is listed in Appendix B.

\section{\label{sec:mass_matrix}Neutrino mass matrices for the THF mixings}

Given the mixing parameters $\sin^2\theta_{13}=(2-\sqrt{3})/6$, $\sin^2\theta_{12}=(8-2\sqrt{3})/13$, $\sin^2\theta_{23}=(5+2\sqrt{3})/13$ and $\delta_{CP}=\pi$ for the THF1 mixing, in a particular phase convention, the corresponding Pontecorvo-Maki-Nakagawa-Sakata (PMNS) matrix is given by
\begin{equation}
\label{eq2}U_{TFH1}=\left(\begin{array}{ccc}
\frac{1}{6}(3+\sqrt{3})  &  \frac{1}{\sqrt{3}}  & \frac{1}{6}(-3+\sqrt{3})\\
\frac{1}{6}(-3+\sqrt{3}) & \frac{1}{\sqrt{3}}   & \frac{1}{6}(3+\sqrt{3}) \\
-\frac{1}{\sqrt{3}}      &  \frac{1}{\sqrt{3}}  & -\frac{1}{\sqrt{3}}
\end{array}\right),
\end{equation}
where we absorb the Majorana phases into the light neutrino mass eigenvalues. In the basis where the charged lepton mass matrix is diagonal, the TFH1 light neutrino mass matrix can be constructed by acting with $U_{TFH1}$ on a generic diagonal neutrino mass matrix.
\begin{eqnarray}
\nonumber m^{TFH1}_{\nu}&=&U^{*}_{TFH1}\mathrm{diag}(m_1,m_2,m_3)U^{\dagger}_{TFH1}\\
\nonumber&=&\frac{m_1}{6}\left(\begin{array}{ccc}2+\sqrt{3} & -1  &  -1-\sqrt{3}\\
-1 & 2-\sqrt{3}  & -1+\sqrt{3} \\
-1-\sqrt{3}  & -1+\sqrt{3}  & 2
\end{array}\right)+\frac{m_2}{3}\left(\begin{array}{ccc}1&1&1 \\
1&1&1 \\
1&1&1
\end{array}\right)\\
\label{eq3}&+&\frac{m_3}{6}\left(\begin{array}{ccc}
2-\sqrt{3}  & -1  & -1+\sqrt{3}  \\
-1  &  2+\sqrt{3}  &  -1-\sqrt{3} \\
-1+\sqrt{3}  & -1-\sqrt{3}  & 2
\end{array}\right),
\end{eqnarray}
where $m_{1,2,3}$ are the light neutrino masses, they are complex to account for the Majorana phases. We notice that the neutrino mass matrix $m^{TFH1}_{\nu}$ is characterized by the following relations
\begin{eqnarray}
\nonumber&&(m^{TFH1}_{\nu})_{23}+(m^{TFH1}_{\nu})_{13}=2(m^{TFH1}_{\nu})_{12}\\
\nonumber&&2(m^{TFH1}_{\nu})_{12}+(m^{TFH1}_{\nu})_{22}=(m^{TFH1}_{\nu})_{11}+2(m^{TFH1}_{\nu})_{13}\\
\label{eq4}&&(m^{TFH1}_{\nu})_{12}+(m^{TFH1}_{\nu})_{33}=(m^{TFH1}_{\nu})_{11}+(m^{TFH1}_{\nu})_{13}\,.
\end{eqnarray}
This means that there should be some underlying symmetries in $m^{TFH1}_{\nu}$, i.e., $G^T_{\nu}m^{TFH1}_{\nu}G_{\nu}=m^{TFH1}_{\nu}$. After some straightforward calculations, we find that, up to an overall sign, the symmetry transformation $G_{\nu}$ takes the following forms
\begin{eqnarray}
\nonumber&&G_1=\frac{1}{3}\left(\begin{array}{ccc}
-1+\sqrt{3}  &  -1  & -1-\sqrt{3} \\
-1  & -1-\sqrt{3}  & -1+\sqrt{3} \\
-1-\sqrt{3}  & -1+\sqrt{3}  & -1
\end{array}\right),~~~G_2=\frac{1}{3}\left(\begin{array}{ccc}
-1&2&2\\
2&-1&2\\
2&2&-1
\end{array}\right)\\
\label{eq5}&&G_3=\frac{1}{3}\left(\begin{array}{ccc}
-1-\sqrt{3}  &-1  & -1+\sqrt{3}  \\
-1  & -1+\sqrt{3}  & -1-\sqrt{3}  \\
-1+\sqrt{3}  &  -1-\sqrt{3}  &  -1
\end{array}\right),
\end{eqnarray}
beside the identity transformation. It is easy to check that the above symmetry matrices satisfy the relations
\begin{equation}
\label{eq6} G^2_{i}=\mathbf{1},~~~~~~~G_iG_j=G_jG_i=G_{k}~~{\rm
with}~~ i\neq j\neq k\;.
\end{equation}
Consequently the symmetry group of the neutrino mass matrix $m^{TFH1}_{\nu}$ is $K_4$, this is a general result for the mass independent leptonic mixing patterns \cite{Lam:2007qc}. Conversely, if a light neutrino mass matrix is invariant under the unitary transformations shown in Eq.(\ref{eq5}), it is exactly diagonalized by the TFH1 mixing matrix.

The TFH2 mixing matrix can be obtained by exchanging the second and third rows of the TFH1 texture,
\begin{equation}
\label{eq7}U_{TFH2}=\left(\begin{array}{ccc}
\frac{1}{6}(3+\sqrt{3})  &  \frac{1}{\sqrt{3}}   &  \frac{1}{6}(-3+\sqrt{3}) \\
-\frac{1}{\sqrt{3}}   &  \frac{1}{\sqrt{3}}  & -\frac{1}{\sqrt{3}}  \\
\frac{1}{6}(-3+\sqrt{3})  &  \frac{1}{\sqrt{3}}   & \frac{1}{6}(3+\sqrt{3})
\end{array}\right),
\end{equation}
which leads to the mixing angles $\sin^2\theta_{13}=(2-\sqrt{3})/6$ and $\sin^2\theta_{12}=\sin^2\theta_{23}=(8-2\sqrt{3})/13$ together with $\delta_{CP}=0$. In the flavor basis, the general neutrino mass matrix diagonalized by $U_{TFH2}$, is of the following form
\begin{eqnarray}
\nonumber m^{TFH2}_{\nu}&=&\frac{m_1}{6}\left(\begin{array}{ccc}
2+\sqrt{3}  &  -1-\sqrt{3}  &  -1  \\
-1-\sqrt{3}  &  2   &  -1+\sqrt{3}  \\
-1   &  -1+\sqrt{3}  &  2-\sqrt{3}
\end{array}\right)+\frac{m_2}{3}\left(\begin{array}{ccc}
1&1&1\\
1&1&1\\
1&1&1
\end{array}\right)\\
\label{eq8}&+&\frac{m_3}{6}\left(\begin{array}{ccc}
2-\sqrt{3}  &   -1+\sqrt{3}   &  -1 \\
-1+\sqrt{3}  & 2  & -1-\sqrt{3}   \\
-1  &  -1-\sqrt{3}  & 2+\sqrt{3}
\end{array}\right).
\end{eqnarray}
We note that $m^{TFH2}_{\nu}$ is related with $m^{TFH1}_{\nu}$ by permutating the second and third rows and columns simultaneously:
\begin{equation}
\label{eq9}m^{TFH2}_{\nu}=T_{23}m^{TFH1}_{\nu}T_{23}\,,
\end{equation}
where
\begin{equation}
\label{eq10}T_{23}=\left(\begin{array}{ccc}1&0&0\\
0&0&1\\
0&1&0
\end{array}\right)\,.
\end{equation}
A symmetric matrix has 6 independent elements in general, whereas there are only 3 parameters $m_1$, $m_2$ and $m_3$ in the mass matrix $m^{TFH2}_{\nu}$. As a result, 3 additional constraints should be satisfied between the elements of $m^{TFH2}_{\nu}$.
\begin{eqnarray}
\nonumber&&(m^{TFH2}_{\nu})_{12}+(m^{TFH2}_{\nu})_{23}=2(m^{TFH2}_{\nu})_{13}\\
\nonumber&&(m^{TFH2}_{\nu})_{13}+(m^{TFH2}_{\nu})_{22}=(m^{TFH2}_{\nu})_{11}+(m^{TFH2}_{\nu})_{12}\\
\label{eq11}&&2(m^{TFH2}_{\nu})_{13}+(m^{TFH2}_{\nu})_{33}=(m^{TFH2}_{\nu})_{11}+2(m^{TFH2}_{\nu})_{12}\,.
\end{eqnarray}
Following the same procedure as above, we find that $m^{TFH2}_{\nu}$ is invariant under the action of the following three unitary transformations
\begin{eqnarray}
\nonumber&&G'_1=\frac{1}{3}\left(\begin{array}{ccc}
-1+\sqrt{3}  & -1-\sqrt{3}   &  -1  \\
-1-\sqrt{3}  &  -1    & -1+\sqrt{3}  \\
-1  & -1+\sqrt{3}   &  -1-\sqrt{3}
\end{array}\right),~~~G'_2=\frac{1}{3}\left(\begin{array}{ccc}
-1 &  2  &  2  \\
2 & -1 & 2  \\
2 & 2 & -1
\end{array}\right),\\
\label{eq12}&&G'_3=\frac{1}{3}\left(\begin{array}{ccc}
-1-\sqrt{3}  &  -1+\sqrt{3}  & -1 \\
-1+\sqrt{3}  &  -1  &  -1-\sqrt{3}  \\
-1  & -1-\sqrt{3}  & -1+\sqrt{3}
\end{array}\right)\,.
\end{eqnarray}
These symmetry matrices $G'_{1,2,3}$ satisfy the relations given in Eq.(\ref{eq6}), they form a $K_4$ group as well. It is interesting to note that $G'_2=G_2$, the reason is that $(1,1,1)^{T}/\sqrt{3}$ is the second column of both $U_{TFH1}$ and $U_{TFH2}$.

\section{The $\Delta(96)$ group}

$\Delta(96)$ belongs to the group series $\Delta(6n^2)$ with $n=4$, the mathematical aspects of the groups of type $\Delta(6n^2)$ have been worked out in detail \cite{D96}. In the case of $n=2$, $\Delta(24)$ is isomorphic to the widely studied family symmetry $S_4$. The $\Delta(96)$ is a non-abelian finite subgroup of SU(3) of order 96, it is isomorphic to $(Z_4\times Z_4)\rtimes S_3$, and it can be conveniently defined by four generators $a$, $b$, $c$ and $d$ obeying the relations:
\begin{eqnarray}
\nonumber&a^3 ~=~ b^2 ~=~ (ab)^2 ~=~ c^4 ~=~ d^4 ~=~1\\
\nonumber&cd ~=~ dc\\
\nonumber&a c a^{-1}=c^{-1} d^{-1},\quad a d a^{-1} =c\\
\label{eq13}& b c b^{-1}=d^{-1}, \quad b d b^{-1}=c^{-1}\,.
\end{eqnarray}
Note that the generator $d$ is not independent, it can be expressed in terms of the generators $b$ and $c$. The 96 group elements are divided into 10 conjugate classes as follows:
\begin{eqnarray}
\nonumber&&{\cal C}_1: 1\\
\nonumber&&{\cal C}_2: cd^2, cd^3, c^2d^3\\
\nonumber&&{\cal C}_3: c^2, d^2, c^2d^2\\
\nonumber&&{\cal C}_4: c^2d, c^3d, c^3d^2\\
\nonumber&&{\cal C}_5: c, d, cd, c^3, d^3, c^3d^3\\
\nonumber&&{\cal C}_6: a, ac, ac^2, ac^3, ad, ad^2, ad^3, acd, acd^2, acd^3, ac^2d, ac^2d^2, ac^2d^3, ac^3d, ac^3d^2, ac^3d^3, a^2, a^2c,\\
\nonumber&&~~\quad a^2c^2, a^2c^3, a^2d, a^2d^2, a^2d^3, a^2cd, a^2cd^2, a^2cd^3, a^2c^2d, a^2c^2d^2, a^2c^2d^3, a^2c^3d, a^2c^3d^2, a^2c^3d^3\\
\nonumber&&{\cal C}_7: ab, abc, abc^2, abc^3, a^2b, a^2bd, a^2bd^2, a^2bd^3, b, bcd, bc^2d^2, bc^3d^3\\
\nonumber&&{\cal C}_8: abd, abcd, abc^2d, abc^3d, a^2bc^3, a^2bc^3d, a^2bc^3d^2, a^2bc^3d^3, bc, bc^2d, bc^3d^2, bd^3\\
\nonumber&&{\cal C}_9: abd^2, abcd^2, abc^2d^2, abc^3d^2, a^2bc^2, a^2bc^2d, a^2bc^2d^2, a^2bc^2d^3, bc^2, bc^3d, bd^2, bcd^3\\
\label{eq14}&&{\cal C}_{10}: abd^3, abcd^3, abc^2d^3, abc^3d^3, a^2bc, a^2bcd, a^2bcd^2, a^2bcd^3, bc^3, bd, bcd^2, bc^2d^3
\end{eqnarray}
Since the number of the un-equivalent irreducible representation is equal to the number of class, $\Delta(96)$ group has 10 irreducible representations: two singlets $\mathbf{1}$ and $\mathbf{1'}$, one doublet $\mathbf{2}$, six triplets $\mathbf{3_1}$, $\mathbf{3'_1}$, $\mathbf{\overline{3}_1}$, $\mathbf{\overline{3}'_1}$, $\mathbf{3_2}$ and $\mathbf{3'_2}$, one sextet $\mathbf{6}$. We note that $\mathbf{\overline{3}_1}$ and $\mathbf{\overline{3}'_1}$ are the complex conjugate representations of $\mathbf{3_1}$ and $\mathbf{3'_1}$ respectively, and the representations $\mathbf{3_1}$, $\mathbf{3'_1}$, $\mathbf{\overline{3}_1}$, $\mathbf{\overline{3}'_1}$ and $\mathbf{6}$ are the faithful representations of the group, while $\mathbf{3_2}$ and $\mathbf{3'_2}$ are not. The character table of $\Delta(96)$ is shown in Table \ref{tab:character}, then the Kronecker product rules for two $\Delta(96)$ irreducible representations follow immediately:
\begin{eqnarray}
\nonumber&&\mathbf{1'}\otimes\mathbf{2}=\mathbf{2},~~~\mathbf{1'}\otimes\mathbf{3_1}=\mathbf{3'_1},~~~\mathbf{1'\otimes\mathbf{3'_1}}=\mathbf{3_1},
~~~\mathbf{1'}\otimes\mathbf{\overline{3}_1}=\mathbf{\overline{3}'_1},~~~\mathbf{1'}\otimes\mathbf{\overline{3}'_1}=\mathbf{\overline{3}_1},\\
\nonumber&&\mathbf{1'}\otimes\mathbf{3_2}=\mathbf{3'_2},~~~\mathbf{1'}\otimes\mathbf{3'_2}=\mathbf{3_2},~~~\mathbf{1'}\otimes\mathbf{6}=\mathbf{6},~~~
\mathbf{2}\otimes\mathbf{2}=\mathbf{1}\oplus\mathbf{1'}\oplus\mathbf{2},~~~\mathbf{2}\otimes\mathbf{3_1}=\mathbf{3_1}\oplus\mathbf{3'_1},\\
\nonumber&&\mathbf{2}\otimes\mathbf{3'_1}=\mathbf{3_1}\oplus\mathbf{3'_1},~~~\mathbf{2}\otimes\mathbf{\overline{3}_1}=\mathbf{\overline{3}_1}\oplus\mathbf{\overline{3}'_1},~~~
\mathbf{2}\otimes\mathbf{\overline{3}'_1}=\mathbf{\overline{3}_1}\oplus\mathbf{\overline{3}'_1},~~~\mathbf{2}\otimes\mathbf{3_2}=\mathbf{3_2}\oplus\mathbf{3'_2},\\
\nonumber&&\mathbf{2}\otimes\mathbf{3'_2}=\mathbf{3_2}\oplus\mathbf{3'_2},~~~\mathbf{2}\otimes\mathbf{6}=\mathbf{6_1}\oplus\mathbf{6_2},~~~\mathbf{3_1}\otimes\mathbf{3_1}=\mathbf{\overline{3}_1}\oplus\mathbf{\overline{3}'_1}\oplus\mathbf{3'_2},
~~~\mathbf{3_1}\otimes\mathbf{3'_1}=\mathbf{\overline{3}_1}\oplus\mathbf{\overline{3}'_1}\oplus\mathbf{3_2},\\
\nonumber&&\mathbf{3_1}\otimes\mathbf{\overline{3}_1}=\mathbf{1}\oplus\mathbf{2}\oplus\mathbf{6},~~~\mathbf{3_1}\otimes\mathbf{\overline{3}'_1}=\mathbf{1'}\oplus\mathbf{2}\oplus\mathbf{6},
~~~\mathbf{3_1}\otimes\mathbf{3_2}=\mathbf{\overline{3}'_1}\oplus\mathbf{6},~~~\mathbf{3_1}\otimes\mathbf{3'_2}=\mathbf{\overline{3}_1}\oplus\mathbf{6},\\
\nonumber&&\mathbf{3_1}\otimes\mathbf{6}=\mathbf{3_1}\oplus\mathbf{3'_1}\oplus\mathbf{3_2}\oplus\mathbf{3'_2}\oplus\mathbf{6},~~~\mathbf{3'_1}\otimes\mathbf{3'_1}=\mathbf{\overline{3}_1}\oplus\mathbf{\overline{3}'_1}\oplus\mathbf{3'_2},~~~\mathbf{3'_1}\otimes\mathbf{\overline{3}_1}=\mathbf{1'}\oplus\mathbf{2}\oplus\mathbf{6},\\
\nonumber&&\mathbf{3'_1}\otimes\mathbf{\overline{3}'_1}=\mathbf{1}\oplus\mathbf{2}\oplus\mathbf{6},~~~\mathbf{3'_1}\otimes\mathbf{3_2}=\mathbf{\overline{3}_1}\oplus\mathbf{6},~~~\mathbf{3'_1}\otimes\mathbf{3'_2}=\mathbf{\overline{3}'_1}\oplus\mathbf{6},\\
\nonumber&&\mathbf{3'_1}\otimes\mathbf{6}=\mathbf{3_1}\oplus\mathbf{3'_1}\oplus\mathbf{3_2}\oplus\mathbf{3'_2}\oplus\mathbf{6},~~~\mathbf{\overline{3}_1}\otimes\mathbf{\overline{3}_1}=\mathbf{3_1}\oplus\mathbf{3'_1}\oplus\mathbf{3'_2},~~~\mathbf{\overline{3}_1}\otimes\mathbf{\overline{3}'_1}=\mathbf{3_1}\oplus\mathbf{3'_1}\oplus\mathbf{3_2},\\
\nonumber&&\mathbf{\overline{3}_1}\otimes\mathbf{3_2}=\mathbf{3'_1}\oplus\mathbf{6},~~~\mathbf{\overline{3}_1}\otimes\mathbf{3'_2}=\mathbf{3_1}\oplus\mathbf{6},~~~\mathbf{\overline{3}_1}\otimes\mathbf{6}=\mathbf{\overline{3}_1}\oplus\mathbf{\overline{3}'_1}\oplus\mathbf{3_2}\oplus\mathbf{3'_2}\oplus\mathbf{6},\\
\nonumber&&\mathbf{\overline{3}'_1}\otimes\mathbf{\overline{3}'_1}=\mathbf{3_1}\oplus\mathbf{3'_1}\oplus\mathbf{3'_2},~~~\mathbf{\overline{3}'_1}\otimes\mathbf{3_2}=\mathbf{3_1}\oplus\mathbf{6},~~~\mathbf{\overline{3}'_1}\otimes\mathbf{3'_2}=\mathbf{3'_1}\oplus\mathbf{6},\\
\nonumber&&\mathbf{\overline{3}'_1}\otimes\mathbf{6}=\mathbf{\overline{3}_1}\oplus\mathbf{\overline{3}'_1}\oplus\mathbf{3_2}\oplus\mathbf{3'_2}\oplus\mathbf{6},~~~\mathbf{3_2}\otimes\mathbf{3_2}=\mathbf{1}\oplus\mathbf{2}\oplus\mathbf{3_2}\oplus\mathbf{3'_2},\\
\nonumber&&\mathbf{3_2}\otimes\mathbf{3'_2}=\mathbf{1'}\oplus\mathbf{2}\oplus\mathbf{3_2}\oplus\mathbf{3'_2},~~~\mathbf{3_2}\otimes\mathbf{6}=\mathbf{3_1}\otimes\mathbf{3'_1}\oplus\mathbf{\overline{3}_1}\oplus\mathbf{\overline{3}'_1}\oplus\mathbf{6},\\
\nonumber&&\mathbf{3'_2}\otimes\mathbf{3'_2}=\mathbf{1}\oplus\mathbf{2}\oplus\mathbf{3_2}\oplus\mathbf{3'_2},~~~\mathbf{3'_2}\otimes\mathbf{6}=\mathbf{3_1}\oplus\mathbf{3'_1}\oplus\mathbf{\overline{3}_1}\oplus\mathbf{\overline{3}'_1}\oplus\mathbf{6},\\
\label{eq15}&&\mathbf{6}\otimes\mathbf{6}=\mathbf{1}\oplus\mathbf{1'}\oplus\mathbf{2}_S\oplus\mathbf{2}_A\oplus\mathbf{3_1}\oplus\mathbf{3'_1}\oplus\mathbf{\overline{3}_1}\oplus\mathbf{\overline{3}'_1}\oplus\mathbf{3_2}\oplus\mathbf{3'_2}\oplus\mathbf{6}_S\oplus\mathbf{6}_A\,,
\end{eqnarray}
where the subscripts $``S"$ and $``A"$ indicate the symmetric and antisymmetric properties of the corresponding representation respectively. The product between the singlet $\mathbf{1}$ and any representation $R$ gives $R$. The explicit representation matrices for the generators have been presented in the Escobar-Luhn basis \cite{D96}, they are rather simple. However, the resulting charged lepton mass matrix is non-diagonal for the TFH mixing patterns. We find useful to work in the basis where the representation matrix of the element $a^2cd$ is always diagonal for various representations. The representation matrices for the generators $a$, $b$, $c$ and $d$ in our basis are listed in Appendix A, and the relation with the Escobar-Luhn basis is discussed. The Clebsch-Gordan coefficients for the decomposition of the representation products in our basis are reported in Appendix B.

\begin{table}[hptb!]
\begin{center}
\begin{tabular}{|c|c|c|c|c|c|c|c|c|c|c|}\hline\hline
   &\multicolumn{10}{c|}{\tt Conjugate Classes}\\\cline{2-11}
   &${\cal C}_1$&${\cal C}_2$&${\cal C}_3$&${\cal C}_4$&${\cal
C}_5$ &${\cal C}_6$ &${\cal C}_7$  &${\cal C}_8$  &${\cal C}_9$  &${\cal C}_{10}$  \\\hline

$n_{{\cal C}_i}$&1&3&3&3&6 & 32  & 12 & 12 & 12 & 12 \\\hline

$h_{{\cal C}_i}$&1&4&2&4&4  & 3  & 2 & 8& 4 & 8\\\hline

$\mathbf{1}$&1&1&1&1&1 &1&1&1&1&1\\\hline

$\mathbf{1}'$ &1&1&1&1&1 &1&$-1$&$-1$&$-1$&$-1$ \\\hline

$\mathbf{2}$ & 2 & 2 & 2 & 2 & 2 & $-1$  & 0  & 0  &   0  & 0  \\\hline

$\mathbf{3_1}$& 3 & $-1 + 2 i$ & $-1$ & $-1 - 2 i$ & 1 & 0 & $-1$ & $i$  & 1 & $-i$
\\\hline

$\mathbf{3'_1}$ & 3&$-1 + 2i$ & $-1$ & $-1- 2i$ & 1 & 0 & 1 & $-i$ & $-1$ & $i$ \\\hline

$\mathbf{\overline{3}_1}$& 3 & $-1 - 2 i$  & $-1$  & $-1 + 2 i$  & 1  & 0  & $-1$  & $-i$  & 1  & $i$   \\\hline

$\mathbf{\overline{3}'_1}$& 3 & $-1 - 2 i$  & $-1$ & $-1 + 2 i$ & 1  & 0  & 1  & $i$  & $-1$  & $-i$ \\\hline

$\mathbf{3_2}$& 3  & $-1$  & 3  & $-1$  & $-1$   & 0  & $-1$  & 1  & $-1$  & 1  \\\hline

$\mathbf{3'_2}$& 3 &  $-1$  & 3  & $-1$  & $-1$  & 0  & 1  & $-1$  & 1  & $-1$  \\\hline

$\mathbf{6}$& 6  & 2  & $-2$  & 2  & $-2$  & 0  & 0  & 0  & 0  & 0  \\\hline\hline

\end{tabular}
\caption{\label{tab:character}Character table of the $\Delta(96)$ group,
where $n_{{\cal C}_i}$ denotes the number of the elements contained in the class ${\cal
C}_i$, and $h_{{\cal C}_i}$ is the order of the elements of ${\cal
C}_i$.}
\end{center}
\end{table}

\section{\label{sec:pathway}Pathway to TFH mixing within $\Delta(96)$ in effective theory}

Since the order of $\Delta(96)$ is somewhat larger, it is more complex than the popular flavor symmetries $A_4$, $S_4$ etc. As a result, there are numerous ways to produce the TFH mixing within $\Delta(96)$. In this section, we adopt an effective field theory approach within the so-called minimalist framework \cite{Zee:2005ut}, where the charged lepton masses are generated by the operator of the following form
\begin{equation}
\label{eq16}\mathcal{O}_{\ell}=E^c\ell h_d\phi_{\ell}\,,
\end{equation}
where $E^c$ is the right-handed charged lepton field, $\ell$ is the
lepton doublet field, $h_d$ is the down-type Higgs doublet, and
$\phi_{\ell}$ is the flavon field which breaks the flavor symmetry
$\Delta(96)$ in the charged lepton sector at LO. Neutrino masses are
generated by the high-dimensional Weinberg operator
\begin{equation}
\label{eq17}\mathcal{O}_{\nu}=\ell h_u \ell h_u\phi_{\nu}\,,
\end{equation}
where $h_u$ is the up-type Higgs doublet, and $\phi_{\nu}$ is the flavon field in the neutrino sector. In order to constrain our systematic search, we choose the Higgs fields $h_{u,d}$ are $\Delta(96)$ singlet $\mathbf{1}$, further assume that there is no coupling between the flavon field in the charged lepton sector and that in the neutrino sector at LO, this requirement can be satisfied by imposing an auxiliary $Z_n$ symmetry. We note that the three generation lepton doublets have to transform as a triplet irreducible representation \cite{Lam:2007qc}. If we assign all of them to singlet, the resulting PMNS matrix would be block-diagonal, consequently only two of the three leptons can mix. For the assignment of one doublet plus a singlet, the PMNS matrix would be block-diagonal or has a zero entry. As in the standard procedure, we assign the fields $E^c$, $\ell$, $\phi_{\ell}$ and $\phi_{\nu}$ to various representations of $\Delta(96)$, then write down all the flavor symmetry allowed forms of the operators $\mathcal{O}_{\ell}$ and $\mathcal{O}_{\nu}$. The possible assignments leading to TFH1 mixing are listed in Table \ref{tab:pathway}, it is remarkable that the lepton doublet $\ell$ can not be assigned to the triplet $\mathbf{3_2}$ or $\mathbf{3'_2}$. To generate TFH mixing, the vacuum expectation value (VEV) of $\phi_{\ell}$ should be aligned as follows:
\begin{equation}
\label{eq18}\langle\phi_{\ell}\rangle=\left\{\begin{array}{ll}
(0,0,v),&~~\quad\quad\phi_{\ell}\sim\mathbf{3_1}, \mathbf{3'_1}, \mathbf{\overline{3}_1}, \mathbf{\overline{3}'_1}, \mathbf{3_2}, \mathbf{3'_2}\\
(0,0,v_3,0,0,v_6),&~~\quad\quad\phi_{\ell}\sim\mathbf{6}\,.
\end{array}
\right.
\end{equation}
The vacuum configuration of $\phi_{\nu}$ is
\begin{equation}
\label{eq19}\langle\phi_{\nu}\rangle=\left\{
\begin{array}{ll}
(1,1,1)u,&~~\quad\quad\phi_{\nu}\sim\mathbf{3'_1}, \mathbf{\overline{3}'_1}\\
(u_1,u_2,(u_1+u_2)/2),&~~\quad\quad\phi_{\nu}\sim\mathbf{3'_2}
\end{array}
\right.
\end{equation}
We note that the above vacuum of $\phi_{\ell}$ breaks $\Delta(96)$ into the $Z_3$ subgroup generated by $a^2cd$, and the VEV of $\phi_{\nu}$ breaks $\Delta(96)$ into the $K_4$ subgroup generated by the elements $a^2bd$ and $d^2$. Due to the existence of these residual symmetries, the required vacuum alignment could be consistently realized by the now-standard supersymmetric driving field method \footnote{We omit the related vacuum alignment issue here because of the fine-tuning needed to obtain the phenomenologically acceptable charged lepton mass hierarchies.} \cite{Altarelli:2005yx}.
All the possible assignments shown in Table \ref{tab:pathway} lead to a diagonal charged lepton mass matrix, and the neutrino mass matrix is exactly diagonalized by the TFH1 mixing matrix in Eq.(\ref{eq2}). As a result, the statements of Feruglio et al. in Refs. \cite{Toorop:2011jn,deAdelhartToorop:2011re} are confirmed. In these realizations, the three charged lepton masses are given in terms of three independent parameters. However, in order to match the observed masses $m_e$, $m_{\mu}$ and $m_{\tau}$, we need tuning the parameters such that some sort of cancellation between them happens. Further fine tuning is required if subleading corrections are included. The same situation is encountered in the $A_5$ model building \cite{Ding:2011cm}. To overcome this defect, we should further break the remnant $Z_3$ symmetry in the charged lepton sector. For the TFH2 mixing, we only need to exchange both the position of $\mu^{c}$ and $\tau^{c}$ and the position of the second and third generation lepton doublets, then the assignments in Table \ref{tab:pathway} still work. Finally we comment on the situation in which $\ell$ transforms as triplet and $E^c$ are $\Delta(96)$ singlets, we need to introduce three triplet flavons with the vacuum alignment along $(1,0,0)$, $(0,1,0)$ and $(0,0,1)$ in the charged lepton sector. This type of alignment breaks $\Delta(96)$ completely, therefore it is not easy to be realized consistently.

\begin{table}[t!]
\begin{center}
\begin{tabular}{|c|c|c|c|}  \hline\hline

$\ell$       &      $E^c$      &     $\phi_{\ell}$    &      $\phi_{\nu}$ \\  \hline

\multirow{8}{*}{$\mathbf{3_1}$}  & $\tau^{c}\sim\mathbf{1}$, $(\mu^c, e^c)\sim\mathbf{2}$  &  $\mathbf{\overline{3}_1}$, $\mathbf{\overline{3}'_1}$  &  \multirow{8}{*}{$\mathbf{3'_1}$ , $\mathbf{3'_2}$}  \\ \cline{2-3}

   & $\tau^{c}\sim\mathbf{1'}$, $(\mu^c, e^c)\sim\mathbf{2}$   &    $\mathbf{\overline{3}_1}$, $\mathbf{\overline{3}'_1}$  &   \\  \cline{2-3}
   & $(\mu^c, e^c, \tau^c)\sim\mathbf{3_1}$   &  $\mathbf{3_1}$, $\mathbf{3'_1}$, $\mathbf{3'_2}$   &    \\  \cline{2-3}
   & $(\mu^c, e^c, \tau^c)\sim\mathbf{3'_1}$  &  $\mathbf{3_1}$, $\mathbf{3'_1}$, $\mathbf{3_2}$    &    \\   \cline{2-3}
   & $(e^c, \mu^c, \tau^c)\sim\mathbf{\overline{3}_1}$   &   $\mathbf{1}$, $\mathbf{6}$  &    \\   \cline{2-3}
   & $(e^c, \mu^c, \tau^c)\sim \mathbf{\overline{3}'_1}$  &  $\mathbf{1'}$, $\mathbf{6}$  &    \\   \cline{2-3}
   &  $(\mu^c, e^c, \tau^c)\sim\mathbf{3_2}$    &   $\mathbf{3'_1}$, $\mathbf{6}$  &   \\    \cline{2-3}
   &  $(\mu^c, e^c, \tau^c)\sim\mathbf{3'_2}$    &   $\mathbf{3_1}$, $\mathbf{6}$  &   \\    \hline

\multirow{8}{*}{$\mathbf{3'_1}$}  &  $\tau^{c}\sim\mathbf{1}$, $(\mu^c, e^c)\sim\mathbf{2}$  &  $\mathbf{\overline{3}_1}$, $\mathbf{\overline{3}'_1}$  & \multirow{8}{*}{$\mathbf{3'_1}$ , $\mathbf{3'_2}$}  \\  \cline{2-3}

 & $\tau^{c}\sim\mathbf{1'}$, $(\mu^c, e^c)\sim\mathbf{2}$   &    $\mathbf{\overline{3}_1}$, $\mathbf{\overline{3}'_1}$  &   \\  \cline{2-3}
 & $(\mu^c, e^c, \tau^c)\sim\mathbf{3_1}$   &  $\mathbf{3_1}$, $\mathbf{3'_1}$, $\mathbf{3_2}$   &    \\  \cline{2-3}
 & $(\mu^c, e^c, \tau^c)\sim\mathbf{3'_1}$  &  $\mathbf{3_1}$, $\mathbf{3'_1}$, $\mathbf{3'_2}$    &    \\   \cline{2-3}
 & $(e^c, \mu^c, \tau^c)\sim\mathbf{\overline{3}_1}$   &   $\mathbf{1'}$, $\mathbf{6}$  &    \\   \cline{2-3}
 & $(e^c, \mu^c, \tau^c)\sim \mathbf{\overline{3}'_1}$  &  $\mathbf{1}$, $\mathbf{6}$  &    \\   \cline{2-3}
 &  $(\mu^c, e^c, \tau^c)\sim\mathbf{3_2}$    &   $\mathbf{3_1}$, $\mathbf{6}$  &   \\    \cline{2-3}
 &  $(\mu^c, e^c, \tau^c)\sim\mathbf{3'_2}$    &   $\mathbf{3'_1}$, $\mathbf{6}$  &   \\    \hline

\multirow{8}{*}{$\mathbf{\overline{3}_1}$}  & $\tau^{c}\sim\mathbf{1}$, $(e^c, \mu^c)\sim\mathbf{2}$  &  $\mathbf{3_1}$, $\mathbf{3'_1}$  & \multirow{8}{*}{$\mathbf{\overline{3}'_1}$ , $\mathbf{3'_2}$}  \\  \cline{2-3}

  & $\tau^{c}\sim\mathbf{1'}$, $(e^c, \mu^c)\sim\mathbf{2}$  &  $\mathbf{3_1}$, $\mathbf{3'_1}$  &   \\  \cline{2-3}
  & $(e^c, \mu^c, \tau^c)\sim\mathbf{3_1}$   &  $\mathbf{1}$, $\mathbf{6}$   &    \\  \cline{2-3}
  & $(e^c, \mu^c, \tau^c)\sim\mathbf{3'_1}$   &  $\mathbf{1'}$, $\mathbf{6}$   &    \\  \cline{2-3}
  & $(\mu^c, e^c, \tau^c)\sim\mathbf{\overline{3}_1}$   &   $\mathbf{\overline{3}_1}$, $\mathbf{\overline{3}'_1}$, $\mathbf{3'_2}$  &    \\   \cline{2-3}
  & $(\mu^c, e^c, \tau^c)\sim\mathbf{\overline{3}'_1}$   &   $\mathbf{\overline{3}_1}$, $\mathbf{\overline{3}'_1}$, $\mathbf{3_2}$  &    \\   \cline{2-3}
  &  $(e^c, \mu^c, \tau^c)\sim\mathbf{3_2}$    &   $\mathbf{\overline{3}'_1}$, $\mathbf{6}$  &   \\    \cline{2-3}
  &  $(e^c, \mu^c, \tau^c)\sim\mathbf{3'_2}$    &   $\mathbf{\overline{3}_1}$, $\mathbf{6}$  &   \\    \hline

\multirow{8}{*}{$\mathbf{\overline{3}'_1}$}  & $\tau^{c}\sim\mathbf{1}$, $(e^c, \mu^c)\sim\mathbf{2}$  &  $\mathbf{3_1}$, $\mathbf{3'_1}$  & \multirow{8}{*}{$\mathbf{\overline{3}'_1}$ , $\mathbf{3'_2}$}  \\  \cline{2-3}

  & $\tau^{c}\sim\mathbf{1'}$, $(e^c, \mu^c)\sim\mathbf{2}$  &  $\mathbf{3_1}$, $\mathbf{3'_1}$  &  \\  \cline{2-3}
  & $(e^c, \mu^c, \tau^c)\sim\mathbf{3_1}$   &  $\mathbf{1'}$, $\mathbf{6}$   &    \\  \cline{2-3}
  & $(e^c, \mu^c, \tau^c)\sim\mathbf{3'_1}$   &  $\mathbf{1}$, $\mathbf{6}$   &    \\  \cline{2-3}
  & $(\mu^c, e^c, \tau^c)\sim\mathbf{\overline{3}_1}$   &   $\mathbf{\overline{3}_1}$, $\mathbf{\overline{3}'_1}$, $\mathbf{3_2}$  &    \\   \cline{2-3}
  & $(\mu^c, e^c, \tau^c)\sim\mathbf{\overline{3}'_1}$   &   $\mathbf{\overline{3}_1}$, $\mathbf{\overline{3}'_1}$, $\mathbf{3'_2}$  &    \\   \cline{2-3}
  &  $(e^c, \mu^c, \tau^c)\sim\mathbf{3_2}$    &   $\mathbf{\overline{3}_1}$, $\mathbf{6}$  &   \\    \cline{2-3}
  &  $(e^c, \mu^c, \tau^c)\sim\mathbf{3'_2}$    &   $\mathbf{\overline{3}'_1}$, $\mathbf{6}$  &   \\    \hline\hline
\end{tabular}
\caption{\label{tab:pathway}Possible assignments of the fields $E^c$, $\ell$, $\phi_{\ell}$ and $\phi_{\nu}$ for the TFH1 mixing, where $\ell=(\ell_1,\ell_2,\ell_3)$ is the lepton doublet field.}
\end{center}
\end{table}

\section{\label{sec:TFH1} Model for THF1 mixing}

\begin{table}[hptb]
\begin{center}
\begin{tabular}{|c|c|c|c|c|c|c||c|c|c|c|c|c|c||c|c|c|c|c|c|c||c|c|c|c|c|c|}\hline\hline
{\tt Fields} & $\ell$ & $e^c$  &  $\mu^c$ &  $\tau^c$  & $\nu^{c}$ &$h_{u,d}$ &  $\chi$  & $\phi$ & $\eta$ &$\xi$  &  $\rho$ & $\varphi$ & $\psi$ & $\sigma^0$ & $\zeta^0$  &  $\chi^0$ &  $\phi^0$  & $\rho^{0}$ &$\varphi^0$
\\\hline

$\Delta(96)$  & $\mathbf{3_1}$ &  $\mathbf{1}$  &  $\mathbf{1'}$ & $\mathbf{1}$  & $\mathbf{\overline{3}_1}$ & $\mathbf{1}$ & $\mathbf{3_1}$ &  $\mathbf{\overline{3}_1}$  &  $\mathbf{2}$ & $\mathbf{3'_1}$  &  $\mathbf{2}$  &  $\mathbf{\overline{3}'_1}$  & $\mathbf{3'_2}$  & $\mathbf{1}$ &  $\mathbf{2}$  &  $\mathbf{3_1}$ & $\mathbf{3'_2}$  & $\mathbf{2}$ & $\mathbf{3_1}$ \\

$Z_3$ & 0 & 2  & 2& 2 & 0  & 0  &  1 & 1 & 0 & 0  &  0  &  0  & 0  & 0 & 1  & 2 & 1  & 0  & 0   \\

$Z_3$ & 0 & 1  &  2  & 0 & 0 & 0 &  0 &  0  &  1  & 1 &  0 &   0  & 0  & 1  & 0  &  2  & 0 & 0 &  0  \\\hline\hline

\end{tabular}
\caption{\label{tab:field1} The transformation properties of the matter fields, the electroweak Higgs doublets, the flavon fields and the driving fields under the flavor symmetry $\Delta(96)\times Z_3\times Z_3$.}
\end{center}
\end{table}

As has been demonstrated in the previous section, the TFH1 mixing can be achieved without fine-tuning. However, these models have a drawback in that a separate fine-tuning of the charged lepton mass parameters is needed at leading order to yield the phenomenologically acceptable mass hierarchies, and further fine tuning is required if the subleading corrections are included. To improve upon this situation, we constructed a $\Delta(96)$ model which overcomes this defect by breaking the residual $Z_3$ symmetry in the charged lepton sector. We formulate our model in the framework of type I see-saw mechanism \cite{SeeSaw}, and supersymmetry (SUSY) is introduced to simplify the discussion of the vacuum alignment. Obviously this model doesn't belong to the possible realizations of TFH mixing listed in Table \ref{tab:pathway}. We assign the three generations of left-handed (LH) lepton doublets $\ell$ and of right-handed (RH) neutrinos $\nu^c$ to $\Delta(96)$ triplets $\mathbf{3_1}$ and $\mathbf{\overline{3}_1}$, while the RH charged leptons $e^c$, $\mu^c$ and $\tau^c$ transform as $\mathbf{1}$, $\mathbf{1'}$ and $\mathbf{1}$ respectively. The flavon sector consists of two set of fields: $\chi\sim\mathbf{3_1}$, $\phi\sim\mathbf{\overline{3}_1}$, $\eta\sim\mathbf{2}$ and $\xi\sim\mathbf{3'_1}$ which couple to the charged leptons at LO, and $\rho\sim\mathbf{2}$, $\varphi\sim\mathbf{\overline{3}'_1}$ and $\psi\sim\mathbf{3'_2}$ which couple to the neutrino sector. Additional symmetries are needed, in general, to prevent unwanted couplings, to ensure the needed vacuum alignment and to reproduce the observed charged lepton mass hierarchies. In our model, the full flavor symmetry is $\Delta(96)\times Z_3\times Z_3$. The fields in the model and their classifications under the flavor symmetry are summarized in Table \ref{tab:field1}. A $U(1)_R$ symmetry related to $R-$parity and the presence of driving fields in the flavon superpotential are common features of supersymmetric formulations. The flavon and Higgs fields are uncharged under $U(1)_R$, the matter fields have $R=1$, the so-called driving fields carry 2 units of $R$ charge. Consequently the driving fields enter linearly into the superpotential. The suitable driving fields and their transformation properties are shown in Table \ref{tab:field1}. The LO driving superpotential $w_d$, which is linear in the driving fields and invariant under the flavor symmetry $\Delta(96)\times Z_3\times Z_3$, is given by
\begin{eqnarray}
\nonumber&&w_d=f_1\sigma^0(\eta\eta)+f_2(\zeta^0(\chi\phi)_{\mathbf{2}})+f_3(\chi^0(\eta\phi)_{\mathbf{\overline{3}_1}})+f_4(\chi^0(\chi\xi)_{\mathbf{\overline{3}_1}})
+f_5(\phi^0(\chi\chi)_{\mathbf{3'_2}})\\
\label{eq20}&&\quad\quad+f_6(\phi^0(\phi\phi)_{\mathbf{3'_2}})+M_{\rho}(\rho^0\rho)+g_1(\rho^0(\rho\rho)_{\mathbf{2}})+g_2(\rho^0(\psi\psi)_{\mathbf{2}})+g_3(\varphi^0(\rho\varphi)_{\mathbf{\overline{3}_1}})\,,
\end{eqnarray}
where we indicate with $(\ldots)$ the singlet $\mathbf{1}$
contraction, with $(\ldots)'$ the singlet $\mathbf{1'}$ and with
$(\ldots)_R$ ($R=\mathbf{2}, \mathbf{3_1}, \mathbf{3'_1}$ etc.) the
representation $R$. The charge assignments in Table \ref{tab:field1}
allow us to separate the above driving superpotential $w_d$ into two
decoupled sets: one is associated with the fields $\chi$, $\phi$,
$\eta$ and $\xi$ which are relevant to the charged lepton sector,
and another one is associated with $\rho$, $\varphi$ and $\psi$
which are relevant to the neutrino masses. In the SUSY limit, the
vacuum configuration is determined by the vanishing of the $F$ term
associated with each component of the driving fields. In the charged
lepton sector, the minimization conditions are as follows,
\begin{eqnarray}
\nonumber&&\frac{\partial w_d}{\partial\sigma^0}=2f_1\eta_1\eta_2=0  \\
\nonumber&&\frac{\partial w_d}{\partial \zeta^0_1}=\omega f_2(\chi_1\phi_2+\chi_2\phi_3+\chi_3\phi_1)=0  \\
\nonumber&&\frac{\partial w_d}{\partial\zeta^0_2}=f_2(\chi_1\phi_3+\chi_2\phi_1+\chi_3\phi_2)=0 \\
\nonumber&&\frac{\partial w_d}{\partial\chi^0_1}=f_3(\eta_1\phi_2+\omega^2\eta_2\phi_3)+f_4(-2\chi_1\xi_1+\chi_2\xi_3+\chi_3\xi_2)=0 \\
\nonumber&&\frac{\partial w_d}{\partial \chi^0_2}=f_3(\eta_1\phi_3+\omega^2\eta_2\phi_1)+f_4(\chi_1\xi_3-2\chi_2\xi_2+\chi_3\xi_1)=0\\
\nonumber&&\frac{\partial w_d}{\partial\chi^0_3}=f_3(\eta_1\phi_1+\omega^2\eta_2\phi_2)+f_4(\chi_1\xi_2+\chi_2\xi_1-2\chi_3\xi_3)=0  \\
\nonumber&&\frac{\partial w_d}{\partial\phi^0_1}=f_5(\chi^2_1+2\chi_2\chi_3)+f_6(\phi^2_2+2\phi_1\phi_3)=0  \\
\nonumber&&\frac{\partial w_d}{\partial\phi^0_2}=f_5(\chi^2_2+2\chi_1\chi_3)+f_6(\phi^2_1+2\phi_2\phi_3)=0  \\
\label{eq21}&&\frac{\partial w_d}{\partial\phi^0_3}=f_5(\chi^2_3+2\chi_1\chi_2)+f_6(\phi^2_3+2\phi_1\phi_2)=0\,,
\end{eqnarray}
where $\omega=e^{2\pi i/3}$ is the cube root of unit. This set of equations admit the solution
\begin{eqnarray}
\label{eq22}&&\langle\chi\rangle=(0,0,v_{\chi}),~~~\langle\phi\rangle=(0,0,v_{\phi}),~~~\langle\eta\rangle=(v_{\eta},0),~~~\langle\xi\rangle=(v_{\xi},0,0)\,,
\end{eqnarray}
where the VEVs obey the relations
\begin{equation}
\label{eq23}v^2_{\chi}=-\frac{f_6}{f_5}v^2_{\phi},~~~v^2_{\xi}=-\frac{f^2_3f_5}{f^2_4f_6}v^2_{\eta}\,,
\end{equation}
with $v_{\phi}$ and $v_{\eta}$ undetermined. We note that the VEVs of $\chi$ and $\phi$ are invariant under the action of $a^2cd$, they break the $\Delta(96)$ flavor symmetry into $Z_3$, while the VEVs of $\eta$ and $\xi$ break $\Delta(96)$ completely. The vacuum configuration of $\rho$, $\varphi$ and $\psi$, which gives rise to TFH1 mixing in the neutrino sector, is determined by the following minimization equations
\begin{eqnarray}
\nonumber&&\frac{\partial w_d}{\partial\rho^0_1}=M_{\rho}\rho_2+g_1\rho^2_1+\omega g_2(\psi^2_1+2\psi_2\psi_3)=0  \\
\nonumber&&\frac{\partial w_d}{\partial\rho^0_2}=M_{\rho}\rho_1+g_1\rho^2_2+g_2(\psi^2_2+2\psi_1\psi_3)=0   \\
\nonumber&&\frac{\partial w_d}{\partial\varphi^0_1}=g_3(\rho_1\varphi_2-\omega^2\rho_2\varphi_3)=0  \\
\nonumber&&\frac{\partial w_d}{\partial\varphi^0_2}=g_3(\rho_1\varphi_3-\omega^2\rho_2\varphi_1)=0  \\
\label{eq24}&&\frac{\partial w_d}{\partial\varphi^0_3}=g_3(\rho_1\varphi_1-\omega^2\rho_2\varphi_2)=0
\end{eqnarray}
The non-trivial solution to the above equations is
\begin{equation}
\label{eq25}\langle\rho\rangle=(1,\omega)v_{\rho},~~~\langle\varphi\rangle=(1,1,1)v_{\varphi},~~~\langle\psi\rangle=(v_1,v_2,(v_1+v_2)/2)
\end{equation}
with the condition
\begin{equation}
\label{eq26}M_{\rho}v_{\rho}+\omega^2g_1v^2_{\rho}+g_2(v^2_1+v^2_2+v_1v_2)=0
\end{equation}
The non-vanishing VEVs of $\rho$, $\varphi$ and $\psi$ in
Eq.(\ref{eq25}) is the most general vacuum configuration invariant
under the $K_4$ subgroup comprised of $a^2bd$, $d^2$, $a^2bd^3$ and
$1$. Starting from the field configurations of Eq.(\ref{eq22}) and
Eq.(\ref{eq25}) and acting on them with elements of the flavor
symmetry group $\Delta(96)$, we can generate other minima of the
scalar potential. However, these new minima are physically
equivalent to the original one, they all lead to the same physics,
i.e., lepton masses and flavor mixing. Without loss of generality,
we can analyze the model by choosing the vacuum in Eq.(\ref{eq22})
and Eq.(\ref{eq25}) as the local minimum. It is important to check
the stability of this LO vacuum configuration, if we introduce small
perturbations to the VEVs of the flavon fields as follows,
\begin{eqnarray}
\nonumber&&\langle\chi\rangle=(\delta\chi_1,\delta\chi_2,v_{\chi}+\delta\chi_3),~~~\langle\phi\rangle=(\delta\phi_1,\delta\phi_2,v_{\phi}),~~~\langle\eta\rangle=(v_{\eta},\delta\eta_2),\\
\nonumber&&\langle\xi\rangle=(v_{\xi}+\delta\xi_1,\delta\xi_2,\delta\xi_3),~~~\langle\rho\rangle=(v_{\rho}+\delta\rho_1,\omega v_{\rho}+\delta\rho_2),\\
\label{eq27}&&\langle\varphi\rangle=(v_{\varphi}+\delta\varphi_1,v_{\varphi}+\delta\varphi_2,v_{\varphi}),~~~\langle\psi\rangle=(v_1,v_2,v_1/2+v_2/2+\delta\psi_3)\,.
\end{eqnarray}
After some straightforward algebra, we find that the only solution to the minimization equations is
\begin{eqnarray}
\nonumber&&\delta\chi_1=\delta\chi_2=\delta\chi_3=0,~~~\delta\phi_1=\delta\phi_2=0,~~~\delta\eta_2=0,~~~\delta\xi_1=\delta\xi_2=\delta\xi_3=0,\\
\label{eq28}&&\delta\rho_1=\delta\rho_2=0,~~~\delta\varphi_1=\delta\varphi_2=0,~~~\delta\psi_3=0
\end{eqnarray}
Therefore the above LO vacuum alignment is stable. We turn now to the Yukawa superpotential for the charged leptons, which is given by
\begin{eqnarray}
\nonumber&&w_{\ell}=\frac{y_{\tau}}{\Lambda}\tau^c(\ell\phi)h_d+\frac{y_{\mu_1}}{\Lambda^2}\mu^c(\ell(\chi\xi)_{\mathbf{\overline{3}'_1}})'h_d
+\frac{y_{\mu_2}}{\Lambda^2}\mu^{c}(\ell(\eta\phi)_{\mathbf{\overline{3}'_1}})'h_d+\frac{y_{e_1}}{\Lambda^3}e^c(\ell\phi)(\eta\eta)h_d \\
\nonumber&&\quad\quad+\frac{y_{e_2}}{\Lambda^3}e^c(\ell((\eta\eta)_{\mathbf{2}}\phi)_{\mathbf{\overline{3}_1}})h_d+\frac{y_{e_3}}{\Lambda^3}e^c(\ell(\chi(\eta\xi)_{\mathbf{3_1}})_{\mathbf{\overline{3}_1}})h_d
+\frac{y_{e_4}}{\Lambda^3}e^c(\ell(\chi(\eta\xi)_{\mathbf{3'_1}})_{\mathbf{\overline{3}_1}})h_d\\
\label{eq29}&&\quad\quad+\frac{y_{e_5}}{\Lambda^3}e^c(\ell(\chi(\xi\xi)_{\mathbf{3'_2}})_{\mathbf{\overline{3}_1}})h_d+\ldots\,,
\end{eqnarray}
where dots stand for higher-dimensional operators which will be
discussed later. It is remarkable that the electron, muon and tau
mass terms are suppressed by $1/\Lambda$, $1/\Lambda^2$ and
$1/\Lambda^3$ respectively. At LO, only the tau mass is generated,
the flavor symmetry $\Delta(96)$ is broken into $Z_3$ by the VEV of
$\phi$, the remaining terms further break this $Z_3$ symmetry
completely. With the vacuum alignment in Eq.(\ref{eq22}), $w_{\ell}$
leads to a diagonal charged lepton mass matrix:
\begin{equation}
\label{eq30}m_{\ell}=\left(\begin{array}{ccc}
\omega^2y_{e_2}\frac{v^2_{\eta}v_{\phi}}{\Lambda^3}+(y_{e_4}-y_{e_3})\frac{v_{\eta}v_{\xi}v_{\chi}}{\Lambda^3}+y_{e_5}\frac{v^2_{\xi}v_{\chi}}{\Lambda^3} &0  &0 \\
0 & y_{\mu_1}\frac{v_{\xi}v_{\chi}}{\Lambda^2}+y_{\mu_2}\frac{v_{\eta}v_{\phi}}{\Lambda^2}  &  0  \\
0 & 0  & y_{\tau}\frac{v_{\phi}}{\Lambda}
\end{array}\right)v_{d}\,,
\end{equation}
where $v_d=\langle h_d\rangle$. Clearly the mass hierarchies of the charged leptons are naturally recovered if the VEVs $v_{\chi}/\Lambda$, $v_{\phi}/\Lambda$, $v_{\eta}/\Lambda$ and $v_{\xi}/\Lambda$ are of order $\lambda^2_c$. We note that the mass hierarchies are determined by the flavor symmetry itself without invoking the Froggatt-Nielsen mechanism \cite{FN}. The superpotential for the neutrino sector can be written as
\begin{eqnarray}
\label{eq31}&&w_{\nu}=y(\nu^c\ell)h_u+x_{\nu_1}((\nu^c\nu^c)_{\mathbf{3'_1}}\varphi)+x_{\nu_2}((\nu^c\nu^c)_{\mathbf{3'_2}}\psi)+\ldots
\end{eqnarray}
The Dirac neutrino mass matrix as obtained from the first term of Eq. (\ref{eq31}) is given by
\begin{equation}
\label{eq32}m_D=yv_u\mathbb{1}\,,
\end{equation}
with $v_u=\langle h_u\rangle$. Given the flavon VEVs of Eq.(\ref{eq25}), the remaining two terms in Eq.(\ref{eq31}) give rise to the Majorana neutrino mass matrix as follows
\begin{equation}
\label{eq33}m_M=\left(\begin{array}{ccc}
-4x_{\nu_1}v_{\varphi}+2x_{\nu_2}v_2  & 2x_{\nu_1}v_{\varphi}+x_{\nu_2}(v_1+v_2)   &   2x_{\nu_1}v_{\varphi}+2x_{\nu_2}v_1  \\
2x_{\nu_1}v_{\varphi}+x_{\nu_2}(v_1+v_2)  &  -4x_{\nu_1}v_{\varphi}+2x_{\nu_2}v_1   &  2x_{\nu_1}v_{\varphi}+2x_{\nu_2}v_2 \\
2x_{\nu_1}v_{\varphi}+2x_{\nu_2}v_1   &  2x_{\nu_1}v_{\varphi}+2x_{\nu_2}v_2    &  -4x_{\nu_1}v_{\varphi}+x_{\nu_2}(v_1+v_2)
\end{array}\right)\,.
\end{equation}
Integrating out the heavy degrees of freedom, we obtain the light neutrino mass matrix, which is given by the see-saw formula
\begin{equation}
\label{eq34}m_{\nu}=-m^T_Dm^{-1}_Mm_D=U_{TFH1}\mathrm{diag}(m_1,m_2,m_3)U^{T}_{TFH1}
\end{equation}
and the light neutrino masses are
\begin{eqnarray}
\label{eq35}m_1=\frac{y^2v^2_u}{6x_{\nu_1}v_{\varphi}+\sqrt{3}\,x_{\nu_2}(v_1-v_2)},~m_2=-\frac{y^2v^2_u}{3x_{\nu_2}(v_1+v_2)},~m_3=\frac{y^2v^2_u}{6x_{\nu_1}v_{\varphi}-\sqrt{3}\,x_{\nu_2}(v_1-v_2)}
\end{eqnarray}
There are no special relations among the above three light neutrino masses, thus we have no prediction for the neutrino mass spectrum, which can be both normal and inverted order hierarchy.

\subsection{Beyond the leading order}

In this section, we address the next-to-leading corrections to the
LO results presented above. The NLO corrections are indicated by the
subleading higher-dimensional operators in the $1/\Lambda$
expansion, which are compatible with the symmetries of the model. We
start with the corrections to $w_d$ which is modified into
\begin{equation}
\label{eq36}w_d=w^0_d+\delta w_d\,,
\end{equation}
where $w^0_d$ is the leading order contribution given in Eq.(\ref{eq20}), and $\delta w_d$ denotes the NLO terms, it is suppressed by one additional power of $1/\Lambda$ with respect to $w^0_d$. The correction terms included in $\delta w_d$ consist of the most general quartic, $\Delta(96)\times Z_3\times Z_3$ invariant polynomial linear in the driving fields. Since the flavons $\rho$, $\varphi$ and $\psi$ are neutral under the auxiliary symmetry $Z_3\times Z_3$, we can insert one of them into the LO terms. Concretely, $\delta w_d$ can be written as :
\begin{equation}
\label{eq37}\delta w_d=\frac{1}{\Lambda}\sum^3_{i=1}s_iI^{\sigma^0}_i+\frac{1}{\Lambda}\sum^{6}_{i=1}z_iI^{\zeta^0}_i+\frac{1}{\Lambda}\sum^9_{i=1}k_iI^{\chi^0}_i
+\frac{1}{\Lambda}\sum^{8}_{i=1}p_iI^{\phi^0}_i+\frac{1}{\Lambda}\sum^{22}_{i=1}r_iI^{\rho^0}_i+\frac{1}{\Lambda}\sum^{19}_{i=1}w_iI^{\varphi^0}_i
\end{equation}
where $s_i$, $z_i$, $k_i$, $p_i$, $r_i$ and $w_i$ are complex coefficients with absolute value of order 1, $I^{\sigma^0}_i$, $I^{\zeta^0}_i$, $I^{\chi^0}_i$, $I^{\phi^0}_i$, $I^{\rho^0}_i$ and $I^{\varphi^0}_i$ denote a basis of independent quartic invariants:
\begin{equation}
\label{eq38}\hskip-2.5cm\begin{array}{lll} I^{\sigma^0}_1=\sigma^0((\eta\eta)_{\mathbf{2}}\rho),~~~ & I^{\sigma^0}_2=\sigma^0((\xi\xi)_{\mathbf{3'_2}}\psi),~~~ &I^{\sigma^0}_3=\sigma^0((\eta\xi)_{\mathbf{3'_1}}\varphi)
\end{array}
\end{equation}\vskip-1.0cm

\begin{equation}
\label{eq39}\begin{array}{lll}
I^{\zeta^0}_1=(\zeta^0((\phi\phi)_{\mathbf{3_1}}\varphi)_{\mathbf{2}}),~~~ & I^{\zeta^0}_2=(\zeta^0((\phi\phi)_{\mathbf{3'_1}}\varphi)_{\mathbf{2}}),
~~~ & I^{\zeta^0}_3=(\zeta^0((\phi\phi)_{\mathbf{3'_2}}\psi)_{\mathbf{2}}),\\
I^{\zeta^0}_4=(\zeta^0((\chi\chi)_{\mathbf{3'_2}}\psi)_{\mathbf{2}}),~~~ & I^{\zeta^0}_5=(\zeta^0\rho)(\chi\phi),~~~ & I^{\zeta^0}_6=(\zeta^0((\chi\phi)_{\mathbf{2}}\rho)_{\mathbf{2}})
\end{array}
\end{equation}\vskip-0.8cm

\begin{equation}
\label{eq40}\hskip1.0cm\begin{array}{lll}
I^{\chi^0}_1=(\chi^0((\eta\chi)_{\mathbf{3_1}}\psi)_{\mathbf{\overline{3}_1}}),~~~ & I^{\chi^0}_2=(\chi^0(\rho(\eta\phi)_{\mathbf{\overline{3}_1}})_{\mathbf{\overline{3}_1}}),~~~ &
I^{\chi^0}_3=(\chi^0(\rho(\chi\xi)_{\mathbf{\overline{3}_1}})_{\mathbf{\overline{3}_1}}),\\
I^{\chi^0}_4=(\chi^0(\rho(\eta\phi)_{\mathbf{\overline{3}'_1}})_{\mathbf{\overline{3}_1}}),~~~ & I^{\chi^0}_5=(\chi^0(\rho(\chi\xi)_{\mathbf{\overline{3}'_1}})_{\mathbf{\overline{3}_1}}),
~~~ & I^{\chi^0}_6=(\chi^0\varphi)'(\xi\phi)',\\
I^{\chi^0}_7=(\chi^0((\xi\phi)_{\mathbf{2}}\varphi)_{\mathbf{\overline{3}_1}}),~~~ & I^{\chi^0}_8=(\chi^0(\varphi(\xi\phi)_{\mathbf{6}})_{\mathbf{\overline{3}_1}}),
~~~ & I^{\chi^0}_9=(\chi^0(\psi(\xi\phi)_{\mathbf{6}})_{\mathbf{\overline{3}_1}})
\end{array}
\end{equation}\vskip-0.8cm

\begin{eqnarray}
\label{eq41}\hskip0.3cm\begin{array}{lll}
I^{\phi^0}_1=(\phi^0\psi)(\chi\phi),~~~ & I^{\phi^0}_2=(\phi^0((\chi\phi)_{\mathbf{2}}\psi)_{\mathbf{3'_2}}),~~~ &
I^{\phi^0}_3=(\phi^0((\chi\chi)_{\mathbf{\overline{3}'_1}}\varphi)_{\mathbf{3'_2}}),\\
I^{\phi^0}_4=(\phi^0(\rho(\chi\chi)_{\mathbf{3'_2}})_{\mathbf{3'_2}}),~~~ & I^{\phi^0}_5=(\phi^0(\rho(\phi\phi)_{\mathbf{3'_2}})_{\mathbf{3'_2}}),
~~~ & I^{\phi^0}_6=(\phi^0(\psi(\chi\chi)_{\mathbf{3'_2}})_{\mathbf{3'_2}}),\\
I^{\phi^0}_7=(\phi^0(\psi(\phi\phi)_{\mathbf{3'_2}})_{\mathbf{3'_2}}),~~~ & I^{\phi^0}_8=(\phi^0(\varphi(\chi\phi)_{\mathbf{6}})_{\mathbf{3'_2}}) &
\end{array}
\end{eqnarray}\vskip-0.8cm

\begin{eqnarray}
\label{eq42}\hskip-0.2cm\begin{array}{lll}
I^{\rho^0}_1=(\rho^0\rho)(\rho\rho),~~~ & I^{\rho^0}_2=(\rho^0\rho)(\psi\psi),~~~ & I^{\rho^0}_3=(\rho^0\rho)'(\rho\rho)',\\
I^{\rho^0}_4=(\rho^0(\rho(\rho\rho)_{\mathbf{2}})_{\mathbf{2}}),~~~ & I^{\rho^0}_5=(\rho^0(\rho(\psi\psi)_{\mathbf{2}})_{\mathbf{2}}),~~~ & I^{\rho^0}_6=(\rho^0((\varphi\varphi)_{\mathbf{3_1}}\varphi)_{\mathbf{2}}),  \\
I^{\rho^0}_7=(\rho^0((\varphi\varphi)_{\mathbf{3'_1}}\varphi)_{\mathbf{2}}),~~~ & I^{\rho^0}_8=(\rho^0((\varphi\psi)_{\mathbf{3'_1}}\varphi)_{\mathbf{2}}),~~~ & I^{\rho^0}_9=(\rho^0((\psi\psi)_{\mathbf{3_2}}\psi)_{\mathbf{2}}), \\
I^{\rho^0}_{10}=(\rho^0((\rho\psi)_{\mathbf{3_2}}\psi)_{\mathbf{2}}),~~~ & I^{\rho^0}_{11}=(\rho^0((\varphi\varphi)_{\mathbf{3'_2}}\psi)_{\mathbf{2}}),~~~ & I^{\rho^0}_{12}=(\rho^0((\psi\psi)_{\mathbf{3'_2}}\psi)_{\mathbf{2}}),   \\
I^{\rho^0}_{13}=(\rho^0((\rho\psi)_{\mathbf{3'_2}}\psi)_{\mathbf{2}}),~~~ & I^{\rho^0}_{14}=(\rho^0(\chi(\chi\chi)_{\mathbf{\overline{3}_1}})_{\mathbf{2}}),~~~ & I^{\rho^0}_{15}=(\rho^0(\chi(\chi\chi)_{\mathbf{\overline{3}'_1}})_{\mathbf{2}}),  \\
I^{\rho^0}_{16}=(\rho^0((\phi\phi)_{\mathbf{3_1}}\phi)_{\mathbf{2}}),~~~ & I^{\rho^0}_{17}=(\rho^0((\phi\phi)_{\mathbf{3'_1}}\phi)_{\mathbf{2}}),~~~ & I^{\rho^0}_{18}=(\rho^0\eta)(\eta\eta),\\
I^{\rho^0}_{19}=(\rho^0\eta)'(\eta\eta)',~~~ &  I^{\rho^0}_{20}=(\rho^0(\eta(\eta\eta)_{\mathbf{2}})_{\mathbf{2}}),~~~ & I^{\rho^0}_{21}=(\rho^0(\xi(\xi\xi)_{\mathbf{\overline{3}_1}})_{\mathbf{2}}),\\
I^{\rho^0}_{22}=(\rho^0(\xi(\xi\xi)_{\mathbf{\overline{3}'_1}})_{\mathbf{2}}) &  &
\end{array}
\end{eqnarray}\vskip-0.8cm

\begin{equation}
\label{eq43}\hskip0.9cm\begin{array}{lll}
I^{\varphi^0}_1=(\varphi^0(\rho(\rho\varphi)_{\mathbf{\overline{3}_1}})_{\mathbf{\overline{3}_1}}),~~~ &I^{\varphi^0}_2=(\varphi^0(\rho(\rho\varphi)_{\mathbf{\overline{3}'_1}})_{\mathbf{\overline{3}_1}}),
~~~ & I^{\varphi^0}_3=(\varphi^0\varphi)'(\rho\rho)', \\
I^{\varphi^0}_4=(\varphi^0((\rho\rho)_{\mathbf{2}}\varphi)_{\mathbf{\overline{3}_1}}),~~~ & I^{\varphi^0}_5=(\varphi^0((\psi\psi)_{\mathbf{2}}\varphi)_{\mathbf{\overline{3}_1}}),
~~~ & I^{\varphi^0}_6=(\varphi^0(\varphi(\varphi\psi)_{\mathbf{6}})_{\mathbf{\overline{3}_1}}), \\
I^{\varphi^0}_7=(\varphi^0((\varphi\varphi)_{\mathbf{3_1}}\psi)_{\mathbf{\overline{3}_1}}),~~~ & I^{\varphi^0}_8=(\varphi^0(\psi(\varphi\psi)_{\mathbf{6}})_{\mathbf{\overline{3}_1}}),
~~~ & I^{\varphi^0}_9=(\varphi^0(\chi(\phi\phi)_{\mathbf{3_1}})_{\mathbf{\overline{3}_1}}),  \\
I^{\varphi^0}_{10}=(\varphi^0(\chi(\phi\phi)_{\mathbf{3'_1}})_{\mathbf{\overline{3}_1}}),~~~ & I^{\varphi^0}_{11}=(\varphi^0(\chi(\chi\chi)_{\mathbf{3'_2}})_{\mathbf{\overline{3}_1}}),
~~~ & I^{\varphi^0}_{12}=(\varphi^0(\chi(\phi\phi)_{\mathbf{3'_2}})_{\mathbf{\overline{3}_1}}),  \\
I^{\varphi^0}_{13}=(\varphi^0\phi)(\chi\phi),~~~ & I^{\varphi^0}_{14}=(\varphi^0((\chi\phi)_{\mathbf{2}}\phi)_{\mathbf{\overline{3}_1}}),
~~~ & I^{\varphi^0}_{15}=(\varphi^0(\phi(\chi\phi)_{\mathbf{6}})_{\mathbf{\overline{3}_1}}), \\
I^{\varphi^0}_{16}=(\varphi^0(\eta(\xi\xi)_{\mathbf{\overline{3}_1}})_{\mathbf{\overline{3}_1}}),~~~ & I^{\varphi^0}_{17}=(\varphi^0(\eta(\xi\xi)_{\mathbf{\overline{3}'_1}})_{\mathbf{\overline{3}_1}}), ~~~ & I^{\varphi^0}_{18}=(\varphi^0((\eta\xi)_{\mathbf{3_1}}\xi)_{\mathbf{\overline{3}_1}}), \\
I^{\varphi^0}_{19}=(\varphi^0((\eta\xi)_{\mathbf{3'_1}}\xi)_{\mathbf{\overline{3}_1}}) & &
\end{array}
\end{equation}
The new vacuum configuration is obtained by imposing the vanishing of the first derivative of $w^0_d+\delta w_d$ with respect to the driving fields. We seek a solution that perturbs the LO minima in Eq.(22) and Eq.(25) to first order in $1/\Lambda$. After lengthy and straightforward algebra, we find that the LO vacuum is modified into
\begin{eqnarray}
\nonumber&&\langle\chi\rangle=(\delta v_{\chi_1},\delta v_{\chi_2},v_{\chi}+\delta v_{\chi_3}),~~~\langle\phi\rangle=(\delta v_{\phi_1},\delta v_{\phi_2},v_{\phi}),~~~\langle\eta\rangle=(v_{\eta},\delta v_{\eta_2}),\\
\nonumber&&\langle\xi\rangle=(v_{\xi}+\delta v_{\xi_1},\delta v_{\xi_2},\delta v_{\xi_3}),~~~\langle\rho\rangle=(v_{\rho}+\delta v_{\rho_1},\omega v_{\rho}+\delta v_{\rho_2}), \\
\label{eq44}&&\langle\varphi\rangle=(v_{\varphi}+\delta v_{\varphi},v_{\varphi}-\delta v_{\varphi},v_{\varphi}),~~~\langle\psi\rangle=(v_1,v_2,v_1/2+v_2/2+\delta v_{\psi_3})\,.
\end{eqnarray}
All the shifts can be expressed in terms of the coefficients $s_i$,
$z_i$, $k_i$, $p_i$, $r_i$, $w_i$ and VEVs of the flavon fields. We
don't present the exact form of these expressions which are
overlong. Note that the shifts associated with the first and the
second component of $\varphi$ are equal except an overall sign. As
is common in discrete flavor symmetry model building, we assume that
all VEVs scaled by the cutoff $\Lambda$ are approximately of the
same order of magnitude about $\lambda^2_c$, we denote that ratio
$\mathrm{VEV}/\Lambda$ by the small parameter $\varepsilon$. Then
all the shifts are suppressed by the factor $\varepsilon$ with
respect to LO VEVs.

\begin{figure}[hptb]
\begin{center}
\begin{tabular}{cc}
\includegraphics[scale=.28]{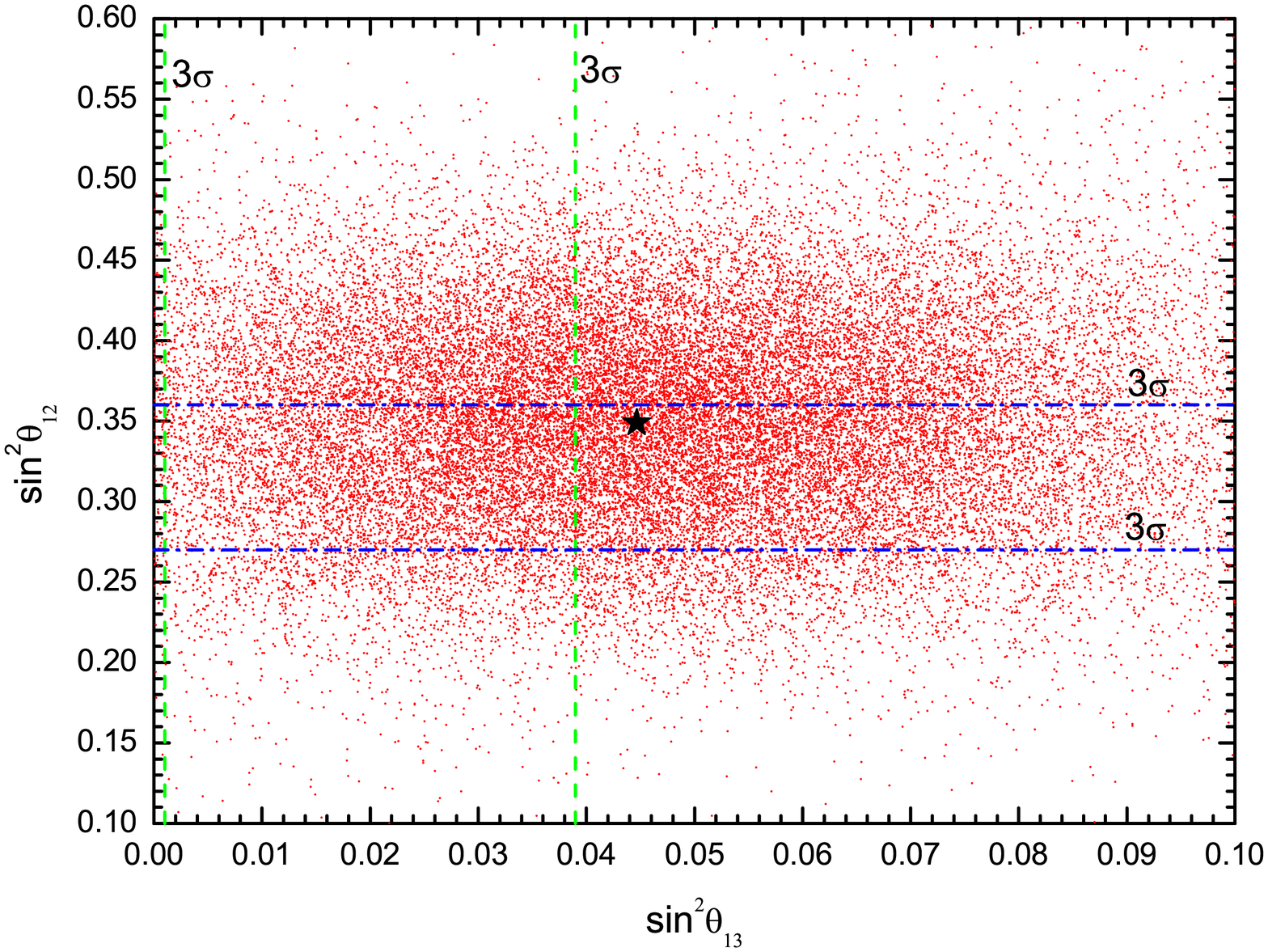} & \includegraphics[scale=.28]{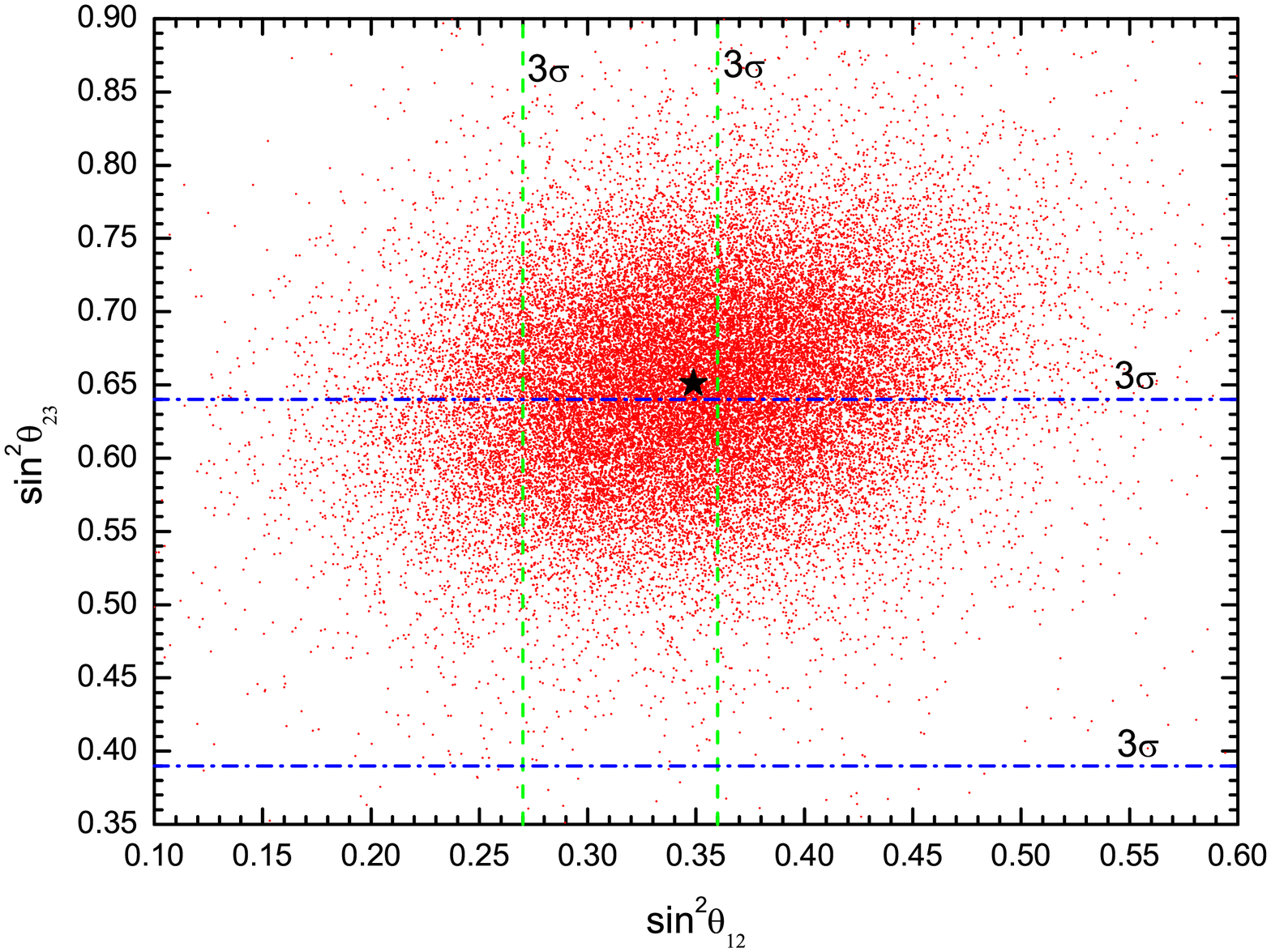}
\end{tabular}
\caption{\label{fig:mixing_angles_TFH1} The scatter plot of $\sin^2\theta_{12}$ against $\sin^2\theta_{13}$ and $\sin^2\theta_{23}$ against $\sin^2\theta_{12}$. The horizontal and vertical lines correspond to the $3\sigma$ bounds on the mixing angles, which are taken from Ref. \cite{Schwetz:2011zk}, the pentagram indicates the values predicted by the TFH1 mixing pattern.}
\end{center}
\end{figure}

The LO predictions for the  charged lepton and neutrino mass matrices are corrected by both the modified vacuum alignment and the subleading operators in the superpotentials $w_{\ell}$ and $w_{\nu}$. The NLO operators contributing to charged lepton masses can be obtained by inserting $\rho$, $\varphi$ or $\psi$ in all possible ways into the LO operators. For convenience, we denote $\chi$ and $\phi$ with $\Phi_{\ell_1}$, $\eta$ and $\xi$ with $\Phi_{\ell_2}$, and $\rho$, $\varphi$ and $\psi$ with $\Phi_{\nu}$. The NLO operators take the form:
\begin{eqnarray}
\label{eq45}&&\tau^{c}(\ell\Phi_{\ell_1}\Phi_{\nu})h_d/\Lambda^2,~~~\mu^c(\ell\Phi_{\ell_1}\Phi_{\ell_2}\Phi_{\nu})'h_d/\Lambda^3,~~~e^{c}(\ell\Phi_{\ell_1}\Phi_{\ell_2}\Phi_{\ell_2}\Phi_{\nu})h_d/\Lambda^4\,,
\end{eqnarray}
where all possible contractions among fields are understood. With the LO vacuum in Eq.(\ref{eq22}) and Eq.(\ref{eq25}), we find that each element of charged lepton mass matrix gets a small correction. Concretely the corrections to the $e$ row, $\mu$ row and $\tau$ row are of order $\varepsilon^4v_d$, $\varepsilon^3v_d$ and $\varepsilon^2v_d$ respectively. As a result, the charged lepton mass matrix with subleading corrections
can be parameterized as
\begin{equation}
\label{eq46}m_{\ell}=\left(\begin{array}{ccc}
a^{\ell}_{11}\varepsilon^2  & a^{\ell}_{12}\varepsilon^3  &  a^{\ell}_{13}\varepsilon^3 \\
a^{\ell}_{21}\varepsilon^2  & a^{\ell}_{22}\varepsilon  &  a^{\ell}_{23}\varepsilon^2  \\
a^{\ell}_{31}\varepsilon   &  a^{\ell}_{32}\varepsilon  & a^{\ell}_{33}
\end{array}\right)\varepsilon v_d\,,
\end{equation}
where the order one coefficients $a^{\ell}_{ij}(i,j=1,2,3)$ are unconstrained by the family symmetry. Including the contributions originating from the LO superpotential $w_{\ell}$ evaluated with the NLO VEVs of Eq.(\ref{eq44}), we see that this correction only amounts to a redefinition
of the $a_{ij}$ in Eq.(\ref{eq46}). Therefore, the unitary matrix $U_{\ell}$, which corresponds to the transformation of the charged leptons used to diagonalize the hermitian matrix $m^{\dagger}_{\ell}m_{\ell}$, is given by
\begin{equation}
\label{eq47}U_{\ell}\approx\left(\begin{array}{ccc}
1  & (\frac{a^{\ell}_{21}}{a^{\ell}_{22}}\varepsilon)^{*}  &  (\frac{a^{\ell}_{31}}{a^{\ell}_{33}}\varepsilon)^{*}  \\
-\frac{a^{\ell}_{21}}{a^{\ell}_{22}}\varepsilon   & 1   & (\frac{a^{\ell}_{32}}{a^{\ell}_{33}}\varepsilon)^{*}  \\
-\frac{a^{\ell}_{31}}{a^{\ell}_{33}}\varepsilon  &  -\frac{a^{\ell}_{32}}{a^{\ell}_{33}}\varepsilon  & 1
\end{array}\right)\,.
\end{equation}
We note that the charged lepton masses are corrected by terms of relative order $\varepsilon$, thus the LO mass hierarchies are not spoiled. Then we move to consider the corrections to the neutrino masses. The NLO operators contributing to the neutrino masses are given by
\begin{eqnarray}
\nonumber&&\delta w_{\nu}=x_{\nu_1}((\nu^c\nu^c)_{\mathbf{3'_1}}\delta\varphi)+x_{\nu_2}((\nu^c\nu^c)_{\mathbf{3'_2}}\delta\psi)
+\frac{\tilde{y}_{\nu_1}}{\Lambda}((\nu^c\ell)_{\mathbf{2}}\rho)h_u+\frac{\tilde{x}_{\nu_3}}{\Lambda}((\nu^c\nu^c)_{\mathbf{3'_1}}(\rho\varphi)_{\mathbf{\overline{3}'_1}})  \\
\label{eq48}&&\quad\quad+\frac{\tilde{x}_{\nu_4}}{\Lambda}((\nu^c\nu^c)_{\mathbf{3'_2}}(\varphi\varphi)_{\mathbf{3'_2}})+\frac{\tilde{x}_{\nu_5}}{\Lambda}((\nu^c\nu^c)_{\mathbf{3'_2}}(\rho\psi)_{\mathbf{3'_2}}) +\frac{\tilde{x}_{\nu_6}}{\Lambda}((\nu^c\nu^c)_{\mathbf{3'_2}}(\psi\psi)_{\mathbf{3'_2}})\,,
\end{eqnarray}
where $\delta\varphi$ and $\delta\psi$ denote the shifted vacuum of the flavons $\varphi$ and $\psi$ respectively. Taking into account the possibility of absorbing the corrections partly into the LO parameters, the corrections to the neutrino Dirac and Majorana mass matrices can be expressed as
\begin{eqnarray}
\label{eq49}&&\delta m_{D}=\tilde{y}\left(\begin{array}{ccc}
0   &  1  &  1  \\
1   & 0  &  1  \\
1  &  1 & 0
\end{array}\right)\varepsilon v_u,~~~~
\delta m_M=\left(\begin{array}{ccc}
-2\tilde{x}_1  & \tilde{x}_2  &  -\tilde{x}_1  \\
\tilde{x}_2   & 2\tilde{x}_1   &  \tilde{x}_1  \\
-\tilde{x}_1   & \tilde{x}_1   &  \tilde{x}_2
\end{array}\right)\varepsilon v_{\varphi}\,,
\end{eqnarray}
where $\tilde{y}=\omega\tilde{y}_{\nu_1}v_{\rho}/(\Lambda\varepsilon)$, $\tilde{x}_1=2x_{\nu_1}\delta v_{\varphi}/(\varepsilon v_{\varphi})$ and $\tilde{x}_2=2x_{\nu_2}\delta v_{\psi_3}/(\varepsilon v_{\varphi})$. The resulting light neutrino mass matrix $m_{\nu}$ is diagonalized by
the unitary transformation
\begin{equation}
\label{eq50}U_{\nu}=U_{TFH1}U'_{\nu}\,,
\end{equation}
where $U'_{\nu}$ is close to an identity matrix with small corrections on the off-diagonal elements, it of the form
\begin{equation}
\label{eq51}U'_{\nu}\approx\left(
\begin{array}{ccc}
1  & (a^{\nu}_{1}\varepsilon)^{*} & (a^{\nu}_{2}\varepsilon)^{*}  \\
-a^{\nu}_{1}\varepsilon   & 1   &  (a^{\nu}_{3}\varepsilon)^{*}  \\
-a^{\nu}_{2}\varepsilon   & -a^{\nu}_{3}\varepsilon   & 1
\end{array}
\right)\,,
\end{equation}
where the parameters $a^{\nu}_1$, $a^{\nu}_2$ and $a^{\nu}_3$ have order one magnitudes, and they can be reconstructed from the LO couplings in Eq.(\ref{eq31}) and the NLO couplings in Eq.(\ref{eq48}). The PMNS matrix is given by $U_{PMNS}=U^{\dagger}_{\ell}U_{\nu}$, then the leptonic mixing angles are modified as
\begin{eqnarray}
\nonumber&&\sin^2\theta_{13}=\frac{2-\sqrt{3}}{6}+\frac{1}{6}\Big[(a^{\ell}_1-a^{\nu}_2)\varepsilon+(1-\sqrt{3})(a^{\ell}_2+a^{\nu}_3)\varepsilon+c.c.\Big] \\
\nonumber&&\sin^2\theta_{12}=\frac{8-2\sqrt{3}}{13}+\frac{1}{169}\Big[(10\sqrt{3}-66)a^{\ell}_1\varepsilon-(18+28\sqrt{3})a^{\ell}_2\varepsilon+13(1+3\sqrt{3})a^{\nu}_1\varepsilon \\
\nonumber&&\quad\quad+(16\sqrt{3}-38)a^{\nu}_2\varepsilon+(11\sqrt{3}-5)a^{\nu}_3\varepsilon+c.c.\Big]  \\
\nonumber&&\sin^2\theta_{23}=\frac{5+2\sqrt{3}}{13}+\frac{1}{169}\Big[(16\sqrt{3}-38)a^{\ell}_1\varepsilon+(5-11\sqrt{3})a^{\ell}_2\varepsilon+13(1+3\sqrt{3})a^{\ell}_3\varepsilon \\
\label{eq52}&&\quad\quad+(10\sqrt{3}-66)a^{\nu}_2\varepsilon+(18+28\sqrt{3})a^{\nu}_{3}\varepsilon+c.c.\Big]\,,
\end{eqnarray}
where we define $a^{\ell}_1=a^{\ell}_{21}/a^{\ell}_{22}$, $a^{\ell}_2=a^{\ell}_{31}/a^{\ell}_{33}$ and $a^{\ell}_3=a^{\ell}_{32}/a^{\ell}_{33}$. Due to the presence of numerous unknown parameters of order one, we can not make quantitative predictions for the mixing angles. We perform a numerical analysis by treating all the LO and NLO coefficients as random complex numbers of absolute value between 1/3 and 3, the expansion parameter $\varepsilon$ is fixed at the indicative value 0.04. The resulting scattering plots for $\sin^2\theta_{13}-\sin^2\theta_{12}$ and $\sin^2\theta_{12}-\sin^2\theta_{23}$ are shown in Fig. \ref{fig:mixing_angles_TFH1}. We see that a considerable part of points fall into the region where the three mixing angles $\theta_{13}$, $\theta_{12}$ and $\theta_{23}$ are in the $3\sigma$ interval, and a relatively large reactor angle can be accommodated.

\section{\label{sec:TFH2}Model for TFH2 mixing}

\begin{table}[hptb]
\begin{center}
\begin{tabular}{|c|c|c|c|c|c|c||c|c|c|c|c|c||c|c|c|c|c|c|}\hline\hline
{\tt Fields} & $\ell$ & $e^c$  &  $\mu^c$ &  $\tau^c$  & $\nu^{c}$ &$h_{u,d}$  &  $\varphi$   &$\xi$   & $\phi$  &  $\zeta$ &  $\chi$   & $\Delta$ & $\sigma^0$ &  $\xi^0$   &  $\varphi^0$  &  $\phi^0$ &$\zeta^0$  &  $\chi^0$
\\\hline

$\Delta(96)$  & $\mathbf{3_1}$ &  $\mathbf{1'}$  &  $\mathbf{1'}$    &  $\mathbf{1}$   &  $\mathbf{3_1}$   &  $\mathbf{1}$  &  $\mathbf{2}$  &  $\mathbf{3'_1}$  &  $\mathbf{\overline{3}_1}$     &  $\mathbf{2}$  &  $\mathbf{3'_1}$  &  $\mathbf{3'_2}$   &  $\mathbf{1}$     &  $\mathbf{1'}$      &  $\mathbf{2}$    &   $\mathbf{3'_2}$   &   $\mathbf{2}$   &  $\mathbf{\overline{3}_1}$  \\

$Z_5$ &  1  &  4    &   1   &  3  & 1   &   0   &  2   &  1   &  1  &  3   &  3   &  3  &  1   &  3   &  3   &   3      &   4   &  4    \\

$Z_2$ &  0  &   1   &  1   & 1    &  0   &  0 &  0  &  1  &  1  &  0  & 0   &  0  &  0    &   0  &   0   &   0    &  0   &  0    \\\hline\hline

\end{tabular}
\caption{\label{tab:field2} Fields and their transformation properties under the flavor symmetry $\Delta(96)\times Z_5\times Z_2$, where $\ell=(\ell_1,\ell_3,\ell_2)$ is the left-handed lepton doublet field.}
\end{center}
\end{table}
In this section, we present another flavor model based on $\Delta(96)$, which naturally leads to the TFH2 mixing at LO. Both the left-handed lepton doublets and right-handed neutrinos are assigned to triplet $\mathbf{3_1}$ in this model, note that $\ell_2$ and $\ell_3$ are the third and second component of the triplet respectively. The full family symmetry is $\Delta(96)\times Z_5\times Z_2$, and the associated field content is given in Table \ref{tab:field2}. The LO driving superpotential takes the form
\begin{eqnarray}
\nonumber&&w_d=h_1\sigma^0(\varphi\varphi)+h_2\xi^0(\xi\phi)'+M_{\varphi}(\varphi^0\varphi)+h_3(\varphi^0(\xi\phi)_{\mathbf{2}})+h_4(\phi^{0}(\xi\xi)_{\mathbf{3'_2}})
+h_5(\phi^0(\phi\phi)_{\mathbf{3}'_2})\\
\label{eq53}&&\quad\quad+g_1(\zeta^0(\zeta\zeta)_{\mathbf{2}})+g_2(\zeta^0(\Delta\Delta)_{\mathbf{2}})+g_3(\chi^0(\zeta\chi)_{\mathbf{3_1}})
\end{eqnarray}
where the flavon fields $\varphi$ and $\phi$ break the flavor symmetry in the charged lepton sector at LO, $\chi$ and $\Delta$ couple to the neutrino sector, their vacuum configurations are determined by the terms in the first line and the second line of Eq.(\ref{eq53}) respectively, while the flavons $\xi$ and $\zeta$ are present for consistent vacuum alignment and higher order corrections. The equations for the minimum of the scalar potential are given by
\begin{eqnarray}
\nonumber&&\frac{\partial w_d}{\partial\sigma^0}=2h_1\varphi_1\varphi_2=0  \\
\nonumber&&\frac{\partial w_d}{\partial\xi^0}=h_2(\xi_1\phi_1+\xi_2\phi_2+\xi_3\phi_3)=0  \\
\nonumber&&\frac{\partial w_d}{\partial\varphi^0_1}=M_{\varphi}\varphi_2-\omega h_3(\xi_1\phi_2+\xi_2\phi_3+\xi_3\phi_1)=0  \\
\nonumber&&\frac{\partial w_d}{\partial\varphi^0_2}=M_{\varphi}\varphi_1+h_3(\xi_1\phi_3+\xi_2\phi_1+\xi_3\phi_2)=0   \\
\nonumber&&\frac{\partial w_d}{\partial\phi^0_1}=h_4(\xi^2_1+2\xi_2\xi_3)+h_5(\phi^2_2+2\phi_1\phi_3)=0    \\
\nonumber&&\frac{\partial w_d}{\partial\phi^0_2}=h_4(\xi^2_2+2\xi_1\xi_3)+h_5(\phi^2_1+2\phi_2\phi_3)=0  \\
\label{eq54}&&\frac{\partial w_d}{\partial\phi^0_3}=h_4(\xi^2_3+2\xi_1\xi_2)+h_5(\phi^2_3+2\phi_1\phi_2)=0\,.
\end{eqnarray}
These equations are satisfied by the alignment
\begin{equation}
\label{eq55}\langle\varphi\rangle=(0,v_{\varphi}),~~~\langle\xi\rangle=(v_{\xi},0,0),~~~\langle\phi\rangle=(0,v_{\phi},0)\,,
\end{equation}
with the relations
\begin{equation}
\label{eq56}v^2_{\phi}=-h_4v^2_{\xi}/h_5,~~~v_{\varphi}=\omega h_3v_{\xi}v_{\phi}/M_{\varphi},~~~v_{\xi}~\mathrm{undetermined}\,.
\end{equation}
Under the action of the element $a^2cd$, we have $a^2cd\langle\varphi\rangle=\omega\langle\varphi\rangle$, $a^2cd\langle\xi\rangle=\omega^2\langle\xi\rangle$ and $a^2cd\langle\phi\rangle=\omega^2\langle\phi\rangle$. The flavor symmetry $\Delta(96)$ is broken completely by the non-vanishing VEVs of $\varphi$, $\xi$ and $\phi$. For the neutrino sector, the vacuum of the flavons $\zeta$, $\chi$ and $\Delta$ is determined by the following minimization equations
\begin{eqnarray}
\nonumber&&\frac{\partial w_d}{\partial\zeta^0_1}=g_1\zeta^2_1+\omega g_2(\Delta^2_1+2\Delta_2\Delta_3)=0 \\
\nonumber&&\frac{\partial w_d}{\partial\zeta^0_2}=g_1\zeta^2_2+g_2(\Delta^2_2+2\Delta_1\Delta_3)=0  \\
\nonumber&&\frac{\partial w_d}{\partial\chi^0_1}=g_3(\zeta_1\chi_3-\omega^2\zeta_2\chi_2)=0  \\
\nonumber&&\frac{\partial w_d}{\partial\chi^0_2}=g_3(\zeta_1\chi_1-\omega^2\zeta_2\chi_3)=0 \\
\label{eq57}&&\frac{\partial w_d}{\partial\chi^0_3}=g_3(\zeta_1\chi_2-\omega^2\zeta_2\chi_1)=0\,,
\end{eqnarray}
which leads to the vacuum configuration
\begin{equation}
\label{eq58}\langle\chi\rangle=(1,1,1)v_{\chi},~~~\langle\zeta\rangle=(1,\omega)v_{\zeta},~~~\langle\Delta\rangle=(v_1,v_2,(v_1+v_2)/2)\,,
\end{equation}
where the various VEVs are related by
\begin{equation}
\label{eq59}g_1v^2_{\zeta}+\omega g_{2}(v^2_1+v^2_2+v_1v_2)=0\,.
\end{equation}
The superpotential $w_{\ell}$, which is responsible for the charged lepton masses, takes the form
\begin{equation}
\label{eq60}w_{\ell}=\frac{y_{\tau}}{\Lambda}\tau^c(\ell\phi)h_d+\frac{y_{\mu}}{\Lambda^2}\mu^c(\ell(\varphi\phi)_{\mathbf{\overline{3}'_1}})'h_d+
\frac{y_e}{\Lambda^3}e^c(\ell((\varphi\varphi)_{\mathbf{2}}\phi)_{\mathbf{\overline{3}'_1}})'h_d+\ldots\,,
\end{equation}
where we omit the term $e^c(\varphi\varphi)'(\ell\phi)h_d$, which vanishes due to the antisymmetric property of the contraction $(\varphi\varphi)'$. After flavor and electroweak symmetry breaking, $w_{\ell}$ results in the following diagonal charged lepton mass matrix:
\begin{equation}
\label{eq61}m_{\ell}=\left(\begin{array}{ccc}
y_{e}v_{\phi}v^2_{\varphi}/\Lambda^3  &  0  &  0  \\
0  & -\omega^2y_{\mu}v_{\phi}v_{\varphi}/\Lambda^2  &  0 \\
0 & 0  &  y_{\tau}v_{\phi}/\Lambda
\end{array}\right)v_d\,.
\end{equation}
It is obvious that the electron, muon and tau masses are controlled by the first, second and third power of $v_{\phi}/\Lambda$ and $v_{\varphi}/\Lambda$ respectively. Therefore the observed mass hierarchies among the charged leptons are naturally produced if $v_{\phi}/\Lambda$ and $v_{\varphi}/\Lambda$ are of order $\lambda^2_c$. The neutrino masses are described by the superpotential $w_{\nu}$ as follows
\begin{eqnarray}
\label{eq62}&&w_{\nu}=\frac{y_1}{\Lambda}((\nu^c\ell)_{\mathbf{\overline{3}'_1}}\chi)h_u+\frac{y_2}{\Lambda}((\nu^c\ell)_{\mathbf{3'_2}}\Delta)h_u+x_1((\nu^c\nu^c)_{\mathbf{\overline{3}'_1}}\chi)
+x_2((\nu^c\nu^c)_{\mathbf{3'_2}}\Delta)+\ldots
\end{eqnarray}
As a consequence, the neutrino Dirac and Majorana mass matrices read as
\begin{eqnarray}
\nonumber&&m_D=\left(\begin{array}{ccc}
-2A+2B   &  A+2C  &  A+B+C   \\
A+B+C   &  A+2B  &  -2A+2C  \\
A+2C   & -2A+B+C   & A+2B
\end{array}\right)v_u \\
\label{eq63}&&m_M=\left(\begin{array}{ccc}
-2M_1+2M_2  &  M_1+M_2+M_3  & M_1+2M_3  \\
M_1+M_2+M_3   & -2M_1+2M_3  &  M_1+2M_2   \\
M_1+2M_3  &  M_1+2M_2   &   -2M_1+M_2+M_3
\end{array}\right)\,,
\end{eqnarray}
where
\begin{eqnarray}
\nonumber&& A=y_1\frac{v_{\chi}}{\Lambda},~~~B=y_2\frac{v_1}{2\Lambda},~~~C=y_2\frac{v_2}{2\Lambda}   \\
\label{eq64}&& M_1=2x_1v_{\chi},~~~M_2=x_2v_1,~~~M_3=x_2v_2\,.
\end{eqnarray}
The light neutrino mass matrix is given by the see-saw relation $m_{\nu}=-m^{T}_Dm^{-1}_Mm_D$, which has the same structure as $m^{TFH2}_{\nu}$ in Eq.(\ref{eq8}). Consequently it is diagonalized by the TFH2 mixing matrix and the eigenvalues are
\begin{equation}
\label{eq65}m_1=\frac{\sqrt{3}(\sqrt{3}A-B+C)^2v^2_u}{\sqrt{3}M_1-M_2+M_3},~m_2=-\frac{3(B+C)^2v^2_u}{M_2+M_3},~m_3=\frac{\sqrt{3}(\sqrt{3}A+B-C)^2v^2_u}{\sqrt{3}M_1+M_2-M_3}\,.
\end{equation}
We see that the light neutrino masses depend on six unrelated parameters, which offer more freedom to tune mass differences and then recover the phenomenology associated to neutrino oscillation. Note that no sum rule among light neutrino masses can be found in this case.

\subsection{Next to leading order corrections}

The vacuum alignment discussed in the previous section is modified
by the higher-dimensional operators allowed by the symmetry of the
model. We denote the NLO corrections to the driving superpotential
by $\delta w_d$, which is suppressed by one power of $1/\Lambda$
with respect to the leading order terms presented in
Eq.(\ref{eq53}). The full expression of $\delta w_d$ is the
following form:
\begin{equation}
\label{eq66}\delta w_d=\frac{1}{\Lambda}\Big(\sum^{6}_{i=1}s_iI^{\sigma^{0}}_i+k_1\xi^{0}((\varphi\varphi)_{\mathbf{2}}\zeta)'+\sum^{3}_{i=1}w_iI^{\varphi^0}_i
+\sum^{3}_{i=1}p_iI^{\phi^{0}}_i+\sum^{3}_{i=1}z_iI^{\zeta^0}_i\Big)\,,
\end{equation}
where $s_i$, $k_1$, $w_i$, $p_i$ and $z_i$ are order one coefficients, $\{I^{\sigma^0}_i, I^{\varphi^0}_i, I^{\phi^0}_i, I^{\zeta^0}_i\}$ denote the
complete set of subleading contractions invariant under $\Delta(96)\times Z_5 \times Z_2$.
\begin{equation}
\label{eq67}\begin{array}{lll}
I^{\sigma^{0}}_1=\sigma^0(\varphi(\xi\phi)_{\mathbf{2}}),\quad\quad & I^{\sigma^{0}}_2=\sigma^0(\zeta(\zeta\zeta)_{\mathbf{2}}),\quad\quad &  I^{\sigma^{0}}_3= \sigma^{0}(\zeta(\Delta\Delta)_{\mathbf{2}}),  \\
I^{\sigma^{0}}_4=\sigma^0(\chi(\chi\chi)_{\mathbf{\overline{3}'_1}}),\quad\quad & I^{\sigma^{0}}_5=\sigma^0((\chi\chi)_{\mathbf{3'_2}}\Delta),\quad\quad  & I^{\sigma^{0}}_6=\sigma^{0}(\Delta(\Delta\Delta)_{\mathbf{3'_2}})
\end{array}
\end{equation}

\begin{equation}
\label{eq68}\begin{array}{lll}
I^{\varphi^0}_1=(\varphi^0\zeta)(\varphi\varphi),\quad\quad & I^{\varphi^0}_2=(\varphi^0\zeta)'(\varphi\varphi)',\quad\quad & I^{\varphi^0}_3=(\varphi^0(\zeta(\varphi\varphi)_{\mathbf{2}})_{\mathbf{2}})
\end{array}
\end{equation}

\begin{equation}
\label{eq69}\begin{array}{lll}
I^{\phi^0}_1=(\phi^0\Delta)(\varphi\varphi),\quad\quad  & I^{\phi^0}_2=(\phi^0\Delta)'(\varphi\varphi)',\quad\quad  & I^{\phi^0}_3=(\phi^0((\varphi\varphi)_{\mathbf{2}}\Delta)_{\mathbf{3'_2}})
\end{array}
\end{equation}

\begin{equation}
\label{eq70}\begin{array}{lll}
I^{\zeta^0}_1=(\zeta^0\varphi)(\varphi\varphi),\quad\quad & I^{\zeta^0}_2=(\zeta^0\varphi)'(\varphi\varphi)',\quad\quad  & I^{\zeta^0}_3=(\zeta^0(\varphi(\varphi\varphi)_{\mathbf{2}})_{\mathbf{2}})\,.
\end{array}
\end{equation}
Note that the corrections to the $\chi^0$ relevant superpotential appear at the next to next to leading order (NNLO) due to the strong constraints of the flavor symmetry. The new vacuum is obtained as usual by solving the vanishing $F-$term constraints. After lengthy calculation, we find that the alignment of $\zeta$, $\chi$ and $\Delta$ given in Eq.(\ref{eq58}) is unchanged at this order. In the charged lepton sector, the vacuum configuration of the flavons $\varphi$, $\xi$ and $\phi$ is given by
\begin{equation}
\label{eq71}\langle\varphi\rangle=(\delta v_{\varphi_1},v_{\varphi}+\delta v_{\varphi_2}),~~~\langle\xi\rangle=(v_{\xi},\delta v_{\xi_2},\delta v_{\xi_3}),~~~\langle\phi\rangle=(\delta v_{\phi_1},v_{\phi}+\delta v_{\phi_2},\delta v_{\phi_3})\,.
\end{equation}
Concretely the shifts are as follows
\begin{eqnarray}
\nonumber&&\delta v_{\varphi_1}=\frac{d_1v^3_{\zeta}}{2h_1\Lambda v_{\varphi}},~\delta v_{\varphi_2}=\frac{\omega^2w_3v_{\zeta}v^3_{\varphi}}{h_3\Lambda v_{\xi}v_{\phi}}+\frac{p_3v_1v^3_{\varphi}}{2h_5\Lambda v^2_{\phi}},~\delta v_{\xi_2}=\frac{\omega k_1v_{\zeta}v^2_{\varphi}}{2h_2\Lambda v_{\phi}}+\frac{p_3v_2v^2_{\varphi}}{4h_4\Lambda v_{\xi}},\\
\nonumber&&\delta v_{\xi_3}=-\frac{\omega d_1v_{\xi}v^3_{\zeta}}{4h_1\Lambda v^2_{\varphi}}+\frac{p_3(v_1+v_2)v^2_{\varphi}}{8h_4\Lambda v_{\xi}},~\delta v_{\phi_2}=\frac{p_3v_1v^2_{\varphi}}{2h_5\Lambda v_{\phi}},~\delta v_{\phi_3}=-\frac{\omega d_1v_{\phi}v^3_{\zeta}}{4h_1\Lambda v^2_{\varphi}}+\frac{p_3(v_1+v_2)v^2_{\varphi}}{8h_5\Lambda v_{\phi}}
\end{eqnarray}
where the parameter $d_1$ is
\begin{equation}
\label{add1}d_1=2s_2+2\omega s_3(v^2_1+v_1v_2+v^2_2)/v^2_{\zeta}+9s_5(v_1+v_2)v^2_{\chi}/(2v^3_{\zeta})-9s_6(v_1+v_2)(v_1-v_2)^2/(4v^3_{\zeta})\,.
\end{equation}
Parametrizing the ratio of a generic flavon VEV to the cutoff scale $\Lambda$ by the small quantity $\epsilon$, we see that all the shifts are of order $\epsilon$ with respect to LO VEVs.

\begin{figure}[hptb]
\begin{center}
\begin{tabular}{cc}
\includegraphics[scale=.28]{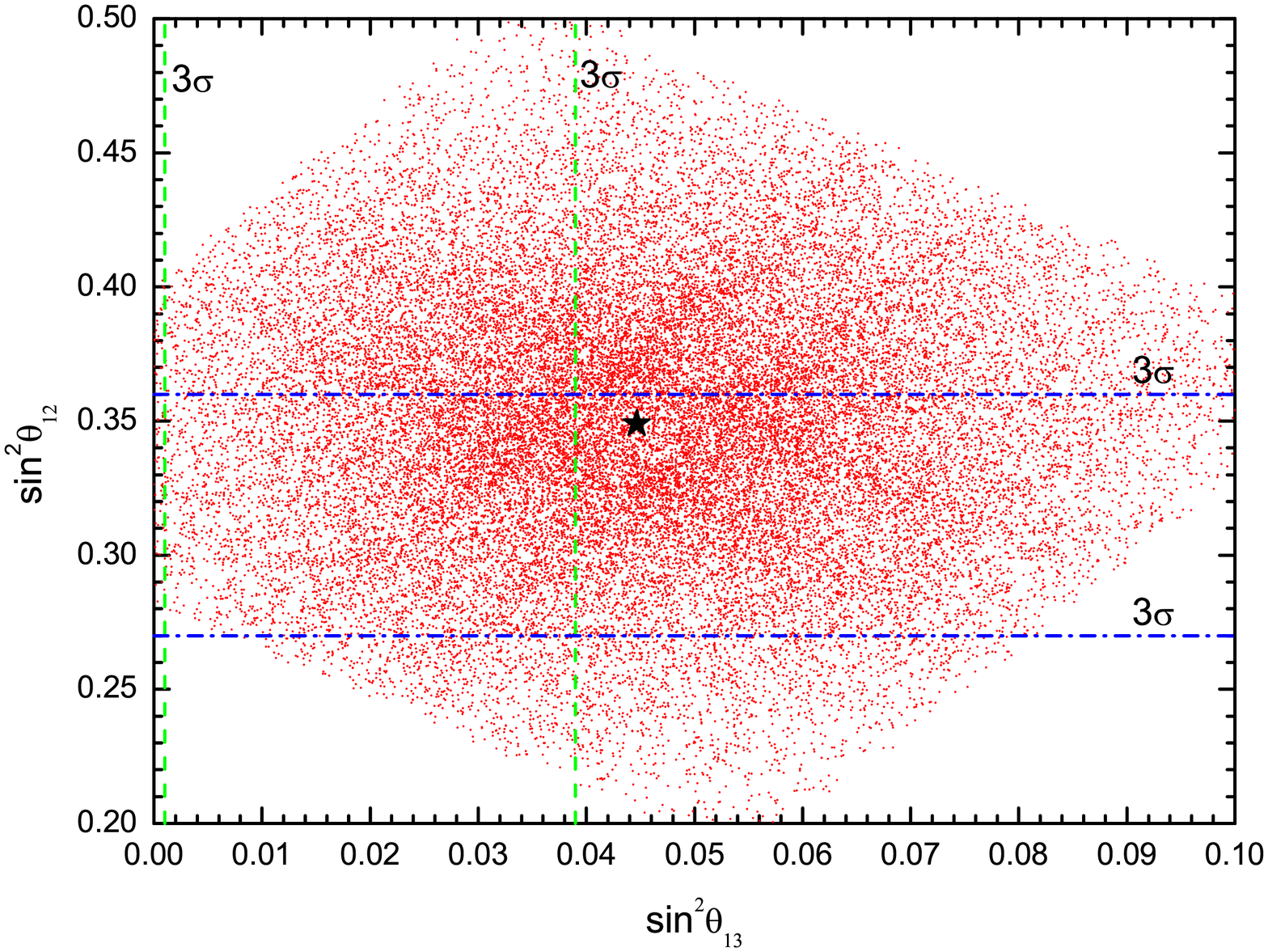} & \includegraphics[scale=.28]{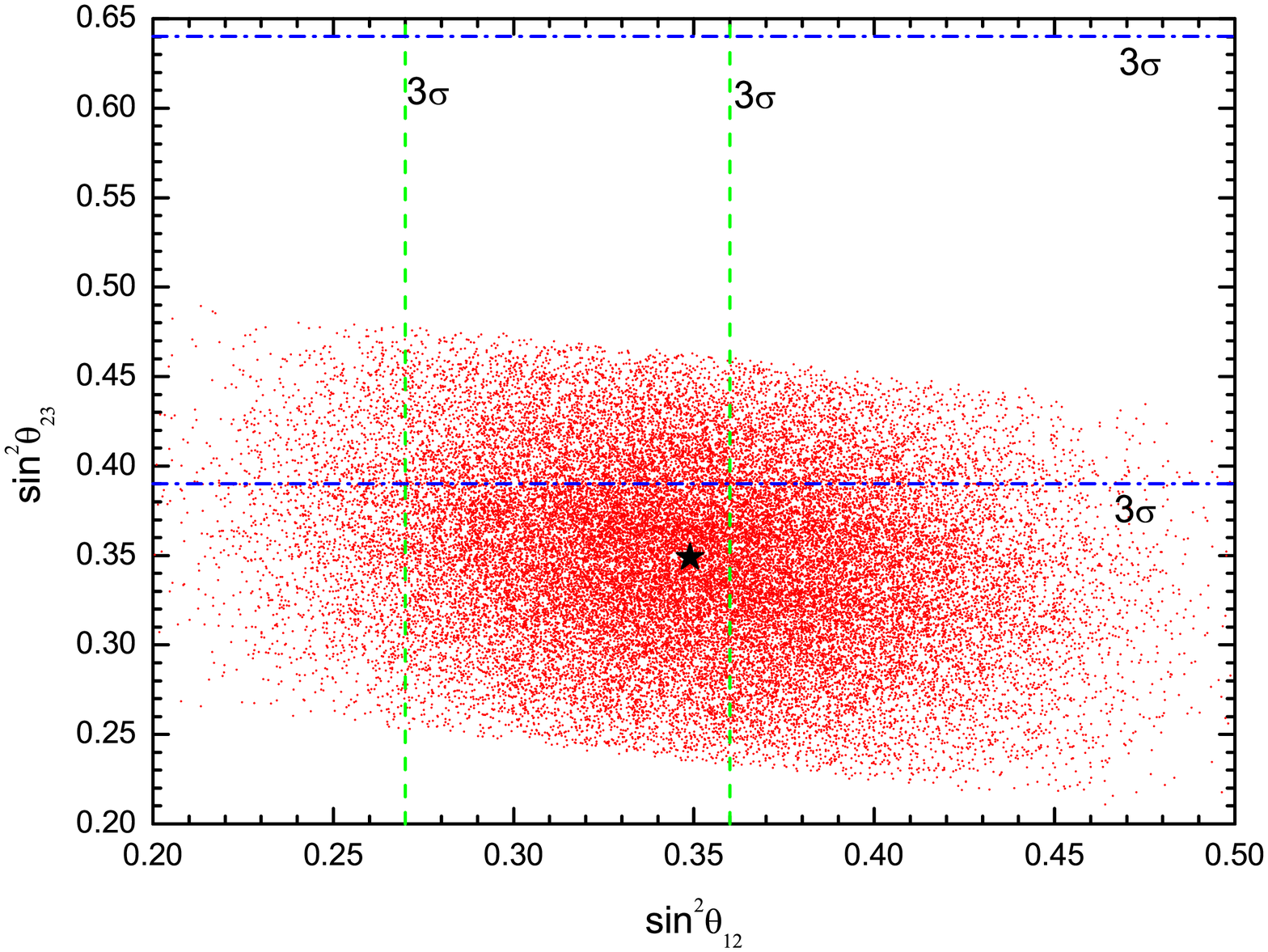}
\end{tabular}
\caption{\label{fig:mixing_angles_TFH2} The scatter plot of $\sin^2\theta_{12}$ against $\sin^2\theta_{13}$ and $\sin^2\theta_{23}$ against $\sin^2\theta_{12}$. The horizontal and vertical lines correspond to the $3\sigma$ bounds on the mixing angles, which are taken from Ref. \cite{Schwetz:2011zk}, the pentagram indicates the values predicted by the TFH2 mixing pattern.}
\end{center}
\end{figure}

We move now to discuss the NLO corrections to the lepton masses and
flavor mixing, we denote $\xi$ and $\phi$ with $\Psi_{\ell}$, and
the flavons $\zeta$, $\chi$ and $\Delta$ with $\Psi_{\nu}$. In the
neutrino sector, imposing the invariance under the flavor symmetry,
the neutrino masses are corrected by the subleading operators of the
structures $(\nu^c\ell\varphi\Psi_{\nu}\Psi_{\nu})h_u/\Lambda^3$ and
$(\nu^c\nu^c\varphi\Psi_{\nu}\Psi_{\nu})/\Lambda^2$, they are
suppressed by $1/\Lambda^2$ compared to the LO results, thus the
corrections to the neutrino Dirac and Majorana mass matrices from
the higher-dimensional operators in $w_{\nu}$ arise at NNLO. As we
have already observed above, the LO vacuum structure of the neutrino
flavons $\zeta$, $\chi$ and $\Delta$ are unchanged at NLO, the
corrections to the neutrino mass from the shifted vacuum appear at
NNLO as well. Therefore we conclude that the light neutrino mass
matrix is still diagonalized by the TFH2 mixing matrix at NLO. In
the charged lepton sector, inserting the NLO VEVs of Eq(\ref{eq71})
into the LO superpotential $w_{\ell}$ of Eq.(\ref{eq60}), we find
that each entry of the charged lepton mass matrix is corrected, and
all off-diagonal elements become non-vanishing and of the order of
the diagonal term in each row multiplied by $\epsilon$. In addition,
the charged lepton masses are corrected by the following
higher-dimensional operators
\begin{equation}
\label{eq72}\tau^c(\ell\varphi\Psi_{\ell}\Psi_{\nu})h_d/\Lambda^3,~\mu^c(\ell\Psi_{\ell}\Psi_{\ell}\Psi_{\ell})'h_d/\Lambda^3,~e^{c}(\ell\varphi\Psi_{\ell}\Psi_{\ell}\Psi_{\ell})'h_d/\Lambda^4,
~e^c(\ell\Psi_{\ell}\Psi_{\nu}\Psi_{\nu}\Psi_{\nu})'h_d/\Lambda^4\,,
\end{equation}
where all possible contractions among fields should be considered.
Note that the corrections to the $\tau^c$ row arise at NNLO. Given
the LO vacuum alignment in Eq.(\ref{eq55}) and Eq.(\ref{eq58}), we
find that the operators of types
$\mu^c(\ell\Psi_{\ell}\Psi_{\ell}\Psi_{\ell})'h_d/\Lambda^3$ and
$e^{c}(\ell\varphi\Psi_{\ell}\Psi_{\ell}\Psi_{\ell})'h_d/\Lambda^4$
only lead to corrections to the (22) and (11) entries of the charged
lepton mass matrix respectively, while all the three entries of the
$e^c$ row are corrected by the operators of the structure
$e^c(\ell\Psi_{\ell}\Psi_{\nu}\Psi_{\nu}\Psi_{\nu})'h_d/\Lambda^4$.
Combining the above two contributions from both the modified vacuum
and the higher-dimensional Yukawa couplings, the corrected charged
lepton mass matrix has the following structure
\begin{equation}
\label{eq73}m_{\ell}=\left(\begin{array}{ccc}
O(\epsilon^2)  & O(\epsilon^3)  & O(\epsilon^3)  \\
O(\epsilon^2) & O(\epsilon)  & O(\epsilon^2)  \\
O(\epsilon)  & O(\epsilon)  & O(1)
\end{array}\right)\epsilon v_d\,,
\end{equation}
where only the order of magnitude of each entry is reported. As a result, the unitary matrix $U_{\ell}$ diagonalizing $m^{\dagger}_{\ell}m_{\ell}$ can be written as
\begin{equation}
\label{eq74}U_{\ell}\simeq\left(\begin{array}{ccc}
1  &  (b^{\ell}_1\epsilon)^{*}  & (b^{\ell}_2\epsilon)^{*} \\
-b^{\ell}_1\epsilon  &  1   &  (b^{\ell}_3\epsilon)^{*}   \\
-b^{\ell}_2\epsilon  &  -b^{\ell}_3\epsilon  & 1
\end{array}\right)\,,
\end{equation}
where the coefficients $b^{\ell}_{1,2,3}$ are of order one. Therefore the TFH2 mixing is violated by the corrections from $U_{\ell}$ at NLO, the lepton mixing angles are given by
\begin{eqnarray}
\nonumber&&\hskip-0.95cm\sin^2\theta_{13}=\frac{2-\sqrt{3}}{6}+\frac{1}{6}\Big[(1-\sqrt{3})b^{\ell}_1\epsilon+b^{\ell}_2\epsilon+c.c.\Big] \\
\nonumber&&\hskip-0.95cm\sin^2\theta_{12}=\frac{8-2\sqrt{3}}{13}-\frac{1}{169}\Big[(18+28\sqrt{3})b^{\ell}_1\epsilon+(66-10\sqrt{3})b^{\ell}_2\epsilon+c.c.\Big] \\
\label{eq75}&&\hskip-0.95cm\sin^2\theta_{23}=\frac{8-2\sqrt{3}}{13}+\frac{1}{169}\Big[(-5+11\sqrt{3})b^{\ell}_1\epsilon+(38-16\sqrt{3})b^{\ell}_2\epsilon+13(1+3\sqrt{3})b^{\ell}_3\epsilon+c.c.\Big].
\end{eqnarray}
We see that all the three mixing angles receive corrections of order $\epsilon$, which can pull the LO TFH2 mixing pattern into the experimentally allowed range. In order to not spoil the successful predictions for $\theta_{13}$ and $\theta_{12}$, the parameter $\epsilon$ should not exceed a few percent. As in section \ref{sec:TFH1}, we perform a numerical simulation, the correlations between $\sin^2\theta_{12}$ and $\sin^2\theta_{13}$, $\sin^2\theta_{23}$ and $\sin^2\theta_{12}$ are shown in Fig. \ref{fig:mixing_angles_TFH2}.

As has been pointed out at the end of section \ref{sec:pathway} in effective theory, by exchanging the position of $\mu^{c}$ and $\tau^{c}$ and that of the second and third generation lepton doublets simultaneously, a model where the mixing TFH1(or TFH2) is derived, can be made into a model in which the other mixing TFH2(or TFH1) is present. When we apply this procedure to the model of section \ref{sec:TFH1} which gives rise to TFH1 mixing, the TFH2 mixing can be reproduced, and vice versa for the model of section \ref{sec:TFH2}. However, the predictions for the masses of muon and tau leptons are exchanged at the same time. As a result, the successful mass hierarchies between the charged leptons in both models would be spoiled. It is remarkable that we could keep the assignment of leptonic fields under $\Delta(96)$ and appropriately modify both the auxiliary symmetry and the flavon fields associated with charged leptons in section \ref{sec:TFH1} so that the TFH2 mixing could be derived naturally, the same is true for the model of section \ref{sec:TFH2}. The construction of these scenarios will be reported elsewhere \cite{Ding:progress}.

\section{\label{sec:other_textures}Possible mixing textures within $\Delta(96)$}

The basic idea of discrete flavor symmetry model building is that
the matter fields transform non-trivially under the flavor symmetry
at the high energy scale, then it is spontaneously broken by the
flavon into $G_{\ell}$ and $G_{\nu}$ subgroups in the charged lepton
and neutrino sectors. Generally the residual symmetry $G_{\ell}$
should be different from the $G_{\nu}$ group, this kind of mismatch
between $G_{\ell}$ and $G_{\nu}$, which results in the mismatch
between the neutrino and charged lepton mass matrices, could leads
to some interesting mass-independent textures. There is a direct
group-theoretical connection between lepton mixing and the
horizontal symmetry \cite{Lam:2011ag}. Given the whole flavor
symmetry and the surviving subgroups $G_{\ell}$ and $G_{\nu}$, one
can carry out a purely group-theoretical analysis to obtain all the
possible mixings, without the presence of flavon fields nor the help
of the Lagrangian. As bas been demonstrated in section
\ref{sec:mass_matrix}, the remnant symmetry of the left-handed
neutrinos forms a $K_4$ group for the TFH mixing, if the neutrinos
are Majorana particles. Consequently we choose $G_{\nu}$ to be the
$K_4$ subgroups of $\Delta(96)$. $G_{\ell}$ is taken to be the
cyclic $Z_{N}$ subgroups of $\Delta(96)$ with $N\geq3$ in this work,
since the resulting three charged lepton masses would be completely
or partially degenerate if $G_{\ell}$ is some non-abelian subgroups.
Moreover, $G_{\ell}$ can not be $Z_2$, otherwise at least two
eigenvalues of $G_{\ell}$ are degenerate, then we can not
distinguish among the three generations of charged leptons with the
residual symmetry $G_{\ell}$, therefore the lepton mixing matrix can
not be predicted uniquely in this case. $\Delta(96)$ has seven $K_4$
subgroups, sixteen $Z_3$ subgroups, twelve $Z_4$ subgroups and six
$Z_8$ subgroups. In terms of the generators $a$, $b$, $c$ and $d$,
they can be expressed as follows:
\begin{itemize}
\item{$K_4$ subgroups}
\begin{equation}
\nonumber\begin{array}{lll}
K^{(1)}_4=\{1,c^2,d^2,c^2d^2\},\quad\quad & K^{(2)}_4=\{1,ab,c^2,abc^2\},\quad\quad & K^{(3)}_4=\{1,abc,c^2,abc^3\},\\
K^{(4)}_4=\{1,a^2b,d^2,a^2bd^2\},\quad\quad & K^{(5)}_4=\{1,a^2bd,d^2,a^2bd^3\},\quad\quad & K^{(6)}_4=\{1,b,c^2d^2,bc^2d^2\},\\
K^{(7)}_4=\{1,bcd,c^2d^2,bc^3d^3\}
\end{array}
\end{equation}
\item{$Z_3$ subgroups}
\begin{equation}
\nonumber\begin{array}{lll}
Z^{(1)}_3=\{1,a,a^2\},\quad\quad & Z^{(2)}_3=\{1,ac,a^2cd\},\quad\quad &  Z^{(3)}_3=\{1,ac^2,a^2c^2d^2\}, \\
Z^{(4)}_3=\{1,ac^3,a^2c^3d^3\}, \quad\quad & Z^{(5)}_3=\{1,ad,a^2c^3\}, \quad\quad & Z^{(6)}_3=\{1,ad^2,a^2c^2\}, \\
Z^{(7)}_3=\{1,ad^3,a^2c\}, \quad\quad & Z^{(8)}_3=\{1,acd,a^2d\}, \quad\quad &  Z^{(9)}_3=\{1,acd^2,a^2c^3d\}, \\
Z^{(10)}_3=\{1,acd^3,a^2c^2d\}, \quad\quad & Z^{(11)}_3=\{1,ac^2d,a^2cd^2\}, \quad\quad & Z^{(12)}_3=\{1,ac^2d^2,a^2d^2\}, \\
Z^{(13)}_3=\{1,ac^2d^3,a^2c^3d^2\}, \quad\quad & Z^{(14)}_3=\{1,ac^3d,a^2c^2d^3\}, \quad\quad & Z^{(15)}_3=\{1,ac^3d^2,a^2cd^3\}, \\ Z^{(16)}_3=\{1,ac^3d^3,a^2d^3\} &  &
\end{array}
\end{equation}
\item{$Z_4$ subgroups}
\begin{equation}
\nonumber\begin{array}{lll}
Z^{(1)}_4=\{1,cd^2,c^2,c^3d^2\}, \quad & Z^{(2)}_4=\{1,cd^3,c^2d^2,c^3d\}, \quad & Z^{(3)}_4=\{1,c^2d^3,d^2,c^2d\},\\
Z^{(4)}_4=\{1,c,c^2,c^3\}, \quad & Z^{(5)}_4=\{1,d,d^2,d^3\},\quad & Z^{(6)}_4=\{1,cd,c^2d^2,c^3d^3\},\\
Z^{(7)}_4=\{1,abd^2,c^2,abc^2d^2\}, \quad  & Z^{(8)}_4=\{1,abcd^2,c^2,abc^3d^2\}, \quad & Z^{(9)}_4=\{1,a^2bc^2,d^2,a^2bc^2d^2\},\\
Z^{(10)}_4=\{1,a^2bc^2d,d^2,a^2bc^2d^3\},\quad & Z^{(11)}_4=\{1,bc^2,c^2d^2,bd^2\}, \quad & Z^{(12)}_4=\{1,bc^3d,c^2d^2,bcd^3\}
\end{array}
\end{equation}
\item{$Z_8$ subgroups}
\begin{equation}
\nonumber\begin{array}{l}
Z^{(1)}_8=\{1,abd,cd^2,abcd^3,c^2,abc^2d,c^3d^2,abc^3d^3\},\\
Z^{(2)}_8=\{1,abcd,cd^2,abc^2d^3,c^2,abc^3d,c^3d^2,abd^3\},\\
Z^{(3)}_8=\{1,a^2bc^3,c^2d^3,a^2bcd^3,d^2,a^2bc^3d^2,c^2d,a^2bcd\},\\
Z^{(4)}_8=\{1,a^2bc^3d,c^2d^3,a^2bc,d^2,a^2bc^3d^3,c^2d,a^2bcd^2\},\\
Z^{(5)}_8=\{1,bc,cd^3,bc^2d^3,c^2d^2,bc^3d^2,c^3d,bd\},\\
Z^{(6)}_8=\{1,bc^2d,cd^3,bc^3,c^2d^2,bd^3,c^3d,bcd^2\}
\end{array}
\end{equation}
\end{itemize}
We embed the left-handed lepton doublets in the faithful three-dimensional irreducible representations of $\Delta(96)$, by considering the large number of combinatorial choices of $G_{\nu}$ and $G_{\ell}$, all the possible lepton mixing matrices and the group structures generated by $G_{\nu}$ and $G_{\ell}$ are listed in Table \ref{tab:Z_3} and Table \ref{tab:Z_4_and_Z_8}, where the absolute values $||U_{PMNS}||$ of the mixing matrix are shown since we are interested in the mixing angles. We note that the mixing matrix is determined only up to permutation of rows and columns, because we don't predict the lepton masses in this approach. The generating elements of the $Z_4$ subgroups $Z^{(1)}_4$, $Z^{(2)}_4$ and $Z^{(3)}_4$ have two degenerate eigenvalues, the lepton mixing matrix can not be determined exactly if the flavor symmetry $\Delta(96)$ in broken to $Z^{(1)}_4$, $Z^{(2)}_4$ or $Z^{(3)}_4$ in the charged lepton sector. As a result, we don't consider these cases in the work. This issue is overcame by extending the above mentioned $Z_4$ to the product of cyclic groups $Z_2\times Z_4$ or $Z_4\times Z_4$ in Ref. \cite{deAdelhartToorop:2011re}. From Table \ref{tab:Z_3} and Table \ref{tab:Z_4_and_Z_8}, it is clear that seven mixing patterns including the tri-bimaximal, bimaximal and TFH mixings can be produced within $\Delta(96)$, where the ambiguity of the order of rows and columns certainly exists. If we require that the elements of $G_{\ell}$ and $G_{\nu}$ generate the full group $\Delta(96)$, only the TFH mixing patterns and the bimaximal mixing are admissible. For the former, $G_{\ell}$ belongs to the $Z_3$ subgroups, and $G_{\ell}$ is one of the $Z_8$ subgroups in the latter case. These results are consistent with those obtained in Ref. \cite{deAdelhartToorop:2011re}. It is remarkable that the generating elements of all the $Z_8$ subgroups and the $Z_4$ subgroups $Z^{(7)}_4,\ldots,Z^{(12)}_{4}$ for the faithful irreducible triplets don't have $+1$ eigenvalue, therefore there don't exist exist some vacuum alignment which breaks $\Delta(96)$ into these subgroups. However, the flavor mixing is associated with the hermitian combination $\widetilde{m}_{\ell}\equiv m^{\dagger}_{\ell}m_{\ell}$, where $m_{\ell}$ is the charged lepton mass matrix. The VEV of the flavon field and its complex conjugate enter into $\widetilde{m}_{\ell}$ simultaneously. The flavon's VEV multiplying the complex conjugate could be invariant under the action of the $Z_8$ and $Z_4$ subgroups although the VEV is not, if we properly choose the flavon fields and their vacuum. Consequently the remnant $Z_4$ or $Z_8$ symmetries can be preserved indirectly, if we concentrate on the flavor mixing. A explicit realization of this scenario in $A_4$ is presented in Ref. \cite{Ding:2011gt}.

It is worth noting that we are able to derive the tri-bimaximal mixing matrix, if the neutrino sector in invariant under the $K_4$ subgroup $K^{(2)}_4$ and the charged lepton sector invariant under $Z^{(1)}_3$, one can refer to Table \ref{tab:Z_3} for other possible choices of $G_{\ell}$ and $G_{\nu}$. However, the group generated through the elements of $Z^{(1)}_3$ and $K^{(2)}_4$ are $S_4$ instead of $\Delta(96)$. This result is consistent with the claim that the minimal flavor symmetry capable of yielding the tri-bimaximal mixing without fine tuning is $S_4$ from group theory point of view \cite{Lam:2008rs}.

By exchanging the rows and columns of the tri-bimaximal mixing matrix, we find another interesting mixing pattern,
\begin{equation}
\label{eq76}||U_{PMNS}||=\left(\begin{array}{ccc}
\frac{1}{\sqrt{2}}  &  \frac{1}{\sqrt{3}}  &  \frac{1}{\sqrt{6}}  \\
0 & \frac{1}{\sqrt{3}}  &  \sqrt{\frac{2}{3}}  \\
\frac{1}{\sqrt{2}}  &  \frac{1}{\sqrt{3}}  &  \frac{1}{\sqrt{6}}
\end{array}\right)\,.
\end{equation}
This texture will be called deformed tri-bimaximal (DTB) mixing, and the resulting mixing angles are
\begin{equation}
\label{eq77}\sin^2\theta_{13}=\frac{1}{6}\,,\quad\quad \sin^2\theta_{12}=\frac{2}{5}\,,\quad\quad \sin^2\theta_{23}=\frac{4}{5}\,.
\end{equation}
In order to be compatible with experimental data, all the three mixing angles should undergo large corrections of order $0.1\sim0.2$, which is roughly the size of the Cabibbo angle. This mixing pattern is an interesting alternative for explaining largish $\theta_{13}$ besides the bimaximal mixing \cite{Altarelli:2009gn}. In the case of bimaximal mixing as the LO mixing pattern, we have to tactfully construct the model so that the solar and reactor mixing angles get sizable corrections while the atmospheric mixing angle is protected from too large corrections. In contrast, the three mixing angles require corrections of the same order of magnitude for the DTB mixing. As a result, the corresponding models should be more easily to be constructed. Obviously the appropriate framework to derive DTB mixing is the $S_4$ horizontal symmetry \cite{Ding:progress}.

\begin{table}[hptb]
\begin{center}
\begin{tabular}{|c|c|c|c|}  \hline\hline
& ${\tt ||U_{PMNS}|| } $  &  {\tt Symmetry breaking}  & {\tt Generated }\\

 &   &  &  {\tt group} \\

 \hline

\multirow{6}{*}{I} & \multirow{6}{*}{$\frac{1}{\sqrt{3}}\left(\begin{array}{ccc}
1& 1  & 1\\
1 & 1 & 1\\
1 & 1 & 1
\end{array}\right)$ } &  $\big(Z^{(1)}_3, K^{(1)}_4\big)$, $\big(Z^{(2)}_3, K^{(1)}_4\big)$, $\big(Z^{(3)}_3, K^{(1)}_4\big)$, & \multirow{6}{*}{$A_4$}  \\

  & & $\big(Z^{(4)}_3, K^{(1)}_4\big)$, $\big(Z^{(5)}_3, K^{(1)}_4\big)$, $\big(Z^{(6)}_3, K^{(1)}_4\big)$, &  \\

  & & $\big(Z^{(7)}_3, K^{(1)}_4\big)$, $\big(Z^{(8)}_3, K^{(1)}_4\big)$,  $\big(Z^{(9)}_3, K^{(1)}_4\big)$, &  \\

  & &  $\big(Z^{(10)}_3, K^{(1)}_4\big)$, $\big(Z^{(11)}_3, K^{(1)}_4\big)$, $\big(Z^{(12)}_3, K^{(1)}_4\big)$, &  \\

  & & $\big(Z^{(13)}_3, K^{(1)}_4\big)$, $\big(Z^{(14)}_3, K^{(1)}_4\big)$, $\big(Z^{(15)}_3, K^{(1)}_4\big)$, & \\

  & & $\big(Z^{(16)}_3, K^{(1)}_4\big)$ \quad\quad\hskip4cm\quad  &  \\
 \hline

\multirow{16}{*}{II} &\multirow{16}{*}{$\left(\begin{array}{ccc}
\sqrt{\frac{2}{3}}  & \frac{1}{\sqrt{3}}   &  0  \\
\frac{1}{\sqrt{6}}  &  \frac{1}{\sqrt{3}}  & \frac{1}{\sqrt{2}}  \\
\frac{1}{\sqrt{6}}  &  \frac{1}{\sqrt{3}}  & \frac{1}{\sqrt{2}}
\end{array}\right)$ }  & $\big(Z^{(1)}_3, K^{(2)}_4\big)$,  $\big(Z^{(1)}_3, K^{(4)}_4\big)$, $\big(Z^{(1)}_3, K^{(6)}_4\big)$, & \multirow{16}{*}{$S_4$}   \\

 & & $\big(Z^{(2)}_3, K^{(3)}_4\big)$, $\big(Z^{(2)}_3, K^{(4)}_4\big)$, $\big(Z^{(2)}_3, K^{(7)}_4\big)$, & \\

 & & $\big(Z^{(3)}_3, K^{(2)}_4\big)$, $\big(Z^{(3)}_3, K^{(4)}_4\big)$, $\big(Z^{(3)}_3, K^{(6)}_4\big)$, &  \\

 & & $\big(Z^{(4)}_3, K^{(3)}_4\big)$, $\big(Z^{(4)}_3, K^{(4)}_4\big)$, $\big(Z^{(4)}_3, K^{(7)}_4\big)$, & \\

 & &  $\big(Z^{(5)}_3, K^{(3)}_4\big)$, $\big(Z^{(5)}_3, K^{(5)}_4\big)$, $\big(Z^{(5)}_3, K^{(6)}_4\big)$, & \\

 & & $\big(Z^{(6)}_3, K^{(2)}_4\big)$, $\big(Z^{(6)}_3, K^{(4)}_4\big)$, $\big(Z^{(6)}_3, K^{(6)}_4\big)$, & \\

 & & $\big(Z^{(7)}_3, K^{(3)}_4\big)$, $\big(Z^{(7)}_3, K^{(5)}_4\big)$, $\big(Z^{(7)}_3, K^{(6)}_4\big)$, &  \\

 & & $\big(Z^{(8)}_3, K^{(2)}_4\big)$, $\big(Z^{(8)}_3, K^{(5)}_4\big)$, $\big(Z^{(8)}_3, K^{(7)}_4\big)$,  &  \\

 & & $\big(Z^{(9)}_3, K^{(3)}_4\big)$, $\big(Z^{(9)}_3, K^{(4)}_4\big)$, $\big(Z^{(9)}_3, K^{(7)}_4\big)$,  &  \\

 & & $\big(Z^{(10)}_3, K^{(2)}_4\big)$, $\big(Z^{(10)}_3, K^{(5)}_4\big)$, $\big(Z^{(10)}_3, K^{(7)}_4\big)$, & \\

 & & $\big(Z^{(11)}_3, K^{(3)}_4\big)$, $\big(Z^{(11)}_3, K^{(5)}_4\big)$, $\big(Z^{(11)}_3, K^{(6)}_4\big)$, & \\

 & &  $\big(Z^{(12)}_3, K^{(2)}_4\big)$,  $\big(Z^{(12)}_3, K^{(4)}_4\big)$, $\big(Z^{(12)}_3, K^{(6)}_4\big)$, &  \\

 & & $\big(Z^{(13)}_3, K^{(3)}_4\big)$, $\big(Z^{(13)}_3, K^{(5)}_4\big)$, $\big(Z^{(13)}_3, K^{(6)}_4\big)$,  & \\

 & & $\big(Z^{(14)}_3, K^{(2)}_4\big)$, $\big(Z^{(14)}_3, K^{(5)}_4\big)$, $\big(Z^{(14)}_3, K^{(7)}_4\big)$,  &  \\

 & & $\big(Z^{(15)}_3, K^{(3)}_4\big)$, $\big(Z^{(15)}_3, K^{(4)}_4\big)$, $\big(Z^{(15)}_3, K^{(7)}_4\big)$, &  \\

 & &  $\big(Z^{(16)}_3, K^{(2)}_4\big)$, $\big(Z^{(16)}_3, K^{(5)}_4\big)$,$\big(Z^{(16)}_3, K^{(7)}_4\big)$   &     \\

 \hline

\multirow{16}{*}{III} & \multirow{16}{*}{$\left(\begin{array}{ccc}
\frac{1}{6}(3+\sqrt{3}) &  \frac{1}{\sqrt{3}}  &  \frac{1}{6}(3-\sqrt{3})  \\
\frac{1}{6}(3-\sqrt{3})  &  \frac{1}{\sqrt{3}}  &  \frac{1}{6}(3+\sqrt{3})  \\
\frac{1}{\sqrt{3}}   &  \frac{1}{\sqrt{3}}    &  \frac{1}{\sqrt{3}}
\end{array}\right)$}  &   $\big(Z^{(1)}_3, K^{(3)}_4\big)$, $\big(Z^{(1)}_3, K^{(5)}_4\big)$, $\big(Z^{(1)}_3, K^{(7)}_4\big)$, & \multirow{16}{*}{$\Delta(96)$}  \\

 & & $\big(Z^{(2)}_3, K^{(2)}_4\big)$, $\big(Z^{(2)}_3, K^{(5)}_4\big)$, $\big(Z^{(2)}_3, K^{(6)}_4\big)$,  &  \\

 & & $\big(Z^{(3)}_3, K^{(3)}_4\big)$, $\big(Z^{(3)}_3, K^{(5)}_4\big)$, $\big(Z^{(3)}_3, K^{(7)}_4\big)$, &  \\

 & & $\big(Z^{(4)}_3, K^{(2)}_4\big)$, $\big(Z^{(4)}_3, K^{(5)}_4\big)$, $\big(Z^{(4)}_3, K^{(6)}_4\big)$,  &  \\

 & & $\big(Z^{(5)}_3, K^{(2)}_4\big)$, $\big(Z^{(5)}_3, K^{(4)}_4\big)$, $\big(Z^{(5)}_3, K^{(7)}_4\big)$,  &   \\

 & & $\big(Z^{(6)}_3, K^{(3)}_4\big)$, $\big(Z^{(6)}_3, K^{(5)}_4\big)$, $\big(Z^{(6)}_3, K^{(7)}_4\big)$,   &  \\

 & & $\big(Z^{(7)}_3, K^{(2)}_4\big)$,  $\big(Z^{(7)}_3, K^{(4)}_4\big)$, $\big(Z^{(7)}_3, K^{(7)}_4\big)$, &   \\

 & & $\big(Z^{(8)}_3, K^{(3)}_4\big)$, $\big(Z^{(8)}_3, K^{(4)}_4\big)$, $\big(Z^{(8)}_3, K^{(6)}_4\big)$,  &   \\

 & & $\big(Z^{(9)}_3, K^{(2)}_4\big)$, $\big(Z^{(9)}_3, K^{(5)}_4\big)$, $\big(Z^{(9)}_3, K^{(6)}_4\big)$,  &  \\

 & & $\big(Z^{(10)}_3, K^{(3)}_4\big)$, $\big(Z^{(10)}_3, K^{(4)}_4\big)$, $\big(Z^{(10)}_3, K^{(6)}_4\big)$,  &   \\

 & & $\big(Z^{(11)}_3, K^{(2)}_4\big)$, $\big(Z^{(11)}_3, K^{(4)}_4\big)$, $\big(Z^{(11)}_3, K^{(7)}_4\big)$, &  \\

 & & $\big(Z^{(12)}_3, K^{(3)}_4\big)$, $\big(Z^{(12)}_3, K^{(5)}_4\big)$, $\big(Z^{(12)}_3, K^{(7)}_4\big)$, &   \\

 & & $\big(Z^{(13)}_3, K^{(2)}_4\big)$, $\big(Z^{(13)}_3, K^{(4)}_4\big)$, $\big(Z^{(13)}_3, K^{(7)}_4\big)$,  &   \\

 & & $\big(Z^{(14)}_3, K^{(3)}_4\big)$, ${\big(}Z^{(14)}_3, K^{(4)}_4{\big)}$, ${\big(}Z^{(14)}_3, K^{(6)}_4{\big)}$, &  \\

 & & $\big(Z^{(15)}_3, K^{(2)}_4\big)$, $\big(Z^{(15)}_3, K^{(5)}_4\big)$, $\big(Z^{(15)}_3, K^{(6)}_4\big)$,  &  \\

 & & $\big(Z^{(16)}_3, K^{(3)}_4\big)$, $\big(Z^{(16)}_3, K^{(4)}_4\big)$, $\big(Z^{(16)}_3, K^{(6)}_4\big)$  &   \\

\hline \hline

\end{tabular}
\end{center}
\caption{\label{tab:Z_3}The allowed leptonic mixing patterns and the generated group structures for the $\Delta(96)$ flavor symmetry breaking into $Z_3$ and $K_4$ in the charged lepton and neutrino sectors respectively.}
\end{table}

\begin{table}[hptb]
\begin{center}
\begin{tabular}{|c|c|c|c|}  \hline\hline
 & ${\tt ||U_{PMNS}|| } $  &  {\tt Symmetry breaking} & {\tt Generated} \\

  &   &  &  {\tt group} \\

 \hline

\multirow{3}{*}{IV} & \multirow{3}{*}{$\left(\begin{array}{ccc}
1&0 &0 \\
0&1&0\\
0&0&1
\end{array}\right)$}  &  $\big(Z^{(4)}_4, K^{(1)}_4\big)$, $\big(Z^{(5)}_4, K^{(1)}_4\big)$, $\big(Z^{(6)}_4, K^{(1)}_4\big)$, & \multirow{3}{*}{$Z_2\times Z_4$} \\

 & & $\big(Z^{(7)}_4, K^{(3)}_4\big)$, $\big(Z^{(8)}_4, K^{(2)}_4\big)$, $\big(Z^{(9)}_4, K^{(5)}_4\big)$,   &  \\

 & & $\big(Z^{(10)}_4, K^{(4)}_4\big)$, $\big(Z^{(11)}_4, K^{(7)}_4\big)$, $\big(Z^{(12)}_4, K^{(6)}_4\big)$  &   \\

 \hline

\multirow{12}{*}{V} & \multirow{12}{*}{$\left(\begin{array}{ccc}
1&0&0 \\
0&\frac{1}{\sqrt{2}}  & \frac{1}{\sqrt{2}} \\
0 & \frac{1}{\sqrt{2}} & \frac{1}{\sqrt{2}}
\end{array}\right)$}  &   $\big(Z^{(4)}_4, K^{(2)}_4\big)$, $\big(Z^{(4)}_4, K^{(3)}_4\big)$, $\big(Z^{(5)}_4, K^{(4)}_4\big)$, & \multirow{6}{*}{$D_4$}  \\

  & & $\big(Z^{(5)}_4, K^{(5)}_4\big)$, $\big(Z^{(6)}_4, K^{(6)}_4\big)$, $\big(Z^{(6)}_4, K^{(7)}_4\big)$,  &  \\

  &  & $\big(Z^{(7)}_4, K^{(1)}_4\big)$, $\big(Z^{(7)}_4, K^{(2)}_4\big)$, $\big(Z^{(8)}_4, K^{(1)}_4\big)$,   &  \\

  &  & $\big(Z^{(8)}_4, K^{(3)}_4\big)$, $\big(Z^{(9)}_4, K^{(1)}_4\big)$, $\big(Z^{(9)}_4, K^{(4)}_4\big)$,  &   \\

  & & $\big(Z^{(10)}_4, K^{(1)}_4\big)$, $\big(Z^{(10)}_4, K^{(5)}_4\big)$, $\big(Z^{(11)}_4, K^{(1)}_4\big)$,  &   \\

  &  & $\big(Z^{(11)}_4, K^{(6)}_4\big)$, $\big(Z^{(12)}_4, K^{(1)}_4\big)$, $\big(Z^{(12)}_4, K^{(7)}_4\big)$  &  \\  \cline{3-4}

  &   &  $\big(Z^{(4)}_4, K^{(4)}_4\big)$, $\big(Z^{(4)}_4, K^{(5)}_4\big)$, $\big(Z^{(4)}_4, K^{(6)}_4\big)$,  &  \multirow{4}{*}{$(Z_4\times Z_4)\rtimes Z_2$} \\

  &   & $\big(Z^{(4)}_4, K^{(7)}_4\big)$, $\big(Z^{(5)}_4, K^{(2)}_4\big)$, $\big(Z^{(5)}_4, K^{(3)}_4\big)$,  &  \\

 &   & $\big(Z^{(5)}_4, K^{(6)}_4\big)$, $\big(Z^{(5)}_4, K^{(7)}_4\big)$, $\big(Z^{(6)}_4, K^{(2)}_4\big)$,  &  \\

 &  & $\big(Z^{(6)}_4, K^{(3)}_4\big)$,  $\big(Z^{(6)}_4, K^{(4)}_4\big)$, $\big(Z^{(6)}_4, K^{(5)}_4\big)$,  &  \\
  \cline{3-4}

 & & $\big(Z^{(1)}_8, K^{(1)}_4\big)$, $\big(Z^{(2)}_8, K^{(1)}_4\big)$, $\big(Z^{(3)}_8, K^{(1)}_4\big)$,  & \multirow{2}{*}{$Z_8\rtimes Z_2$}  \\

 & & $\big(Z^{(4)}_8, K^{(1)}_4\big)$, $\big(Z^{(5)}_8, K^{(1)}_4\big)$, $\big(Z^{(6)}_8, K^{(1)}_4\big)$  &  \\  \hline

\multirow{4}{*}{VI} & \multirow{4}{*}{$\left(\begin{array}{ccc}
 1&0&0\\
 0 & \frac{\sqrt{2-\sqrt{2}}}{2}  & \frac{\sqrt{2+\sqrt{2}}}{2} \\
 0 & \frac{\sqrt{2+\sqrt{2}}}{2}  & \frac{\sqrt{2-\sqrt{2}}}{2}
  \end{array}\right)$} & $\big(Z^{(1)}_8, K^{(2)}_4\big)$, $\big(Z^{(1)}_8, K^{(3)}_4\big)$, $\big(Z^{(2)}_8, K^{(2)}_4\big)$, &  \multirow{4}{*}{$(Z_4\times Z_4)\rtimes Z_2$}  \\

  & & $\big(Z^{(2)}_8, K^{(3)}_4\big)$, $\big(Z^{(3)}_8, K^{(4)}_4\big)$, $\big(Z^{(3)}_8, K^{(5)}_4\big)$,  &   \\

  & & $\big(Z^{(4)}_8, K^{(4)}_4\big)$, $\big(Z^{(4)}_8, K^{(5)}_4\big)$, $\big(Z^{(5)}_8, K^{(6)}_4\big)$,  &   \\

  & & $\big(Z^{(5)}_8, K^{(7)}_4\big)$, $\big(Z^{(6)}_8, K^{(6)}_4\big)$, $\big(Z^{(6)}_8, K^{(7)}_4\big)$  &   \\ 

\hline

\multirow{16}{*}{VII} & \multirow{16}{*}{$\left(\begin{array}{ccc}
\frac{1}{\sqrt{2}}  & \frac{1}{\sqrt{2}}  &0 \\
\frac{1}{2}  &  \frac{1}{2}  & \frac{1}{\sqrt{2}} \\
\frac{1}{2}  &  \frac{1}{2}  & \frac{1}{\sqrt{2}} \\
\end{array}\right)$}  &  $\big(Z^{(7)}_4, K^{(4)}_4\big)$, $\big(Z^{(7)}_4, K^{(5)}_4\big)$, $\big(Z^{(7)}_4, K^{(6)}_4\big)$,  & \multirow{8}{*}{$S_4$} \\

 & & $\big(Z^{(7)}_4, K^{(7)}_4\big)$, $\big(Z^{(8)}_4, K^{(4)}_4\big)$, $\big(Z^{(8)}_4, K^{(5)}_4\big)$,  &   \\

 & &  $\big(Z^{(8)}_4, K^{(6)}_4\big)$, $\big(Z^{(8)}_4, K^{(7)}_4\big)$, $\big(Z^{(9)}_4, K^{(2)}_4\big)$,   &  \\

 & & $\big(Z^{(9)}_4, K^{(3)}_4\big)$, $\big(Z^{(9)}_4, K^{(6)}_4\big)$, $\big(Z^{(9)}_4, K^{(7)}_4\big)$,   &   \\

 & & $\big(Z^{(10)}_4, K^{(2)}_4\big)$, $\big(Z^{(10)}_4, K^{(3)}_4\big)$, $\big(Z^{(10)}_4, K^{(6)}_4\big)$,  &   \\

 & & $\big(Z^{(10)}_4, K^{(7)}_4\big)$, $\big(Z^{(11)}_4, K^{(2)}_4\big)$, $\big(Z^{(11)}_4, K^{(3)}_4\big)$,  &   \\

 & & $\big(Z^{(11)}_4, K^{(4)}_4\big)$, $\big(Z^{(11)}_4, K^{(5)}_4\big)$,  $\big(Z^{(12)}_4, K^{(2)}_4\big)$,  &  \\

 & & $\big(Z^{(12)}_4, K^{(3)}_4\big)$, $\big(Z^{(12)}_4, K^{(4)}_4\big)$, $\big(Z^{(12)}_4, K^{(5)}_4\big)$,  &  \\
 \cline{3-4}

 & & $\big(Z^{(1)}_8, K^{(4)}_4\big)$, $\big(Z^{(1)}_8, K^{(5)}_4\big)$, $\big(Z^{(1)}_8, K^{(6)}_4\big)$, &   \multirow{8}{*}{$\Delta(96)$} \\

 & & $\big(Z^{(1)}_8, K^{(7)}_4\big)$, $\big(Z^{(2)}_8, K^{(4)}_4\big)$, $\big(Z^{(2)}_8, K^{(5)}_4\big)$, &   \\

 & & $\big(Z^{(2)}_8, K^{(6)}_4\big)$, $\big(Z^{(2)}_8, K^{(7)}_4\big)$, $\big(Z^{(3)}_8, K^{(2)}_4\big)$,  &  \\

 & & $\big(Z^{(3)}_8, K^{(3)}_4\big)$, $\big(Z^{(3)}_8, K^{(6)}_4\big)$, $\big(Z^{(3)}_8, K^{(7)}_4\big)$,  &   \\

 & & $\big(Z^{(4)}_8, K^{(2)}_4\big)$, $\big(Z^{(4)}_8, K^{(3)}_4\big)$, $\big(Z^{(4)}_8, K^{(6)}_4\big)$,  &   \\

 & & $\big(Z^{(4)}_8, K^{(7)}_4\big)$, $\big(Z^{(5)}_8, K^{(2)}_4\big)$, $\big(Z^{(5)}_8, K^{(3)}_4\big)$,  &   \\

 & & $\big(Z^{(5)}_8, K^{(4)}_4\big)$, $\big(Z^{(5)}_8, K^{(5)}_4\big)$, $\big(Z^{(6)}_8, K^{(2)}_4\big)$,  &   \\

 & & $\big(Z^{(6)}_8, K^{(3)}_4\big)$, $\big(Z^{(6)}_8, K^{(4)}_4\big)$, $\big(Z^{(6)}_8, K^{(5)}_4\big)$   &   \\

\hline \hline

\end{tabular}
\caption{\label{tab:Z_4_and_Z_8}The allowed leptonic mixing patterns and the generated group structures for the $\Delta(96)$ flavor symmetry breaking into $Z_4$ or $Z_8$ in the charged lepton sector, and into $K_4$ in the neutrino sector.}
\end{center}
\end{table}

\section{Summary and conclusions}

The TFH mixing patterns, which are called {\tt M1} and {\tt M2} in Refs. \cite{Toorop:2011jn,deAdelhartToorop:2011re}, are favored by the latest experimental results \cite{Abe:2011sj,minos,double_chooz} and the global data fittings \cite{Schwetz:2011zk,Fogli:2011qn}. In this work, we perform a comprehensive study of the lepton flavor model within the flavor symmetry $\Delta(96)$, where the TFH textures are naturally derived at leading order. The neutrino mass matrix diagonalized by the TFH mixing and its symmetry properties are examined in the charged lepton diagonal basis. Within the so-called minimalist framework, we study the possible ways to produce TFH mixing within $\Delta(96)$, the assignments of the matter fields under $\Delta(96)$ and the associated flavons are presented. In these realizations, the left-handed lepton doublets are assigned to $\Delta(96)$ triplets $\mathbf{3_1}$, $\mathbf{3'_1}$, $\mathbf{\overline{3}_1}$ or $\mathbf{\overline{3}'_1}$, the TFH patterns can be naturally achieved if the flavor symmetry is broken into $Z_3$ in the charged lepton sector and $K_4$ in the neutrino sector. The claims of Feruglio et al. in Refs. \cite{Toorop:2011jn,deAdelhartToorop:2011re} are verified. There are three independent parameters in the charged lepton sector so that we can fit the three charged lepton masses. However, we need to tune the three parameters so that certain cancellation occurs to account for the tiny masses of electron and muon. This defect can be overcame by further breaking the remnant $Z_3$ symmetry of the charged lepton sector.

Two flavor models for TFH1 and TFH2 mixings are constructed, the
family symmetries are $\Delta(96)\times Z_3\times Z_3$ and
$\Delta(96)\times Z_5\times Z_2$ respectively. Both models embed the
left-handed lepton doublets into the triplet $\mathbf{3_1}$, and the
right-handed charged leptons are $\Delta(96)$ singlets, while the
right-handed neutrinos transform as $\mathbf{\overline{3}_1}$ in the
TFH1 model and $\mathbf{3_1}$ in the TFH2 model. As regards to the
spontaneous breaking of the flavor symmetry in the two models, the
neutrino sector is invariant under the $K_4$ subgroup generated by
$a^2bd$ and $d^2$ at leading order, the family symmetry is broken
completely in the charged lepton sector. The charged lepton mass
hierarchies are determined by the flavor symmetry breaking without
resorting to the Froggatt-Nielsen mechanism. Note that a residual
$Z_3$ symmetry generated by $a^2cd$ is preserved in the tau lepton
mass term of the TFH1 model, nevertheless the electron and muon mass
terms break the flavor symmetry $\Delta(96)$ completely. The next to
leading order corrections induced by higher-dimensional operators
are discussed in detail, we find that all the three lepton mixing
angles receive correction of the same order of magnitude. Since the
experimentally allowed departures of $\theta_{13}$ and $\theta_{12}$
from the TFH values $\sin^2\theta_{13}=(2-\sqrt{3})/6$ and
$\sin^2\theta_{12}=(8-2\sqrt{3})/13$ are small, at most of order
$\lambda^2_c$, the deviations of the mixing angles from the values
predicted by TFH mixing are expected to be at most of order
$\lambda^2_c$ in these models. The mixing angles in the
experimentally preferred range can be accommodated by our models, as
is demonstrated by the numerical analysis.

We investigate the possible mixing patterns if the family symmetry $\Delta(96)$ is broken into $K_4$ in the neutrino sector and the cyclic group $Z_N$  with $N\geq3$ in the charged lepton sector. Besides the tri-bimaximal and bimaximal patterns, the TFH mixing is found. If we impose the constraint that the residual symmetries $K_4$ and $Z_N$ generate the whole group $\Delta(96)$ instead of its subgroups, only bimaximal and TFH mixings are acceptable. This conclusion is consistent with that in Refs. \cite{Toorop:2011jn,deAdelhartToorop:2011re}. By permutating the rows and columns of the tri-bimaximal matrix, we can obtain the DTB texture which suggests $\sin^2\theta_{13}=1/6$, $\sin^2\theta_{12}=2/5$ and $\sin^2\theta_{23}=4/5$.
In order to be compatible with the experimental data in particular a sizable $\theta_{13}$ indicated by recent experiments, all the three mixing angles require corrections of order $\lambda_c$ in the case of DTB mixing. If bimaximal mixing is the leading order approximation, both $\theta_{13}$ and $\theta_{23}$ need large corrections of order $\lambda_c$, while so large corrections to $\theta_{23}$ should be forbidden. As a result, it may be better to take DTB rather than bimaximal as the leading order mixing pattern. $S_4$ is the appropriate family symmetry to derive this DTB mixing.

It is remarkable that $\Delta(96)$ has doublet representation which can be utilized to describe the quark sector. Because of the heaviness of the top quark, we could assign the quarks to the doublet and singlet representations of $\Delta(96)$. It is interesting to extend the present framework to the quark sector to give a coherent description of both quark and lepton.

\section*{Acknowledgements}

We are grateful to D.Meloni for his participation in the early stage
of the work. This work is supported by the National Natural Science
Foundation of China under Grant No.~10905053, Chinese Academy
KJCX2-YW-N29 and the 973 project with Grant No.~2009CB825200.

\vskip0.8cm

{\it Note added:} Near the completion of this work, the work \cite{deAdelhartToorop:2011re} appeared in the arXiv. The paper \cite{deAdelhartToorop:2011re} studied the lepton mixing patterns which can be derived from finite modular group $\Gamma_N$ by preserving the subgroups $G_{\nu}$ and $G_{\ell}$ in the neutrino and charged lepton sectors respectively. Section 7 of our work overlaps with section 3.5 of Ref. \cite{deAdelhartToorop:2011re}. We find the same results: none choices of $G_{\nu}=K_4$ and $G_{\ell}=Z_4$ can generate the full group $\Delta(96)$, only bimaximal mixing and TFH mixing are allowable if we require $G_{\nu}$ and $G_{\ell}$ generate $\Delta(96)$. Our work contains many associated details in addition, all possible combinations of $G_{\nu}=K_4$ and $G_{\ell}=Z_N (N\geq3)$ and the generated group structures are displayed in the present paper.

\vskip0.8cm

\appendix

\section*{\label{apd:a} Appendix A: Representation matrices  and other representation of $\Delta(96)$}

We would like to work in the charged lepton diagonal basis, this scenario can be realized by choosing the representation matrix of the element $a^2cd$ to be diagonal for various irreducible representations. To determine this basis, we start from the Escobar-Luhn (EL) basis of $\Delta(96)$ \cite{D96}, where the representation matrices for the generators $c$ and $d$ are diagonal. In the EL basis, the representation matrices for the group generators are particularly simple, their explicit forms are listed in Table \ref{tab:representation_matrix}, where $a_{1,2}$ and $c_{1,2}$ are $3\times3$ matrices,
\begin{eqnarray}
\nonumber&&a_1=\left(\begin{array}{ccc}
0&1&0\\
0&0&1\\
1&0&0
\end{array}\right),\quad\quad a_2=\left(\begin{array}{ccc}
0&0&1\\
1&0&0\\
0&1&0
\end{array}\right),\\
\label{eq:app1_1}&&c_1=\left(\begin{array}{ccc}
i &0 &0 \\
0&i&0\\
0&0&-1
\end{array}\right),\quad c_2=\left(\begin{array}{ccc}
-1&0&0\\
0&-i&0\\
0&0&-i
\end{array}\right)\,.
\end{eqnarray}

\begin{table}[t!]
\begin{center}
\begin{tabular}{|c|c|c|c|} \hline\hline
  &  $a$   &  $b$   &   $c$ \\ \hline

$\mathbf{1}$  &  1  &  1  &  1  \\  \hline

$\mathbf{1'}$  &  1   &   $-1$  & 1   \\  \hline

$\mathbf{2}$ & $ \left(
\begin{array}{cc}
 \omega & 0 \\
 0 & \omega^2
\end{array}
\right)$  &  $\left(
\begin{array}{cc}
 0 & 1 \\
 1 & 0
\end{array}
\right)$  &  $\left(
\begin{array}{cc}
 1 & 0 \\
 0 & 1
\end{array}
\right)$  \\\hline

$\mathbf{3_1}$ &  $\left(
\begin{array}{ccc}
 0 & 1 & 0 \\
 0 & 0 & 1 \\
 1 & 0 & 0
\end{array}
\right)$  &  $\left(
\begin{array}{ccc}
 0 & 0 & -1 \\
 0 & -1 & 0 \\
 -1 & 0 & 0
\end{array}
\right)$  &  $\left(
\begin{array}{ccc}
 i & 0 & 0 \\
 0 & -i & 0 \\
 0 & 0 & 1
\end{array}
\right)$   \\\hline

$\mathbf{3'_1}$  & $\left(
\begin{array}{ccc}
 0 & 1 & 0 \\
 0 & 0 & 1 \\
 1 & 0 & 0
\end{array}
\right) $  &   $\left(
\begin{array}{ccc}
 0 & 0 & 1 \\
 0 & 1 & 0 \\
 1 & 0 & 0
\end{array}
\right)$   &  $\left(
\begin{array}{ccc}
 i & 0 & 0 \\
 0 & -i & 0 \\
 0 & 0 & 1
\end{array}
\right)$  \\\hline

$\mathbf{\overline{3}_1}$ &  $\left(
\begin{array}{ccc}
 0 & 1 & 0 \\
 0 & 0 & 1 \\
 1 & 0 & 0
\end{array}
\right) $  &  $\left(
\begin{array}{ccc}
 0 & 0 & -1 \\
 0 & -1 & 0 \\
 -1 & 0 & 0
\end{array}
\right)$   &  $\left(
\begin{array}{ccc}
 -i & 0 & 0 \\
 0 & i & 0 \\
 0 & 0 & 1
\end{array}
\right)$   \\\hline

$\mathbf{\overline{3}'_1}$  &  $\left(
\begin{array}{ccc}
 0 & 1 & 0 \\
 0 & 0 & 1 \\
 1 & 0 & 0
\end{array}
\right)$  &  $\left(
\begin{array}{ccc}
 0 & 0 & 1 \\
 0 & 1 & 0 \\
 1 & 0 & 0
\end{array}
\right)$  &  $\left(
\begin{array}{ccc}
 -i & 0 & 0 \\
 0 & i & 0 \\
 0 & 0 & 1
\end{array}
\right)$   \\\hline

$\mathbf{3_2}$  & $\left(
\begin{array}{ccc}
 0 & 1 & 0 \\
 0 & 0 & 1 \\
 1 & 0 & 0
\end{array}
\right)$   &  $\left(
\begin{array}{ccc}
 0 & 0 & -1 \\
 0 & -1 & 0 \\
 -1 & 0 & 0
\end{array}
\right)$   &  $\left(
\begin{array}{ccc}
 -1 & 0 & 0 \\
 0 & -1 & 0 \\
 0 & 0 & 1
\end{array}
\right)$  \\\hline

$\mathbf{3'_2}$ &  $\left(
\begin{array}{ccc}
 0 & 1 & 0 \\
 0 & 0 & 1 \\
 1 & 0 & 0
\end{array}
\right)$   &  $\left(
\begin{array}{ccc}
 0 & 0 & 1 \\
 0 & 1 & 0 \\
 1 & 0 & 0
\end{array}
\right)$   &  $\left(
\begin{array}{ccc}
 -1 & 0 & 0 \\
 0 & -1 & 0 \\
 0 & 0 & 1
\end{array}
\right)$  \\\hline

$\mathbf{6}$  & $\left(\begin{array}{cc}
a_1&0\\
0 & a_2
\end{array}\right)$  &  $\left(\begin{array}{cc} 0&\mathbb{1}\\
\mathbb{1}  & 0
\end{array}\right)$  & $\left(\begin{array}{cc}
c_1 &0 \\
0 & c_2
\end{array}\right)$  \\\hline\hline

\end{tabular}
\caption{\label{tab:representation_matrix} The representation matrices for the generators $a$, $b$ and $c$ in the EL basis. Here we omit the generator $d$, as it is not independent of $a$, $b$ and $c$.}
\end{center}
\end{table}

Note that $a^2cd$ is not diagonal in EL basis, we can diagonalize it via a unitary transformation $U$ such that
\begin{equation}
\label{eq:app1_2}a=U^{\dagger}a_{EL}U,\quad\quad b=U^{\dagger}b_{EL}U,\quad\quad c=U^{\dagger}c_{EL}U,\quad\quad d=U^{\dagger}d_{EL}U
\end{equation}
where $a_{EL}$, $b_{EL}$, $c_{EL}$ and $d_{EL}$ denote the representation matrices in the EL basis. For $\mathbf{1}$, $\mathbf{1'}$ and $\mathbf{2}$, the representation matrices in our basis are exactly the same as those in the EL basis, the unitary transformation is identity. For the $\mathbf{3_1}$, we have
\begin{eqnarray}
\nonumber&&a_{\mathbf{3_1}}=\frac{1}{3}\left(
\begin{array}{ccc}
 \omega  & 1+\sqrt{3} & \left(1-\sqrt{3}\right)
   \omega ^2 \\
 1-\sqrt{3} & \omega ^2 & \left(1+\sqrt{3}\right)
   \omega  \\
 \left(1+\sqrt{3}\right) \omega ^2 & (1-\sqrt{3})\omega & 1
\end{array}
\right)\\
\nonumber&&b_{\mathbf{3_1}}=\frac{1}{3}\left(
\begin{array}{ccc}
 -1-\sqrt{3} & -\omega  & \left(\sqrt{3}-1\right)
   \omega ^2 \\
 -\omega ^2 & \sqrt{3}-1 &
   -\left(1+\sqrt{3}\right) \omega  \\
 \left(\sqrt{3}-1\right) \omega  &
   -\left(1+\sqrt{3}\right) \omega ^2 & -1
\end{array}
\right)\\
\label{eq:app1_3}&&c_{\mathbf{3_1}}=\frac{1}{3}\left(
\begin{array}{ccc}
 1 & \left(1-\sqrt{3}\right) \omega ^2 &
   \left(1+\sqrt{3}\right) \omega  \\
 \left(1+\sqrt{3}\right) \omega  & 1 &
   \left(1-\sqrt{3}\right) \omega ^2 \\
 \left(1-\sqrt{3}\right) \omega ^2 &
   \left(1+\sqrt{3}\right) \omega  & 1
\end{array}
\right)\,,
\end{eqnarray}
with
\begin{equation}
\label{eq:app1_4}U_{\mathbf{3_1}}=\frac{1}{\sqrt{3}}\left(
\begin{array}{ccc}
 -i \omega  & -i \omega  & -i \omega  \\
 \omega ^2 & 1 & \omega  \\
 1 & \omega ^2 & \omega
\end{array}
\right)\,,
\end{equation}
where $\omega\equiv e^{2\pi i/3}=-\frac{1}{2}+i\frac{\sqrt{3}}{2}$ is the cube root of unit. It is remarkable that the symmetry transformation of the light neutrino mass matrix given in Eq.(\ref{eq5}) can be expressed as $G_1=a_{\mathbf{3_1}}^2b_{\mathbf{3_1}}d_{\mathbf{3_1}}^3$, $G_2=d^2_{\mathbf{3_1}}$ and $G_3=a_{\mathbf{3_1}}^2b_{\mathbf{3_1}}d_{\mathbf{3_1}}$. For the representations $\mathbf{3'_1}$, $\mathbf{\overline{3}_1}$ and $\mathbf{\overline{3}'_1}$, the  results are
\begin{eqnarray}
\nonumber&&a_{\mathbf{3'_1}}=a_{\mathbf{3_1}},~~~b_{\mathbf{3'_1}}=-b_{\mathbf{3_1}},~~~c_{\mathbf{3'_1}}=c_{\mathbf{3_1}},~~~U_{\mathbf{3'_1}}=U_{\mathbf{3_1}}\\
\nonumber&&a_{\mathbf{\overline{3}_1}}=a^{*}_{\mathbf{3_1}},~~~b_{\mathbf{\overline{3}_1}}=b^{*}_{\mathbf{3_1}},~~~~~c_{\mathbf{\overline{3}_1}}=c^{*}_{\mathbf{3_1}},~~~U_{\mathbf{\overline{3}_1}}=U^{*}_{\mathbf{3}_1} \\
\label{eq:app1_5}&&a_{\mathbf{\overline{3}'_1}}=a^{*}_{\mathbf{3_1}},~~~b_{\mathbf{\overline{3}'_1}}=-b^{*}_{\mathbf{3_1}},~~~c_{\mathbf{\overline{3}'_1}}=c^{*}_{\mathbf{3_1}},~~~U_{\mathbf{\overline{3}'_1}}=U^{*}_{\mathbf{3_1}}\,.
\end{eqnarray}
In the case of $\mathbf{3_2}$ and $\mathbf{3'_2}$, we have
\begin{eqnarray}
\nonumber&a_{\mathbf{3_2}}=\frac{1}{3}\left(
\begin{array}{ccc}
 -\omega  & 2 & 2 \omega ^2 \\
 2 & -\omega ^2 & 2 \omega  \\
 2 \omega ^2 & 2 \omega  & -1
\end{array}
\right),b_{\mathbf{3_2}}=\frac{1}{3}\left(
\begin{array}{ccc}
 -2 & \omega  & -2 \omega ^2 \\
 \omega ^2 & -2 & -2 \omega  \\
 -2 \omega  & -2 \omega ^2 & 1
\end{array}
\right),c_{\mathbf{3_2}}=\frac{1}{3}\left(
\begin{array}{ccc}
 -1 & 2 \omega ^2 & 2 \omega  \\
 2 \omega  & -1 & 2 \omega ^2 \\
 2 \omega ^2 & 2 \omega  & -1
\end{array}
\right)\\
\label{eq:app1_6}&a_{\mathbf{3'_2}}=a_{\mathbf{3_2}},~~~b_{\mathbf{3'_2}}=-b_{\mathbf{3_2}},~~~c_{\mathbf{3'_2}}=c_{\mathbf{3_2}},~~~U_{\mathbf{3'_2}}=U_{\mathbf{3_2}}=\mathrm{diag}(-i,1,1)\cdot U_{\mathbf{3_1}}\,.
\end{eqnarray}
Finally, for the sextet $\mathbf{6}$
\begin{eqnarray}
\label{eq:app1_7}a=\left(\begin{array}{cc}
\mathbb{a}_1 & 0 \\
0  &  \mathbb{a}_2
\end{array}\right),~~~b=\left(\begin{array}{cc}
0  &  \mathbb{b}_1  \\
\mathbb{b}_2  & 0
\end{array}\right),~~~c=\left(\begin{array}{cc}
\mathbb{c}_1  &  0\\
0 &  \mathbb{c}_2
\end{array}\right)\,,
\end{eqnarray}
where
\begin{eqnarray}
\nonumber&&\mathbb{a}_1=\frac{1}{3}\left(
\begin{array}{ccc}
 (-1-2 i) \omega  & -1+i & (-1+i) \omega ^2 \\
 -1+i & (-1-2 i) \omega ^2 & (-1+i) \omega  \\
 (-1+i) \omega ^2 & (-1+i) \omega  & -1-2 i
\end{array}
\right)\\
\nonumber&&\mathbb{a}_2=\frac{1}{3}\left(
\begin{array}{ccc}
 (-1+2 i) \omega  & -1-i & (-1-i) \omega ^2 \\
 -1-i & (-1+2 i) \omega ^2 & (-1-i) \omega  \\
 (-1-i) \omega ^2 & (-1-i) \omega  & -1+2 i
\end{array}
\right)\\
\nonumber&&\mathbb{b}_1=\frac{1}{3}\left(
\begin{array}{ccc}
 (-1+i) \omega ^2 & -1-2 i & (-1+i) \omega  \\
 (-1-2 i) \omega  & (-1+i) \omega ^2 & -1+i \\
 -1+i & (-1+i) \omega  & (-1-2 i) \omega ^2
\end{array}
\right)\\
\nonumber&&\mathbb{b}_2=\frac{1}{3}\left(
\begin{array}{ccc}
 (-1-i) \omega  & (-1+2 i) \omega ^2 & -1-i \\
 -1+2 i & (-1-i) \omega  & (-1-i) \omega ^2 \\
 (-1-i) \omega ^2 & -1-i & (-1+2 i) \omega
\end{array}
\right)\\
\nonumber&&\mathbb{c}_1=\frac{1}{3}\left(
\begin{array}{ccc}
 -1+2 i & (-1-i) \omega ^2 & (-1-i) \omega  \\
 (-1-i) \omega  & -1+2 i & (-1-i) \omega ^2 \\
 (-1-i) \omega ^2 & (-1-i) \omega  & -1+2 i
\end{array}
\right)\\
\label{eq:app1_8}&&\mathbb{c}_2=\frac{1}{3}\left(
\begin{array}{ccc}
 -1-2 i & (-1+i) \omega ^2 & (-1+i) \omega  \\
 (-1+i) \omega  & -1-2 i & (-1+i) \omega ^2 \\
 (-1+i) \omega ^2 & (-1+i) \omega  & -1-2 i
\end{array}
\right)\,.
\end{eqnarray}
The unitary matrix $U_{\mathbf{6}}$ is
\begin{equation}
\label{eq:app1_9}U_{\mathbf{6}}=\left(\begin{array}{cc}
\mathbb{u}_1  &  0\\
0  & \mathbb{u}_2
\end{array}\right)\,,
\end{equation}
with
\begin{eqnarray}
\nonumber&&\mathbb{u}_1=\mathrm{diag}(i,-i,i)\cdot U_{\mathbf{3_1}}\\
\label{eq:app1_10}&&\mathbb{u}_2=\frac{1}{\sqrt{3}}\left(
\begin{array}{ccc}
 -i \omega ^2 & -i \omega  & -i \\
 i \omega  & i \omega ^2 & i \\
 1 & 1 & 1
\end{array}
\right)\,.
\end{eqnarray}

\section*{\label{apd:b}Appendix B: Clebsch-Gordan Coefficients of $\Delta(96)$}

We report here the complete list of the Clebsch-Gordan coefficients for our basis. In the following we use $\alpha_i$ to indicate the elements of the first representation of the product and $\beta_i$ to denote those of the second representation.

\begin{itemize}

\item{$\mathbf{1'}\otimes\mathbf{2}=\mathbf{2}$}

\begin{equation}
\nonumber\mathbf{2}\sim\left(
\begin{array}{c}
 \alpha _1 \beta _1 \\
 -\alpha _1 \beta _2
\end{array}
\right)
\end{equation}


\item{$\mathbf{1'}\otimes\mathbf{3_1}=\mathbf{3'_1}$}

\begin{equation}
\nonumber\mathbf{3'_1}\sim\left(
\begin{array}{c}
 \alpha _1 \beta _1 \\
 \alpha _1 \beta _2 \\
 \alpha _1 \beta _3
\end{array}
\right)
\end{equation}


\item{$\mathbf{1'\otimes\mathbf{3'_1}}=\mathbf{3_1}$}

\begin{equation}
\nonumber\mathbf{3_1}\sim\left(
\begin{array}{c}
 \alpha _1 \beta _1 \\
 \alpha _1 \beta _2 \\
 \alpha _1 \beta _3
\end{array}
\right)
\end{equation}


\item{$\mathbf{1'}\otimes\mathbf{\overline{3}_1}=\mathbf{\overline{3}'_1}$}

\begin{equation}
\nonumber\mathbf{\overline{3}'_1}\sim\left(
\begin{array}{c}
 \alpha _1 \beta _1 \\
 \alpha _1 \beta _2 \\
 \alpha _1 \beta _3
\end{array}
\right)
\end{equation}


\item{$\mathbf{1'}\otimes\mathbf{\overline{3}'_1}=\mathbf{\overline{3}_1}$}

\begin{equation}
\nonumber\mathbf{\overline{3}_1}\sim\left(
\begin{array}{c}
 \alpha _1 \beta _1 \\
 \alpha _1 \beta _2 \\
 \alpha _1 \beta _3
\end{array}
\right)
\end{equation}


\item{$\mathbf{1'}\otimes\mathbf{3_2}=\mathbf{3'_2}$}

\begin{equation}
\nonumber\mathbf{3'_2}\sim\left(
\begin{array}{c}
 \alpha _1 \beta _1 \\
 \alpha _1 \beta _2 \\
 \alpha _1 \beta _3
\end{array}
\right)
\end{equation}


\item{$\mathbf{1'}\otimes\mathbf{3'_2}=\mathbf{3_2}$}

\begin{equation}
\nonumber\mathbf{3_2}\sim\left(
\begin{array}{c}
 \alpha _1 \beta _1 \\
 \alpha _1 \beta _2 \\
 \alpha _1 \beta _3
\end{array}
\right)
\end{equation}


\item{$\mathbf{1'}\otimes\mathbf{6}=\mathbf{6}$}

\begin{equation}
\nonumber\mathbf{6}\sim\left(
\begin{array}{c}
 \alpha _1 \beta _1 \\
 \alpha _1 \beta _2 \\
 \alpha _1 \beta _3 \\
 -\alpha _1 \beta _4 \\
 -\alpha _1 \beta _5 \\
 -\alpha _1 \beta _6
\end{array}
\right)
\end{equation}


\item{$\mathbf{2}\otimes\mathbf{2}=\mathbf{1}\oplus\mathbf{1'}\oplus\mathbf{2}$}

\begin{equation}
\nonumber\mathbf{1}\sim \alpha _1 \beta _2+\alpha _2 \beta _1
\end{equation}

\begin{equation}
\nonumber\mathbf{1'}\sim\alpha _1 \beta _2-\alpha _2 \beta _1
\end{equation}

\begin{equation}
\nonumber\mathbf{2}\sim\left(
\begin{array}{c}
 \alpha _2 \beta _2 \\
 \alpha _1 \beta _1
\end{array}
\right)
\end{equation}


\item{$\mathbf{2}\otimes\mathbf{3_1}=\mathbf{3_1}\oplus\mathbf{3'_1}$}

\begin{equation}
\nonumber\mathbf{3_1}\sim\left(
\begin{array}{c}
 \alpha _1 \beta _3+\omega ^2\alpha _2 \beta _2 \\
 \alpha _1 \beta _1+\omega ^2\alpha _2 \beta _3  \\
 \alpha _1 \beta _2+\omega ^2\alpha _2 \beta _1
\end{array}
\right)
\end{equation}

\begin{equation}
\nonumber\mathbf{3'_1}\sim\left(
\begin{array}{c}
 \alpha _1 \beta _3-\omega ^2\alpha _2 \beta _2  \\
 \alpha _1 \beta _1-\omega ^2\alpha _2 \beta _3  \\
 \alpha _1 \beta _2-\omega ^2\alpha _2 \beta _1
\end{array}
\right)
\end{equation}


\item{$\mathbf{2}\otimes\mathbf{3'_1}=\mathbf{3_1}\oplus\mathbf{3'_1}$}

\begin{equation}
\nonumber\mathbf{3_1}\sim\left(
\begin{array}{c}
 \alpha _1 \beta _3-\omega ^2\alpha _2 \beta _2  \\
 \alpha _1 \beta _1-\omega ^2\alpha _2 \beta _3 \\
 \alpha _1 \beta _2-\omega ^2\alpha _2 \beta _1
\end{array}
\right)
\end{equation}

\begin{equation}
\nonumber\mathbf{3'_1}\sim\left(
\begin{array}{c}
 \alpha _1 \beta _3+\omega ^2\alpha _2 \beta _2  \\
 \alpha _1 \beta _1+\omega ^2\alpha _2 \beta _3  \\
 \alpha _1 \beta _2+\omega ^2\alpha _2 \beta _1
\end{array}
\right)
\end{equation}


\item{$\mathbf{2}\otimes\mathbf{\overline{3}_1}=\mathbf{\overline{3}_1}\oplus\mathbf{\overline{3}'_1}$}

\begin{equation}
\nonumber\mathbf{\overline{3}_1}\sim\left(
\begin{array}{c}
 \alpha _1 \beta _2+\omega ^2\alpha _2 \beta _3  \\
 \alpha _1 \beta _3+\omega ^2\alpha _2 \beta _1  \\
 \alpha _1 \beta _1+\omega ^2\alpha _2 \beta _2
\end{array}
\right)
\end{equation}

\begin{equation}
\nonumber\mathbf{\overline{3}'_1}\sim\left(
\begin{array}{c}
 \alpha _1 \beta _2-\omega ^2\alpha _2 \beta _3  \\
 \alpha _1 \beta _3-\omega ^2\alpha _2 \beta _1  \\
 \alpha _1 \beta _1-\omega ^2\alpha _2 \beta _2
\end{array}
\right)
\end{equation}


\item{$\mathbf{2}\otimes\mathbf{\overline{3}'_1}=\mathbf{\overline{3}_1}\oplus\mathbf{\overline{3}'_1}$}

\begin{equation}
\nonumber\mathbf{\overline{3}_1}\sim\left(
\begin{array}{c}
 \alpha _1 \beta _2-\omega ^2\alpha _2 \beta _3  \\
 \alpha _1 \beta _3-\omega ^2\alpha _2 \beta _1  \\
 \alpha _1 \beta _1-\omega ^2\alpha _2 \beta _2
\end{array}
\right)
\end{equation}

\begin{equation}
\nonumber\mathbf{\overline{3}'_1}\sim\left(
\begin{array}{c}
 \alpha _1 \beta _2+\omega ^2\alpha _2 \beta _3  \\
 \alpha _1 \beta _3+\omega ^2\alpha _2 \beta _1  \\
 \alpha _1 \beta _1+\omega ^2\alpha _2 \beta _2
\end{array}
\right)
\end{equation}


\item{$\mathbf{2}\otimes\mathbf{3_2}=\mathbf{3_2}\oplus\mathbf{3'_2}$}

\begin{equation}
\nonumber\mathbf{3_2}\sim\left(
\begin{array}{c}
 \alpha _1 \beta _3+\omega ^2\alpha _2 \beta _2  \\
 \alpha _1 \beta _1+\omega ^2\alpha _2 \beta _3  \\
 \alpha _1 \beta _2+\omega ^2\alpha _2 \beta _1
\end{array}
\right)
\end{equation}

\begin{equation}
\nonumber\mathbf{3'_2}\sim\left(
\begin{array}{c}
 \alpha _1 \beta _3-\omega ^2\alpha _2 \beta _2  \\
 \alpha _1 \beta _1-\omega ^2\alpha _2 \beta _3  \\
 \alpha _1 \beta _2-\omega ^2\alpha _2 \beta _1
\end{array}
\right)
\end{equation}


\item{$\mathbf{2}\otimes\mathbf{3'_2}=\mathbf{3_2}\oplus\mathbf{3'_2}$}

\begin{equation}
\nonumber\mathbf{3_2}\sim\left(
\begin{array}{c}
 \alpha _1 \beta _3-\omega ^2\alpha _2 \beta _2  \\
 \alpha _1 \beta _1-\omega ^2\alpha _2 \beta _3  \\
 \alpha _1 \beta _2-\omega ^2\alpha _2 \beta _1
\end{array}
\right)
\end{equation}

\begin{equation}
\nonumber\mathbf{3'_2}\sim\left(
\begin{array}{c}
 \alpha _1 \beta _3+\omega ^2\alpha _2 \beta _2  \\
 \alpha _1 \beta _1+\omega ^2\alpha _2 \beta _3  \\
 \alpha _1 \beta _2+\omega ^2\alpha _2 \beta _1
\end{array}
\right)
\end{equation}


\item{$\mathbf{2}\otimes\mathbf{6}=\mathbf{6_1}\oplus\mathbf{6_2}$}

\begin{equation}
\nonumber\mathbf{6_1}\sim\left(
\begin{array}{c}
 \alpha _2 \beta _2 \\
 \alpha _2 \beta _3 \\
 \alpha _2 \beta _1 \\
 \omega\alpha _1 \beta _6   \\
 \omega\alpha _1 \beta _4  \\
 \omega\alpha _1 \beta _5
\end{array}
\right)
\end{equation}

\begin{equation}
\nonumber\mathbf{6_2}\sim\left(
\begin{array}{c}
 \alpha _1 \beta _3 \\
 \alpha _1 \beta _1 \\
 \alpha _1 \beta _2 \\
 \omega ^2\alpha _2 \beta _5  \\
 \omega ^2\alpha _2 \beta _6  \\
 \omega ^2\alpha _2 \beta _4
\end{array}
\right)
\end{equation}


\item{$\mathbf{3_1}\otimes\mathbf{3_1}=\mathbf{\overline{3}_1}\oplus\mathbf{\overline{3}'_1}\oplus\mathbf{3'_2}$}

\begin{equation}
\nonumber\mathbf{\overline{3}_1}\sim\left(
\begin{array}{c}
 \alpha _2 \beta _3-\alpha _3 \beta _2 \\
 \alpha _3 \beta _1-\alpha _1 \beta _3 \\
 \alpha _1 \beta _2-\alpha _2 \beta _1
\end{array}
\right)
\end{equation}

\begin{equation}
\nonumber\mathbf{\overline{3}'_1}\sim\left(
\begin{array}{c}
 -2 \alpha _1 \beta _1+\alpha _2 \beta _3 +\alpha _3 \beta _2\\
 \alpha _1 \beta _3-2 \alpha _2 \beta _2+\alpha _3 \beta _1 \\
 \alpha _1 \beta _2+\alpha _2 \beta _1-2 \alpha _3 \beta _3
\end{array}
\right)
\end{equation}

\begin{equation}
\nonumber\mathbf{3'_2}\sim\left(
\begin{array}{c}
 \alpha _1 \beta _3+\alpha _2 \beta _2+\alpha _3 \beta _1 \\
 \alpha _1 \beta _1+\alpha _2 \beta _3+\alpha _3 \beta _2 \\
 \alpha _1 \beta _2+\alpha _2 \beta _1+\alpha _3 \beta _3
\end{array}
\right)
\end{equation}


\item{$\mathbf{3_1}\otimes\mathbf{3'_1}=\mathbf{\overline{3}_1}\oplus\mathbf{\overline{3}'_1}\oplus\mathbf{3_2}$}

\begin{equation}
\nonumber\mathbf{\overline{3}_1}\sim\left(
\begin{array}{c}
 -2 \alpha _1 \beta _1+\alpha _2 \beta _3+\alpha _3 \beta _2 \\
 \alpha _1 \beta _3 -2 \alpha _2 \beta _2+\alpha _3 \beta _1\\
 \alpha _1 \beta _2+\alpha _2 \beta _1-2 \alpha _3 \beta _3
\end{array}
\right)
\end{equation}

\begin{equation}
\nonumber\mathbf{\overline{3}'_1}\sim\left(
\begin{array}{c}
 \alpha _2 \beta _3-\alpha _3 \beta _2 \\
 \alpha _3 \beta _1-\alpha _1 \beta _3 \\
 \alpha _1 \beta _2-\alpha _2 \beta _1
\end{array}
\right)
\end{equation}

\begin{equation}
\nonumber\mathbf{3_2}\sim\left(
\begin{array}{c}
 \alpha _1 \beta _3+\alpha _2 \beta _2+\alpha _3 \beta _1 \\
 \alpha _1 \beta _1+\alpha _2 \beta _3 +\alpha _3 \beta _2\\
 \alpha _1 \beta _2+\alpha _2 \beta _1+\alpha _3 \beta _3
\end{array}
\right)
\end{equation}


\item{$\mathbf{3_1}\otimes\mathbf{\overline{3}_1}=\mathbf{1}\oplus\mathbf{2}\oplus\mathbf{6}$}

\begin{equation}
\nonumber\mathbf{1}\sim\alpha _1 \beta _1+\alpha _2 \beta _2+\alpha _3 \beta _3
\end{equation}

\begin{equation}
\nonumber\mathbf{2}\sim\left(
\begin{array}{c}
\alpha_1 \beta _3+\alpha _2 \beta _1+\alpha _3 \beta _2 \\
 \omega\left( \alpha _1 \beta _2+\alpha _2 \beta _3+\alpha _3 \beta _1\right)
\end{array}
\right)
\end{equation}

\begin{equation}
\nonumber\mathbf{6}\sim\left(
\begin{array}{c}
\alpha _1 \beta _3  +\omega\alpha _2 \beta _1 +\omega^2\alpha _3 \beta _2 \\
 \omega \alpha _1 \beta _2+\omega^2\alpha_2 \beta _3+\alpha _3 \beta _1   \\
 \omega^2\alpha_1 \beta _1+\alpha _2 \beta _2  +\omega\alpha _3 \beta _3  \\
 \alpha _1 \beta _3  +\omega^2\alpha_2 \beta _1+ \omega\alpha _3 \beta _2  \\
 \omega^2\alpha_1 \beta _2+\omega\alpha _2 \beta _3 +\alpha _3 \beta _1   \\
 \omega\alpha _1 \beta _1 +\alpha _2 \beta _2  +\omega^2\alpha _3 \beta _3
\end{array}
\right)
\end{equation}


\item{$\mathbf{3_1}\otimes\mathbf{\overline{3}'_1}=\mathbf{1'}\oplus\mathbf{2}\oplus\mathbf{6}$}

\begin{equation}
\nonumber\mathbf{1'}\sim\alpha _1 \beta _1+\alpha _2 \beta _2+\alpha _3 \beta _3
\end{equation}

\begin{equation}
\nonumber\mathbf{2}\sim\left(
\begin{array}{c}
 \alpha_1 \beta _3+\alpha _2 \beta _1+\alpha _3 \beta _2 \\
 -\omega \left(\alpha _1 \beta _2+\alpha _2 \beta _3+\alpha _3 \beta _1\right)
\end{array}
\right)
\end{equation}

\begin{equation}
\nonumber\mathbf{6}\sim\left(
\begin{array}{c}
\alpha _1 \beta _3+\omega\alpha _2 \beta _1+\omega^2\alpha _3 \beta _2 \\
 \omega\alpha _1 \beta _2 +\omega^2\alpha _2 \beta _3+\alpha _3 \beta _1  \\
 \omega^2\alpha _1 \beta _1+\alpha _2 \beta _2+\omega\alpha _3 \beta _3   \\
 -\alpha _1 \beta _3 -\omega^2\alpha _2 \beta _1-\omega\alpha _3 \beta _2  \\
 -\omega^2\alpha _1 \beta _2-\omega\alpha _2 \beta _3 -\alpha _3 \beta _1   \\
 -\omega\alpha _1 \beta _1 -\alpha _2 \beta _2 -\omega^2\alpha_3 \beta _3
\end{array}
\right)
\end{equation}


\item{$\mathbf{3_1}\otimes\mathbf{3_2}=\mathbf{\overline{3}'_1}\oplus\mathbf{6}$}

\begin{equation}
\nonumber\mathbf{\overline{3}'_1}\sim\left(
\begin{array}{c}
 \alpha _1 \beta _1+\alpha _2 \beta _3+\alpha _3 \beta _2 \\
  \alpha _1 \beta _3+\alpha _2 \beta _2+\alpha _3 \beta _1\\
 \alpha _1 \beta _2+\alpha _2 \beta _1+\alpha _3 \beta _3
\end{array}
\right)
\end{equation}

\begin{equation}
\nonumber\mathbf{6}\sim\left(
\begin{array}{c}
 \alpha _1 \beta _3 +\omega\alpha _2 \beta _2+\omega^2\alpha _3 \beta _1  \\
 \omega^2\alpha _1 \beta _1+\alpha _2 \beta _3 +\omega\alpha _3 \beta _2 \\
 \omega\alpha_1 \beta _2+\omega^2\alpha _2 \beta _1  + \alpha _3 \beta _3  \\
 \omega\alpha _1 \beta _3+\alpha _2 \beta _2 +\omega^2\alpha _3 \beta _1   \\
 \omega^2\alpha _1 \beta _1  +\omega\alpha _2 \beta _3+\alpha _3 \beta _2  \\
 \alpha _1 \beta _2 +\omega^2\alpha _2 \beta _1+\omega\alpha _3 \beta _3
\end{array}
\right)
\end{equation}


\item{$\mathbf{3_1}\otimes\mathbf{3'_2}=\mathbf{\overline{3}_1}\oplus\mathbf{6}$}

\begin{equation}
\nonumber\mathbf{\overline{3}_1}\sim\left(
\begin{array}{c}
 \alpha _1 \beta _1+\alpha _2 \beta _3+\alpha _3 \beta _2 \\
 \alpha _1 \beta _3+\alpha _2 \beta _2+\alpha _3 \beta _1 \\
 \alpha _1 \beta _2+\alpha _2 \beta _1+\alpha _3 \beta _3
\end{array}
\right)
\end{equation}

\begin{equation}
\nonumber\mathbf{6}\sim\left(
\begin{array}{c}
 \alpha _1 \beta _3+\omega\alpha _2 \beta _2 +\omega^2\alpha _3 \beta _1  \\
 \omega^2\alpha _1 \beta _1 +\alpha _2 \beta _3 +\omega\alpha _3 \beta _2 \\
 \omega\alpha _1 \beta _2 +\omega^2\alpha _2 \beta _1+\alpha _3 \beta _3 \\
 -\omega\alpha_1 \beta _3-\alpha _2 \beta _2 -\omega^2\alpha _3 \beta _1   \\
 -\omega^2\alpha _1 \beta _1-\omega\alpha_2 \beta _3-\alpha _3 \beta _2 \\
 -\alpha _1 \beta _2 -\omega^2\alpha _2 \beta _1  -\omega\alpha_3 \beta _3
\end{array}
\right)
\end{equation}


\item{$\mathbf{3_1}\otimes\mathbf{6}=\mathbf{3_1}\oplus\mathbf{3'_1}\oplus\mathbf{3_2}\oplus\mathbf{3'_2}\oplus\mathbf{6}$}

\begin{equation}
\nonumber\mathbf{3_1}\sim\left(
\begin{array}{c}
 \alpha _1 \beta _3+\omega\alpha _1 \beta _6 +\omega\alpha _2 \beta _2  +\alpha _2 \beta _5+\omega ^2\alpha _3 \beta _1 +\omega^2\alpha _3 \beta _4  \\
 \omega\alpha _1 \beta _1+\alpha _1 \beta _4 +\omega ^2\alpha _2 \beta _3 +\omega^2\alpha _2 \beta _6 +\alpha _3 \beta _2+\omega\alpha _3 \beta _5  \\
 \omega ^2\alpha _1 \beta _2+\omega^2\alpha _1 \beta _5+\alpha _2 \beta _1+\omega\alpha _2 \beta _4+\omega\alpha _3 \beta _3+\alpha _3 \beta _6
\end{array}
\right)
\end{equation}

\begin{equation}
\nonumber\mathbf{3'_1}\sim\left(
\begin{array}{c}
 \alpha _1 \beta _3-\omega\alpha _1 \beta _6+\omega\alpha _2 \beta _2-\alpha _2 \beta _5+\omega ^2\alpha _3 \beta _1 -\omega^2\alpha _3 \beta _4   \\
 \omega\alpha _1 \beta _1-\alpha _1 \beta _4+\omega ^2\alpha _2 \beta _3 -\omega^2\alpha _2 \beta _6+\alpha _3 \beta _2-\omega\alpha _3 \beta _5  \\
 \omega ^2\alpha _1 \beta _2 -\omega^2\alpha _1 \beta _5+\alpha _2 \beta _1 -\omega\alpha _2 \beta _4 +\omega\alpha _3 \beta _3-\alpha _3 \beta _6
\end{array}
\right)
\end{equation}

\begin{equation}
\nonumber\mathbf{3_2}\sim\left(
\begin{array}{c}
 \alpha _1 \beta _3+\omega\alpha_1 \beta _6+\alpha _2 \beta_2+\omega\alpha _2 \beta _5+\alpha _3 \beta _1 +\omega\alpha _3 \beta _4    \\
 \omega ^2\alpha _1 \beta _1 +\omega^2\alpha _1 \beta _4+\omega ^2\alpha _2 \beta _3+\omega ^2\alpha _2 \beta _6+\omega^2\alpha _3 \beta _2 +\omega ^2\alpha _3 \beta _5   \\
  \omega\alpha _1 \beta _2+\alpha _1 \beta_5+\omega\alpha _2 \beta _1+\alpha _2 \beta _4+\omega\alpha_3 \beta _3+\alpha _3 \beta _6
\end{array}
\right)
\end{equation}

\begin{equation}
\nonumber\mathbf{3'_2}\sim\left(
\begin{array}{c}
\alpha _1 \beta _3-\omega\alpha_1 \beta _6+\alpha _2 \beta_2-\omega\alpha _2 \beta _5+\alpha _3 \beta _1 -\omega\alpha _3 \beta _4    \\
 \omega ^2\alpha _1 \beta _1-\omega^2\alpha _1 \beta _4+\omega ^2\alpha _2 \beta _3-\omega ^2\alpha_2 \beta _6+\omega^2\alpha _3 \beta _2 -\omega^2\alpha _3 \beta _5   \\
 \omega\alpha _1 \beta _2-\alpha _1\beta _5 +\omega\alpha _2 \beta _1-\alpha _2 \beta _4+\omega\alpha _3 \beta _3-\alpha _3 \beta _6
\end{array}
\right)
\end{equation}

\begin{equation}
\nonumber\mathbf{6}\sim\left(
\begin{array}{c}
 \alpha _1 \beta _3+\omega^2\alpha _2 \beta _2+\omega\alpha _3 \beta _1  \\
 \omega^2\alpha _1 \beta _1+\omega\alpha _2 \beta _3+\alpha _3 \beta _2   \\
 \omega\alpha _1 \beta _2 +\alpha _2 \beta _1+\omega^2\alpha _3 \beta _3 \\
 -\omega\alpha _1 \beta _6 -\omega^2\alpha_2 \beta _5-\alpha _3 \beta _4   \\
 -\omega^2\alpha_1 \beta _4-\alpha _2 \beta _6 -\omega \alpha _3 \beta _5  \\
 -\alpha _1 \beta _5  -\omega\alpha _2 \beta _4 -\omega^2\alpha _3 \beta _6
\end{array}
\right)
\end{equation}


\item{$\mathbf{3'_1}\otimes\mathbf{3'_1}=\mathbf{\overline{3}_1}\oplus\mathbf{\overline{3}'_1}\oplus\mathbf{3'_2}$}

\begin{equation}
\nonumber\mathbf{\overline{3}_1}\sim\left(
\begin{array}{c}
 \alpha _2 \beta _3-\alpha _3 \beta _2 \\
 \alpha _3 \beta _1-\alpha _1 \beta _3 \\
 \alpha _1 \beta _2-\alpha _2 \beta _1
\end{array}
\right)
\end{equation}

\begin{equation}
\nonumber\mathbf{\overline{3}'_1}\sim\left(
\begin{array}{c}
 -2 \alpha _1 \beta _1+\alpha _2 \beta _3+\alpha _3 \beta _2 \\
 \alpha _1 \beta _3-2 \alpha _2 \beta _2+\alpha _3 \beta _1 \\
 \alpha _1 \beta _2+\alpha _2 \beta _1-2 \alpha _3 \beta _3
\end{array}
\right)
\end{equation}

\begin{equation}
\nonumber\mathbf{3'_2}\sim\left(
\begin{array}{c}
\alpha _1 \beta _3+\alpha _2 \beta _2+\alpha _3 \beta _1 \\
 \alpha _1 \beta _1+\alpha _2 \beta _3 +\alpha _3 \beta _2\\
 \alpha _1 \beta _2+\alpha _2 \beta _1+\alpha _3 \beta _3
\end{array}
\right)
\end{equation}


\item{$\mathbf{3'_1}\otimes\mathbf{\overline{3}_1}=\mathbf{1'}\oplus\mathbf{2}\oplus\mathbf{6}$}

\begin{equation}
\nonumber\mathbf{1'}\sim \alpha _1 \beta _1+\alpha _2 \beta _2+\alpha _3 \beta _3
\end{equation}

\begin{equation}
\nonumber\mathbf{2}\sim\left(
\begin{array}{c}
 \alpha _1 \beta _3+\alpha _2 \beta _1+\alpha _3 \beta _2  \\
 -\omega \left(\alpha _1 \beta _2+\alpha _2 \beta _3+\alpha _3 \beta _1\right)
\end{array}
\right)
\end{equation}

\begin{equation}
\nonumber\mathbf{6}\sim\left(
\begin{array}{c}
 \alpha _1 \beta _3 +\omega\alpha _2 \beta _1+\omega ^2\alpha _3 \beta _2 \\
 \omega\alpha _1 \beta _2 +\omega ^2\alpha _2 \beta _3+\alpha _3 \beta _1  \\
 \omega ^2\alpha _1 \beta _1+\alpha _2 \beta _2 +\omega\alpha _3 \beta _3  \\
 -\alpha _1 \beta _3 -\omega ^2\alpha_2 \beta _1-\omega\alpha _3 \beta _2 \\
 -\omega ^2\alpha_1 \beta _2-\omega\alpha _2 \beta _3 -\alpha _3 \beta _1   \\
 -\omega\alpha _1 \beta _1-\alpha _2 \beta _2-\omega ^2\alpha_3 \beta _3
\end{array}
\right)
\end{equation}


\item{$\mathbf{3'_1}\otimes\mathbf{\overline{3}'_1}=\mathbf{1}\oplus\mathbf{2}\oplus\mathbf{6}$}

\begin{equation}
\nonumber\mathbf{1}\sim \alpha _1 \beta _1+\alpha _2 \beta _2+\alpha _3 \beta _3
\end{equation}

\begin{equation}
\nonumber\mathbf{2}\sim\left(
\begin{array}{c}
 \alpha _1 \beta _3 +\alpha _2 \beta _1 +\alpha _3 \beta _2  \\
 \omega \left(\alpha _1 \beta _2+\alpha _2 \beta _3+\alpha _3 \beta _1\right)
\end{array}
\right)
\end{equation}

\begin{equation}
\nonumber\mathbf{6}\sim\left(
\begin{array}{c}
 \alpha _1 \beta _3  +\omega\alpha _2 \beta _1 +\omega^2\alpha_3 \beta _2 \\
 \omega\alpha _1 \beta _2 +\omega^2\alpha _2 \beta _3+\alpha _3 \beta _1   \\
 \omega^2\alpha_1 \beta _1+\alpha _2 \beta _2+\omega\alpha _3 \beta _3   \\
  \alpha _1 \beta _3+\omega^2\alpha_2 \beta _1+\omega\alpha _3 \beta _2  \\
 \omega^2\alpha_1 \beta _2+\omega\alpha _2 \beta _3 +\alpha _3 \beta _1   \\
 \omega\alpha _1 \beta _1 +\alpha _2 \beta _2  +\omega^2\alpha_3 \beta _3
\end{array}
\right)
\end{equation}


\item{$\mathbf{3'_1}\otimes\mathbf{3_2}=\mathbf{\overline{3}_1}\oplus\mathbf{6}$}

\begin{equation}
\nonumber\mathbf{\overline{3}_1}\sim\left(
\begin{array}{c}
 \alpha _1 \beta _1+\alpha _2 \beta _3+\alpha _3 \beta _2 \\
 \alpha _1 \beta _3+\alpha _2 \beta _2+\alpha _3 \beta _1 \\
 \alpha _1 \beta _2+\alpha _2 \beta _1+\alpha _3 \beta _3
\end{array}
\right)
\end{equation}

\begin{equation}
\nonumber\mathbf{6}\sim\left(
\begin{array}{c}
\alpha _1 \beta _3 +\omega\alpha _2 \beta _2+\omega^2\alpha _3 \beta _1  \\
 \omega^2\alpha _1 \beta _1 +\alpha _2 \beta _3+\omega\alpha _3 \beta _2 \\
 \omega\alpha _1 \beta _2+\omega^2\alpha _2 \beta _1 +\alpha _3 \beta _3  \\
 -\omega\alpha_1 \beta _3-\alpha _2 \beta _2-\omega^2\alpha _3 \beta _1   \\
 -\omega^2\alpha _1 \beta _1-\omega\alpha_2 \beta _3 -\alpha _3 \beta _2   \\
 -\alpha _1 \beta _2 -\omega^2\alpha _2 \beta _1-\omega\alpha_3 \beta _3
\end{array}
\right)
\end{equation}


\item{$\mathbf{3'_1}\otimes\mathbf{3'_2}=\mathbf{\overline{3}'_1}\oplus\mathbf{6}$}

\begin{equation}
\nonumber\mathbf{\overline{3}'_1}\sim\left(
\begin{array}{c}
 \alpha _1 \beta _1+\alpha _2 \beta _3+\alpha _3 \beta _2 \\
 \alpha _1 \beta _3+\alpha _2 \beta _2+\alpha _3 \beta _1 \\
 \alpha _1 \beta _2+\alpha _2 \beta _1+\alpha _3 \beta _3
\end{array}
\right)
\end{equation}

\begin{equation}
\nonumber\mathbf{6}\sim\left(
\begin{array}{c}
 \alpha _1 \beta _3+\omega\alpha_2 \beta _2+\omega^2\alpha _3 \beta _1   \\
 \omega^2\alpha _1 \beta _1+\alpha _2 \beta _3 +\omega\alpha_3 \beta _2 \\
 \omega\alpha_1 \beta _2+\omega^2\alpha _2 \beta _1+\alpha _3 \beta _3   \\
 \omega\alpha_1 \beta _3+\alpha _2 \beta _2+\omega^2\alpha _3 \beta _1   \\
 \omega^2\alpha _1 \beta _1+\omega\alpha_2 \beta _3+\alpha _3 \beta _2  \\
 \alpha _1 \beta _2 +\omega^2\alpha _2 \beta _1+\omega\alpha_3 \beta _3
\end{array}
\right)
\end{equation}


\item{$\mathbf{3'_1}\otimes\mathbf{6}=\mathbf{3_1}\oplus\mathbf{3'_1}\oplus\mathbf{3_2}\oplus\mathbf{3'_2}\oplus\mathbf{6}$}

\begin{equation}
\nonumber\mathbf{3_1}\sim\left(
\begin{array}{c}
 \alpha _1 \beta _3-\omega\alpha _1 \beta _6+\omega\alpha _2 \beta _2-\alpha _2 \beta _5 + \omega ^2\alpha _3 \beta _1-\omega^2\alpha _3 \beta _4  \\
\omega\alpha _1 \beta _1-\alpha _1 \beta _4+\omega ^2\alpha _2 \beta _3-\omega^2\alpha _2 \beta _6+\alpha _3 \beta _2-\omega\alpha _3 \beta _5  \\
\omega ^2\alpha _1 \beta _2 -\omega^2\alpha _1 \beta _5 +\alpha _2 \beta _1-\omega\alpha _2 \beta _4 +\omega\alpha _3 \beta _3-\alpha _3 \beta _6
\end{array}
\right)
\end{equation}

\begin{equation}
\nonumber\mathbf{3'_1}\sim\left(
\begin{array}{c}
 \alpha _1 \beta _3+\omega\alpha _1 \beta _6+\omega\alpha _2 \beta _2 +\alpha _2 \beta _5 +\omega ^2\alpha _3 \beta _1 +\omega^2\alpha _3 \beta _4  \\
 \omega\alpha _1 \beta _1+\alpha _1 \beta _4  +\omega ^2\alpha _2 \beta _3 +\omega^2\alpha _2 \beta _6 +\alpha _3 \beta _2 +\omega\alpha _3 \beta _5  \\
 \omega ^2\alpha _1 \beta _2 +\omega^2\alpha _1 \beta _5 +\alpha _2 \beta _1+\omega\alpha _2 \beta _4 +\omega\alpha _3 \beta _3  +\alpha _3 \beta _6
\end{array}
\right)
\end{equation}

\begin{equation}
\nonumber\mathbf{3_2}\sim\left(
\begin{array}{c}
 \alpha _1 \beta _3-\omega\alpha_1 \beta _6+\alpha _2 \beta_2-\omega\alpha _2 \beta _5+\alpha _3 \beta _1-\omega\alpha _3 \beta _4     \\
 \omega ^2\alpha _1 \beta _1-\omega^2\alpha _1 \beta _4+\omega ^2\alpha _2 \beta _3-\omega ^2\alpha_2 \beta _6 +\omega^2\alpha _3 \beta _2 -\omega ^2\alpha _3 \beta _5  \\
\omega\alpha _1 \beta _2-\alpha _1\beta _5 +\omega \alpha _2 \beta _1-\alpha _2 \beta _4 +\omega\alpha _3 \beta _3-\alpha _3 \beta _6
\end{array}
\right)
\end{equation}

\begin{equation}
\nonumber\mathbf{3'_2}\sim\left(
\begin{array}{c}
\alpha _1 \beta _3+\omega\alpha_1 \beta _6+\alpha _2 \beta _2+\omega\alpha _2 \beta _5+\alpha _3 \beta _1+\omega\alpha _3 \beta _4     \\
\omega ^2\alpha _1 \beta _1+\omega^2\alpha _1 \beta _4+\omega ^2\alpha _2 \beta _3+\omega ^2\alpha _2 \beta _6+\omega^2\alpha _3 \beta _2 +\omega ^2\alpha _3 \beta _5  \\
\omega\alpha _1 \beta _2+\alpha _1 \beta_5+\omega\alpha _2 \beta _1+\alpha _2 \beta _4+\omega\alpha_3 \beta _3+\alpha _3 \beta _6
\end{array}
\right)
\end{equation}

\begin{equation}
\nonumber\mathbf{6}\sim\left(
\begin{array}{c}
 \alpha _1 \beta _3  +\omega^2\alpha_2 \beta _2+\omega\alpha _3 \beta _1  \\
 \omega^2\alpha _1 \beta _1+\omega\alpha _2 \beta _3 +\alpha _3 \beta _2   \\
 \omega \alpha _1 \beta _2 +\alpha _2 \beta _1+\omega^2\alpha_3 \beta _3 \\
 \omega\alpha _1 \beta _6  +\omega^2\alpha_2 \beta _5+\alpha _3 \beta _4 \\
 \omega^2\alpha _1 \beta _4+\alpha _2 \beta _6 +\omega\alpha _3 \beta _5   \\
 \alpha _1 \beta _5  +\omega\alpha _2 \beta _4+\omega^2\alpha_3 \beta _6
\end{array}
\right)
\end{equation}


\item{$\mathbf{\overline{3}_1}\otimes\mathbf{\overline{3}_1}=\mathbf{3_1}\oplus\mathbf{3'_1}\oplus\mathbf{3'_2}$}

\begin{equation}
\nonumber\mathbf{3_1}\sim\left(
\begin{array}{c}
 \alpha _2 \beta _3-\alpha _3 \beta _2 \\
 \alpha _3 \beta _1-\alpha _1 \beta _3 \\
 \alpha _1 \beta _2-\alpha _2 \beta _1
\end{array}
\right)
\end{equation}

\begin{equation}
\nonumber\mathbf{3'_1}\sim\left(
\begin{array}{c}
 -2 \alpha _1 \beta _1+\alpha _2 \beta _3+\alpha _3 \beta _2 \\
 \alpha _1 \beta _3-2 \alpha _2 \beta _2+\alpha _3 \beta _1 \\
 \alpha _1 \beta _2+\alpha _2 \beta _1-2 \alpha _3 \beta _3
\end{array}
\right)
\end{equation}

\begin{equation}
\nonumber\mathbf{3'_2}\sim\left(
\begin{array}{c}
 \alpha _1 \beta _1+\alpha _2 \beta _3+\alpha _3 \beta _2 \\
 \alpha _1 \beta _3+\alpha _2 \beta _2+\alpha _3 \beta _1 \\
 \alpha _1 \beta _2+\alpha _2 \beta _1+\alpha _3 \beta _3
\end{array}
\right)
\end{equation}


\item{$\mathbf{\overline{3}_1}\otimes\mathbf{\overline{3}'_1}=\mathbf{3_1}\oplus\mathbf{3'_1}\oplus\mathbf{3_2}$}

\begin{equation}
\nonumber\mathbf{3_1}\sim\left(
\begin{array}{c}
 -2 \alpha _1 \beta _1+\alpha _2 \beta _3 +\alpha _3 \beta _2\\
 \alpha _1 \beta _3-2 \alpha _2 \beta _2+\alpha _3 \beta _1 \\
 \alpha _1 \beta _2+\alpha _2 \beta _1-2 \alpha _3 \beta _3
\end{array}
\right)
\end{equation}

\begin{equation}
\nonumber\mathbf{3'_1}\sim\left(
\begin{array}{c}
 \alpha _2 \beta _3-\alpha _3 \beta _2 \\
 \alpha _3 \beta _1-\alpha _1 \beta _3 \\
 \alpha _1 \beta _2-\alpha _2 \beta _1
\end{array}
\right)
\end{equation}

\begin{equation}
\nonumber\mathbf{3_2}\sim\left(
\begin{array}{c}
 \alpha _1 \beta _1+\alpha _2 \beta _3+\alpha _3 \beta _2 \\
 \alpha _1 \beta _3+\alpha _2 \beta _2+\alpha _3 \beta _1 \\
 \alpha _1 \beta _2+\alpha _2 \beta _1+\alpha _3 \beta _3
\end{array}
\right)
\end{equation}


\item{$\mathbf{\overline{3}_1}\otimes\mathbf{3_2}=\mathbf{3'_1}\oplus\mathbf{6}$}

\begin{equation}
\nonumber\mathbf{3'_1}\sim\left(
\begin{array}{c}
 \alpha _1 \beta _2+\alpha _2 \beta _3+\alpha _3 \beta _1 \\
 \alpha _1 \beta _3+\alpha _2 \beta _1+\alpha _3 \beta _2\\
 \alpha _1 \beta _1+\alpha _2 \beta _2+\alpha _3 \beta _3
\end{array}
\right)
\end{equation}

\begin{equation}
\nonumber\mathbf{6}\sim\left(
\begin{array}{c}
\alpha _1 \beta _2+\omega\alpha _2 \beta _3+\omega^2\alpha _3 \beta _1 \\
\omega\alpha _1 \beta _3 +\omega^2\alpha _2 \beta _1+\alpha _3 \beta _2   \\
\omega^2\alpha_1 \beta _1+\alpha _2 \beta _2 +\omega\alpha _3 \beta _3   \\
\alpha _1 \beta _2+\omega^2\alpha _2 \beta _3+\omega\alpha _3 \beta _1  \\
\omega^2\alpha_1 \beta _3+\omega\alpha _2 \beta _1+\alpha _3 \beta _2   \\
\omega\alpha _1 \beta _1 +\alpha _2 \beta _2+\omega^2\alpha_3 \beta _3
\end{array}
\right)
\end{equation}


\item{$\mathbf{\overline{3}_1}\otimes\mathbf{3'_2}=\mathbf{3_1}\oplus\mathbf{6}$}

\begin{equation}
\nonumber\mathbf{3_1}\sim\left(
\begin{array}{c}
 \alpha _1 \beta _2+\alpha _2 \beta _3+\alpha _3 \beta _1 \\
 \alpha _1 \beta _3+\alpha _2 \beta _1+\alpha _3 \beta _2 \\
 \alpha _1 \beta _1+\alpha _2 \beta _2+\alpha _3 \beta _3
\end{array}
\right)
\end{equation}

\begin{equation}
\nonumber\mathbf{6}\sim\left(
\begin{array}{c}
\alpha _1 \beta _2 +\omega\alpha _2 \beta _3 +\omega^2\alpha _3 \beta _1 \\
 \omega\alpha _1 \beta _3 +\omega^2\alpha _2 \beta _1 +\alpha _3 \beta _2 \\
 \omega^2\alpha _1 \beta _1+\alpha _2 \beta _2 +\omega\alpha _3 \beta _3  \\
 -\alpha _1 \beta _2-\omega^2\alpha _2 \beta _3-\omega\alpha _3 \beta _1  \\
 -\omega^2\alpha_1 \beta _3-\omega\alpha _2 \beta _1 -\alpha _3 \beta _2   \\
 -\omega\alpha _1 \beta _1-\alpha _2 \beta _2-\omega^2\alpha_3 \beta _3
\end{array}
\right)
\end{equation}


\item{$\mathbf{\overline{3}_1}\otimes\mathbf{6}=\mathbf{\overline{3}_1}\oplus\mathbf{\overline{3}'_1}\oplus\mathbf{3_2}\oplus\mathbf{3'_2}\oplus\mathbf{6}$}

\begin{equation}
\nonumber\mathbf{\overline{3}_1}\sim\left(
\begin{array}{c}
\alpha _1 \beta _3+\omega\alpha _1 \beta _6+\omega\alpha _2 \beta _1+\alpha _2 \beta _4 +\omega^2\alpha _3 \beta _2 +\omega^2\alpha _3 \beta _5  \\
\omega\alpha _1 \beta _2+\alpha _1 \beta _5 +\omega ^2\alpha _2 \beta _3+\omega^2\alpha _2 \beta _6+\alpha _3 \beta _1 +\omega\alpha _3 \beta _4  \\
\omega ^2\alpha _1 \beta _1 +\omega^2\alpha _1 \beta _4+\alpha _2 \beta _2+\omega\alpha _2 \beta _5+\omega\alpha _3 \beta _3+\alpha _3 \beta _6
\end{array}
\right)
\end{equation}

\begin{equation}
\nonumber\mathbf{\overline{3}'_1}\sim\left(
\begin{array}{c}
\alpha _1 \beta _3-\omega\alpha _1 \beta _6+\omega\alpha _2 \beta _1-\alpha _2 \beta _4+\omega ^2\alpha _3 \beta _2 -\omega^2\alpha _3 \beta _5  \\
\omega\alpha _1 \beta _2-\alpha _1 \beta _5 +\omega ^2\alpha _2 \beta _3 -\omega^2\alpha _2 \beta _6+\alpha _3 \beta _1-\omega\alpha _3 \beta _4  \\
\omega ^2\alpha _1 \beta _1-\omega^2\alpha _1 \beta _4+\alpha _2 \beta _2-\omega\alpha _2 \beta _5+\omega\alpha _3 \beta _3-\alpha _3 \beta _6
\end{array}
\right)
\end{equation}

\begin{equation}
\nonumber\mathbf{3_2}\sim\left(
\begin{array}{c}
\alpha _1 \beta _2 +\alpha _1 \beta _5 +\alpha _2 \beta _3 +\alpha _2 \beta _6 +\alpha _3 \beta _1 +\alpha _3 \beta _4  \\
\omega\alpha _1 \beta _3+\omega^2\alpha_1 \beta _6+\omega\alpha _2 \beta _1 +\omega^2\alpha _2 \beta _4+\omega\alpha _3 \beta_2 +\omega^2\alpha _3 \beta _5   \\
\omega^2\alpha _1 \beta _1+\omega\alpha _1 \beta _4  +\omega^2\alpha _2 \beta _2+\omega\alpha _2 \beta_5  +\omega^2\alpha_3 \beta _3 +\omega\alpha _3 \beta _6
\end{array}
\right)
\end{equation}

\begin{equation}
\nonumber\mathbf{3'_2}\sim\left(
\begin{array}{c}
\alpha _1 \beta _2 -\alpha _1 \beta _5+\alpha _2 \beta _3-\alpha _2 \beta _6+\alpha _3 \beta _1-\alpha _3 \beta _4  \\
\omega\alpha _1 \beta _3-\omega^2\alpha_1 \beta _6+\omega\alpha _2 \beta _1 -\omega^2\alpha _2 \beta _4+\omega\alpha _3 \beta_2-\omega^2\alpha _3 \beta _5   \\
\omega^2\alpha _1 \beta _1-\omega\alpha _1 \beta _4 +\omega^2\alpha _2 \beta _2 -\omega\alpha _2\beta _5+\omega^2\alpha _3 \beta _3 -\omega\alpha _3 \beta _6
\end{array}
\right)
\end{equation}

\begin{equation}
\nonumber\mathbf{6}\sim\left(
\begin{array}{c}
\alpha _1 \beta _2 +\omega ^2\alpha _2 \beta _3 +\omega\alpha _3 \beta _1   \\
\omega\alpha _1 \beta _3 +\alpha _2 \beta _1 +\omega ^2\alpha _3 \beta _2  \\
\omega ^2\alpha _1 \beta _1+\omega\alpha _2 \beta _2+\alpha _3 \beta _3 \\
-\alpha_1 \beta _5-\omega\alpha _2 \beta _6-\omega ^2\alpha _3 \beta _4   \\
-\omega ^2\alpha _1 \beta _6 -\alpha_2 \beta _4-\omega\alpha _3 \beta _5   \\
-\omega\alpha _1 \beta _4 - \omega ^2\alpha _2 \beta _5-\alpha_3 \beta _6
\end{array}
\right)
\end{equation}


\item{$\mathbf{\overline{3}'_1}\otimes\mathbf{\overline{3}'_1}=\mathbf{3_1}\oplus\mathbf{3'_1}\oplus\mathbf{3'_2}$}

\begin{equation}
\nonumber\mathbf{3_1}\sim\left(
\begin{array}{c}
 \alpha _2 \beta _3-\alpha _3 \beta _2 \\
 \alpha _3 \beta _1-\alpha _1 \beta _3 \\
 \alpha _1 \beta _2-\alpha _2 \beta _1
\end{array}
\right)
\end{equation}

\begin{equation}
\nonumber\mathbf{3'_1}\sim\left(
\begin{array}{c}
 -2 \alpha _1 \beta _1+\alpha _2 \beta _3+\alpha _3 \beta _2 \\
 \alpha _1 \beta _3-2 \alpha _2 \beta _2+\alpha _3 \beta _1 \\
 \alpha _1 \beta _2+\alpha _2 \beta _1-2 \alpha _3 \beta _3
\end{array}
\right)
\end{equation}

\begin{equation}
\nonumber\mathbf{3'_2}\sim\left(
\begin{array}{c}
 \alpha _1 \beta _1+\alpha _2 \beta _3+\alpha _3 \beta _2 \\
 \alpha _1 \beta _3+\alpha _2 \beta _2+\alpha _3 \beta _1 \\
 \alpha _1 \beta _2+\alpha _2 \beta _1+\alpha _3 \beta _3
\end{array}
\right)
\end{equation}


\item{$\mathbf{\overline{3}'_1}\otimes\mathbf{3_2}=\mathbf{3_1}\oplus\mathbf{6}$}

\begin{equation}
\nonumber\mathbf{3_1}\sim\left(
\begin{array}{c}
 \alpha _1 \beta _2+\alpha _2 \beta _3+\alpha _3 \beta _1 \\
 \alpha _1 \beta _3+\alpha _2 \beta _1+\alpha _3 \beta _2 \\
 \alpha _1 \beta _1+\alpha _2 \beta _2+\alpha _3 \beta _3
\end{array}
\right)
\end{equation}

\begin{equation}
\nonumber\mathbf{6}\sim\left(
\begin{array}{c}
\alpha _1 \beta _2+\omega\alpha _2 \beta _3 +\omega^2\alpha _3 \beta _1 \\
\omega\alpha _1 \beta _3+\omega^2\alpha _2 \beta _1+\alpha _3 \beta _2  \\
\omega^2\alpha _1 \beta _1+\alpha _2 \beta _2 +\omega\alpha _3 \beta _3  \\
-\alpha _1 \beta _2 -\omega^2\alpha_2 \beta _3 -\omega\alpha _3 \beta _1 \\
-\omega^2\alpha _1 \beta _3-\omega\alpha _2 \beta _1 -\alpha _3 \beta _2   \\
-\omega\alpha _1 \beta _1 -\alpha _2 \beta _2 -\omega^2\alpha_3 \beta _3
\end{array}
\right)
\end{equation}


\item{$\mathbf{\overline{3}'_1}\otimes\mathbf{3'_2}=\mathbf{3'_1}\oplus\mathbf{6}$}

\begin{equation}
\nonumber\mathbf{3'_1}\sim\left(
\begin{array}{c}
 \alpha _1 \beta _2+\alpha _2 \beta _3+\alpha _3 \beta _1 \\
 \alpha _1 \beta _3+\alpha _2 \beta _1+\alpha _3 \beta _2 \\
 \alpha _1 \beta _1+\alpha _2 \beta _2+\alpha _3 \beta _3
\end{array}
\right)
\end{equation}

\begin{equation}
\nonumber\mathbf{6}\sim\left(
\begin{array}{c}
\alpha _1 \beta _2  +\omega\alpha _2 \beta _3 +\omega^2\alpha_3 \beta _1 \\
 \omega\alpha _1 \beta _3 +\omega^2\alpha_2 \beta _1+\alpha _3 \beta _2   \\
 \omega^2\alpha_1 \beta _1+\alpha _2 \beta _2 +\omega\alpha _3 \beta _3   \\
 \alpha _1 \beta _2  +\omega^2\alpha_2 \beta _3+\omega\alpha _3 \beta _1  \\
 \omega^2\alpha_1 \beta _3+\omega\alpha _2 \beta _1 +\alpha _3 \beta _2   \\
\omega\alpha _1 \beta _1 +\alpha _2 \beta _2  +\omega^2\alpha_3 \beta _3
\end{array}
\right)
\end{equation}


\item{$\mathbf{\overline{3}'_1}\otimes\mathbf{6}=\mathbf{\overline{3}_1}\oplus\mathbf{\overline{3}'_1}\oplus\mathbf{3_2}\oplus\mathbf{3'_2}\oplus\mathbf{6}$}

\begin{equation}
\nonumber\mathbf{\overline{3}_1}\sim\left(
\begin{array}{c}
\alpha _1 \beta _3 -\omega\alpha _1 \beta _6 +\omega\alpha _2 \beta _1-\alpha _2 \beta _4 +\omega ^2\alpha _3 \beta _2 -\omega^2\alpha _3 \beta _5  \\
\omega\alpha _1 \beta _2-\alpha _1 \beta _5+\omega ^2\alpha _2 \beta _3 -\omega^2\alpha _2 \beta _6+\alpha _3 \beta _1-\omega\alpha _3 \beta _4  \\
\omega ^2\alpha _1 \beta _1-\omega^2\alpha _1 \beta _4+\alpha _2 \beta _2-\omega\alpha _2 \beta _5+\omega\alpha _3 \beta _3 -\alpha _3 \beta _6
\end{array}
\right)
\end{equation}

\begin{equation}
\nonumber\mathbf{\overline{3}'_1}\sim\left(
\begin{array}{c}
\alpha _1 \beta _3+\omega\alpha _1 \beta _6+\omega\alpha _2 \beta _1+\alpha _2 \beta _4 +\omega ^2\alpha _3 \beta _2 +\omega^2\alpha _3 \beta _5  \\
\omega\alpha _1 \beta _2+\alpha _1 \beta _5+\omega ^2\alpha _2 \beta _3 +\omega^2\alpha _2 \beta _6+\alpha _3 \beta _1  +\omega\alpha _3 \beta _4  \\
\omega ^2\alpha _1 \beta _1 +\omega^2\alpha _1 \beta _4+\alpha _2 \beta _2+\omega\alpha _2 \beta _5+\omega\alpha _3 \beta _3+\alpha _3 \beta _6
\end{array}
\right)
\end{equation}

\begin{equation}
\nonumber\mathbf{3_2}\sim\left(
\begin{array}{c}
\alpha _1 \beta _2-\alpha _1 \beta _5+\alpha _2 \beta _3-\alpha _2 \beta _6+\alpha _3 \beta _1 -\alpha _3 \beta _4   \\
\omega\alpha _1 \beta _3-\omega^2\alpha_1 \beta _6+\omega\alpha _2 \beta _1 -\omega^2\alpha _2 \beta _4+\omega\alpha _3 \beta _2 -\omega^2\alpha _3 \beta _5    \\
\omega^2 \alpha _1 \beta _1-\omega\alpha _1 \beta _4+\omega^2\alpha _2 \beta _2-\omega\alpha _2\beta _5 +\omega^2\alpha _3 \beta _3-\omega\alpha _3 \beta _6
\end{array}
\right)
\end{equation}

\begin{equation}
\nonumber\mathbf{3'_2}\sim\left(
\begin{array}{c}
\alpha _1 \beta _2+\alpha _1 \beta _5+\alpha _2 \beta _3+\alpha _2 \beta _6+\alpha _3 \beta _1 +\alpha _3 \beta _4   \\
\omega\alpha _1 \beta _3+\omega^2\alpha_1 \beta _6 +\omega\alpha _2 \beta _1+\omega^2 \alpha _2 \beta _4+\omega\alpha _3 \beta_2  +\omega^2\alpha _3 \beta _5    \\
\omega^2\alpha _1 \beta _1+\omega\alpha _1 \beta _4+\omega^2\alpha _2 \beta _2+\omega\alpha _2 \beta_5+\omega^2\alpha_3 \beta _3+\omega\alpha _3 \beta _6
\end{array}
\right)
\end{equation}

\begin{equation}
\nonumber\mathbf{6}\sim\left(
\begin{array}{c}
 \alpha_1 \beta _2+\omega ^2\alpha _2 \beta _3 +\omega\alpha _3 \beta _1   \\
\omega\alpha _1 \beta _3  +\alpha_2 \beta _1+\omega ^2\alpha _3 \beta _2  \\
\omega ^2 \alpha _1 \beta _1 +\omega\alpha _2 \beta _2  +\alpha_3 \beta _3 \\
 \alpha_1 \beta _5+\omega\alpha _2 \beta _6 +\omega ^2\alpha _3 \beta _4   \\
 \omega ^2\alpha _1 \beta _6 +\alpha_2 \beta _4+\omega\alpha _3 \beta _5   \\
\omega\alpha _1 \beta _4  + \omega ^2\alpha _2 \beta _5 +\alpha_3 \beta _6
\end{array}
\right)
\end{equation}


\item{$\mathbf{3_2}\otimes\mathbf{3_2}=\mathbf{1}\oplus\mathbf{2}\oplus\mathbf{3_2}\oplus\mathbf{3'_2}$}

\begin{equation}
\nonumber\mathbf{1}\sim \alpha _1 \beta _2+\alpha _2 \beta _1+\alpha _3 \beta _3
\end{equation}

\begin{equation}
\nonumber\mathbf{2}\sim\left(
\begin{array}{c}
 \alpha_1 \beta _3+\alpha _2 \beta _2+\alpha _3 \beta _1 \\
 \omega \left(\alpha _1 \beta _1+\alpha _2 \beta _3+\alpha _3 \beta _2\right)
\end{array}
\right)
\end{equation}

\begin{equation}
\nonumber\mathbf{3_2}\sim\left(
\begin{array}{c}
 \alpha _3 \beta _1-\alpha _1 \beta _3 \\
 \alpha _2 \beta _3-\alpha _3 \beta _2 \\
 \alpha _1 \beta _2-\alpha _2 \beta _1
\end{array}
\right)
\end{equation}

\begin{equation}
\nonumber\mathbf{3'_2}\sim\left(
\begin{array}{c}
 \alpha _1 \beta _3-2 \alpha _2 \beta_2+\alpha _3 \beta _1 \\
 -2\alpha _1 \beta _1+\alpha_2 \beta _3+\alpha _3 \beta _2 \\
\alpha _1 \beta _2+\alpha _2 \beta _1-2\alpha _3 \beta _3
\end{array}
\right)
\end{equation}


\item{$\mathbf{3_2}\otimes\mathbf{3'_2}=\mathbf{1'}\oplus\mathbf{2}\oplus\mathbf{3_2}\oplus\mathbf{3'_2}$}

\begin{equation}
\nonumber\mathbf{1'}\sim \alpha _1 \beta _2+\alpha _2 \beta _1+\alpha _3 \beta _3
\end{equation}

\begin{equation}
\nonumber\mathbf{2}\sim\left(
\begin{array}{c}
 \alpha_1 \beta _3+\alpha _2 \beta _2+\alpha _3 \beta _1 \\
-\omega \left( \alpha _1 \beta _1+\alpha _2 \beta _3+\alpha _3 \beta _2\right)
\end{array}
\right)
\end{equation}

\begin{equation}
\nonumber\mathbf{3_2}\sim\left(
\begin{array}{c}
\alpha _1 \beta _3-2 \alpha _2 \beta_2+\alpha _3 \beta _1 \\
-2\alpha _1 \beta _1+\alpha_2 \beta _3+\alpha _3 \beta _2 \\
\alpha _1 \beta _2+\alpha _2 \beta _1-2\alpha _3 \beta _3
\end{array}
\right)
\end{equation}

\begin{equation}
\nonumber\mathbf{3'_2}\sim\left(
\begin{array}{c}
\alpha _3 \beta _1 -\alpha _1 \beta _3 \\
\alpha _2 \beta _3 -\alpha _3 \beta _2 \\
\alpha _1 \beta _2 -\alpha _2 \beta _1
\end{array}
\right)
\end{equation}


\item{$\mathbf{3_2}\otimes\mathbf{6}=\mathbf{3_1}\otimes\mathbf{3'_1}\oplus\mathbf{\overline{3}_1}\oplus\mathbf{\overline{3}'_1}\oplus\mathbf{6}$}

\begin{equation}
\nonumber\mathbf{3_1}\sim\left(
\begin{array}{c}
\alpha _1 \beta _3+\omega\alpha _1 \beta _6 +\omega ^2\alpha _2 \beta _2 +\omega^2\alpha _2 \beta _5 +\omega\alpha _3 \beta _1+\alpha _3 \beta _4 \\
\alpha _1 \beta _1+\omega\alpha _1 \beta _4 +\omega ^2\alpha _2 \beta _3 +\omega^2\alpha _2 \beta _6 +\omega\alpha _3 \beta _2+\alpha _3 \beta _5 \\
\alpha _1 \beta _2+\omega\alpha _1 \beta _5 +\omega ^2\alpha _2 \beta _1 +\omega^2\alpha _2 \beta _4 +\omega\alpha _3 \beta _3+\alpha _3 \beta _6
\end{array}
\right)
\end{equation}

\begin{equation}
\nonumber\mathbf{3'_1}\sim\left(
\begin{array}{c}
\alpha _1 \beta _3-\omega\alpha _1 \beta _6+\omega ^2\alpha _2 \beta _2-\omega^2\alpha _2 \beta _5 +\omega\alpha _3 \beta _1-\alpha _3 \beta _4 \\
\alpha _1 \beta _1-\omega\alpha _1 \beta _4+\omega ^2\alpha _2 \beta _3 -\omega^2\alpha _2 \beta _6 +\omega\alpha _3 \beta _2-\alpha _3 \beta _5 \\
\alpha _1 \beta _2-\omega\alpha _1 \beta _5+\omega ^2\alpha _2 \beta _1-\omega^2\alpha _2 \beta _4 +\omega\alpha _3 \beta _3-\alpha _3 \beta _6
\end{array}
\right)
\end{equation}

\begin{equation}
\nonumber\mathbf{\overline{3}_1}\sim\left(
\begin{array}{c}
\alpha _1 \beta _1 +\alpha _1 \beta _4+\omega\alpha _2 \beta _3+\omega^2\alpha _2 \beta _6+\omega^2\alpha _3 \beta _2+\omega\alpha _3 \beta _5 \\
\alpha _1 \beta _3 +\alpha _1 \beta _6+\omega\alpha _2 \beta _2+\omega^2\alpha _2 \beta _5+\omega^2\alpha _3 \beta _1+\omega\alpha _3 \beta _4 \\
\alpha _1 \beta _2 +\alpha _1 \beta _5+\omega\alpha _2 \beta _1+\omega^2\alpha _2 \beta _4+\omega^2\alpha _3 \beta _3+\omega\alpha _3 \beta _6
\end{array}
\right)
\end{equation}

\begin{equation}
\nonumber\mathbf{\overline{3}'_1}\sim\left(
\begin{array}{c}
\alpha _1 \beta _1-\alpha _1 \beta _4+\omega\alpha _2 \beta _3-\omega^2\alpha _2 \beta _6+\omega^2\alpha _3 \beta _2-\omega\alpha _3 \beta _5 \\
\alpha _1 \beta _3-\alpha _1 \beta _6+\omega\alpha _2 \beta _2-\omega^2\alpha _2 \beta _5+\omega^2\alpha _3 \beta _1-\omega\alpha _3 \beta _4 \\
\alpha _1 \beta _2 -\alpha _1 \beta _5+\omega\alpha _2 \beta _1-\omega^2\alpha _2 \beta _4+\omega^2\alpha _3 \beta _3-\omega\alpha _3 \beta _6
\end{array}
\right)
\end{equation}

\begin{equation}
\nonumber\mathbf{6}\sim\left(
\begin{array}{c}
\alpha _1 \beta _6 +\alpha _2 \beta _5+\alpha _3 \beta _4  \\
\alpha _1 \beta _4+\alpha _2 \beta _6+\alpha _3 \beta _5   \\
\alpha _1 \beta _5+\alpha _2 \beta _4+\alpha _3 \beta _6   \\
-\omega^2\left(\alpha _1 \beta _3+\alpha _2 \beta _2+\alpha _3 \beta _1\right) \\
-\omega^2\left(\alpha _1 \beta _1+\alpha _2 \beta _3+\alpha _3 \beta _2\right) \\
-\omega^2\left(\alpha _1 \beta _2+ \alpha _2 \beta _1+\alpha _3 \beta _3\right)
\end{array}
\right)
\end{equation}


\item{$\mathbf{3'_2}\otimes\mathbf{3'_2}=\mathbf{1}\oplus\mathbf{2}\oplus\mathbf{3_2}\oplus\mathbf{3'_2}$}

\begin{equation}
\nonumber\mathbf{1}\sim \alpha _1 \beta _2+\alpha _2 \beta _1+\alpha _3 \beta _3
\end{equation}

\begin{equation}
\nonumber\mathbf{2}\sim\left(
\begin{array}{c}
 \alpha _1 \beta _3+\alpha _2 \beta _2 +\alpha _3 \beta _1  \\
 \omega \left(\alpha _1 \beta _1+\alpha _2 \beta _3+\alpha _3 \beta _2\right)
\end{array}
\right)
\end{equation}

\begin{equation}
\nonumber\mathbf{3_2}\sim\left(
\begin{array}{c}
\alpha _3 \beta _1 -\alpha _1 \beta _3 \\
\alpha _2 \beta _3 -\alpha _3 \beta _2 \\
\alpha _1 \beta _2 -\alpha _2 \beta _1
\end{array}
\right)
\end{equation}

\begin{equation}
\nonumber\mathbf{3'_2}\sim\left(
\begin{array}{c}
\alpha _1 \beta _3-2 \alpha _2 \beta_2+\alpha _3 \beta _1 \\
 -2\alpha _1 \beta _1+\alpha _2 \beta _3+\alpha _3 \beta _2 \\
\alpha _1 \beta _2+\alpha _2 \beta _1-2\alpha _3 \beta _3
\end{array}
\right)
\end{equation}


\item{$\mathbf{3'_2}\otimes\mathbf{6}=\mathbf{3_1}\oplus\mathbf{3'_1}\oplus\mathbf{\overline{3}_1}\oplus\mathbf{\overline{3}'_1}\oplus\mathbf{6}$}

\begin{equation}
\nonumber\mathbf{3_1}\sim\left(
\begin{array}{c}
\alpha _1 \beta _3 -\omega\alpha _1 \beta _6 +\omega ^2\alpha _2 \beta _2 -\omega^2\alpha _2 \beta _5+\omega\alpha _3 \beta _1-\alpha _3 \beta _4 \\
\alpha _1 \beta _1-\omega\alpha _1 \beta _4+\omega ^2\alpha _2 \beta _3-\omega^2\alpha _2 \beta _6 +\omega\alpha _3 \beta _2 -\alpha _3 \beta _5 \\
\alpha _1 \beta _2 -\omega\alpha _1 \beta _5 +\omega ^2\alpha _2 \beta _1-\omega^2\alpha _2 \beta _4 +\omega\alpha _3 \beta _3 -\alpha _3 \beta _6
\end{array}
\right)
\end{equation}

\begin{equation}
\nonumber\mathbf{3'_1}\sim\left(
\begin{array}{c}
\alpha _1 \beta _3+\omega\alpha _1 \beta _6 +\omega ^2\alpha _2 \beta _2 +\omega^2\alpha _2 \beta _5 +\omega\alpha _3 \beta _1  +\alpha _3 \beta _4 \\
\alpha _1 \beta _1+\omega\alpha _1 \beta _4 +\omega ^2\alpha _2 \beta _3 +\omega^2\alpha _2 \beta _6 +\omega\alpha _3 \beta _2  +\alpha _3 \beta _5 \\
\alpha _1 \beta _2+\omega\alpha _1 \beta _5 +\omega ^2\alpha _2 \beta _1 +\omega^2\alpha _2 \beta _4 +\omega\alpha _3 \beta _3+\alpha _3 \beta _6
\end{array}
\right)
\end{equation}

\begin{equation}
\nonumber\mathbf{\overline{3}_1}\sim\left(
\begin{array}{c}
\alpha _1 \beta _1 -\alpha _1 \beta _4+\omega\alpha _2 \beta _3-\omega^2\alpha _2 \beta _6+\omega^2\alpha _3 \beta _2-\omega\alpha _3 \beta _5 \\
\alpha _1 \beta _3-\alpha _1 \beta _6+\omega\alpha _2 \beta _2-\omega^2\alpha _2 \beta _5+\omega^2\alpha _3 \beta _1-\omega\alpha _3 \beta _4 \\
\alpha _1 \beta _2-\alpha _1 \beta _5+\omega\alpha _2 \beta _1-\omega^2\alpha _2 \beta _4 +\omega^2\alpha _3 \beta _3-\omega\alpha _3 \beta _6
\end{array}
\right)
\end{equation}

\begin{equation}
\nonumber\mathbf{\overline{3}'_1}\sim\left(
\begin{array}{c}
\alpha _1 \beta _1 +\alpha _1 \beta _4+\omega\alpha _2 \beta _3+\omega^2\alpha _2 \beta _6+\omega^2\alpha _3 \beta _2+\omega\alpha _3 \beta _5 \\
\alpha _1 \beta _3 +\alpha _1 \beta _6+\omega\alpha _2 \beta _2+\omega^2\alpha _2 \beta _5+\omega^2\alpha _3 \beta _1+\omega\alpha _3 \beta _4 \\
\alpha _1 \beta _2 +\alpha _1 \beta _5+\omega\alpha _2 \beta _1+\omega^2\alpha _2 \beta _4+\omega^2\alpha _3 \beta _3+\omega\alpha _3 \beta _6
\end{array}%
\right)
\end{equation}


\item{$\mathbf{6}\otimes\mathbf{6}=\mathbf{1}\oplus\mathbf{1'}\oplus\mathbf{2}_S\oplus\mathbf{2}_A\oplus\mathbf{3_1}\oplus\mathbf{3'_1}\oplus\mathbf{\overline{3}_1}\oplus\mathbf{\overline{3}'_1}\oplus\mathbf{3_2}\oplus\mathbf{3'_2}\oplus\mathbf{6}_S\oplus\mathbf{6}_A$}

\begin{equation}
\nonumber\mathbf{1}\sim\alpha _1 \beta _5+\alpha _2 \beta _4+\alpha _3 \beta _6+\alpha _4 \beta _2+\alpha _5 \beta _1+\alpha _6 \beta_3
\end{equation}

\begin{equation}
\nonumber\mathbf{1'}\sim \alpha _1 \beta _5+\alpha _2 \beta _4+\alpha _3 \beta _6-\alpha _4 \beta _2-\alpha _5 \beta _1-\alpha _6 \beta_3
\end{equation}



\begin{equation}
\nonumber\mathbf{2}_S\sim\left(
\begin{array}{c}
\alpha _1 \beta _6+\alpha _2 \beta _5+\alpha _3 \beta _4+\alpha _4 \beta_3+\alpha _5 \beta _2+\alpha _6 \beta _1   \\
\omega\left(\alpha _1 \beta _4+\alpha _2\beta _6+\alpha _3 \beta _5+\alpha _4 \beta _1+\alpha _5\beta _3+\alpha _6 \beta _2\right)
\end{array}
\right)
\end{equation}

\begin{equation}
\nonumber\mathbf{2}_A\sim\left(
\begin{array}{c}
\alpha _1 \beta _6+\alpha _2 \beta _5+\alpha _3 \beta _4-\alpha _4 \beta_3-\alpha _5 \beta _2-\alpha _6 \beta _1 \\
-\omega\left(\alpha _1 \beta _4+\alpha _2\beta _6+\alpha _3 \beta _5-\alpha _4 \beta _1-\alpha _5\beta _3-\alpha _6 \beta _2\right)
\end{array}
\right)
\end{equation}

\begin{equation}
\nonumber\mathbf{3_1}\sim\left(
\begin{array}{c}
\alpha _1 \beta _6+\omega\alpha _2 \beta _5+\omega^2\alpha _3 \beta _4-\omega^2\alpha _4 \beta _3-\omega\alpha _5 \beta _2-\alpha _6 \beta _1   \\
\omega\alpha _1 \beta _4+\omega^2\alpha _2 \beta _6+\alpha _3 \beta _5-\omega\alpha _4 \beta _1-\alpha _5 \beta _3 -\omega^2\alpha _6 \beta _2   \\ \omega^2\alpha _1 \beta _5+\alpha _2 \beta _4+\omega\alpha _3 \beta _6-\alpha _4 \beta _2 -\omega^2\alpha _5 \beta _1-\omega\alpha _6 \beta _3
\end{array}
\right)
\end{equation}

\begin{equation}
\nonumber\mathbf{3'_1}\sim\left(
\begin{array}{c}
\alpha _1 \beta _6+\omega\alpha _2 \beta _5+\omega^2\alpha _3 \beta _4+\omega^2\alpha _4 \beta _3+\omega\alpha _5 \beta _2+\alpha _6 \beta _1   \\
\omega\alpha _1 \beta _4+\omega^2\alpha _2 \beta _6+\alpha _3 \beta _5+\omega\alpha _4 \beta _1+\alpha _5 \beta _3 +\omega^2\alpha _6 \beta _2   \\
\omega^2\alpha _1 \beta _5+\alpha _2 \beta _4+\omega\alpha _3 \beta _6  +\alpha _4 \beta _2 +\omega^2\alpha _5 \beta _1+\omega\alpha _6 \beta _3
\end{array}
\right)
\end{equation}

\begin{equation}
\nonumber\mathbf{\overline{3}_1}\sim\left(
\begin{array}{c}
\alpha _1 \beta _4+\omega^2\alpha _2 \beta _6+\omega\alpha _3 \beta _5-\alpha _4 \beta _1-\omega\alpha _5 \beta _3 -\omega ^2\alpha _6 \beta _2   \\
\omega\alpha _1 \beta _6+\alpha _2 \beta _5+\omega^2\alpha _3 \beta _4-\omega ^2\alpha _4 \beta _3-\alpha _5 \beta _2 -\omega\alpha _6 \beta _1   \\
\omega^2\alpha _1 \beta _5 +\omega\alpha _2 \beta _4+\alpha _3 \beta _6-\omega\alpha _4 \beta _2 -\omega ^2\alpha _5 \beta _1-\alpha _6 \beta _3
\end{array}
\right)
\end{equation}

\begin{equation}
\nonumber\mathbf{\overline{3}'_1}\sim\left(
\begin{array}{c}
\alpha _1 \beta _4+\omega^2\alpha _2 \beta _6+\omega\alpha _3 \beta _5+\alpha _4 \beta _1+\omega\alpha _5 \beta _3 +\omega ^2\alpha _6 \beta _2   \\
\omega\alpha _1 \beta _6+\alpha _2 \beta _5+\omega^2\alpha _3 \beta _4+\omega ^2\alpha _4 \beta _3 +\alpha _5 \beta _2+\omega\alpha _6 \beta _1   \\
\omega^2\alpha _1 \beta _5+\omega\alpha _2 \beta _4+\alpha _3 \beta _6+\omega\alpha _4 \beta _2+\omega ^2\alpha _5 \beta _1  +\alpha _6 \beta _3
\end{array}
\right)
\end{equation}

\begin{equation}
\nonumber\mathbf{3_2}\sim\left(
\begin{array}{c}
\alpha _1 \beta _3+\alpha _2\beta _2+\alpha _3 \beta _1-\omega\alpha _4 \beta _6-\omega\alpha _5\beta _5-\omega\alpha _6 \beta _4 \\
\alpha _1 \beta _1+\alpha _2 \beta _3+\alpha _3 \beta _2-\omega\alpha _4 \beta _4-\omega\alpha _5 \beta _6-\omega\alpha _6\beta _5 \\
\alpha _1 \beta _2 +\alpha _2 \beta _1+\alpha _3 \beta _3 -\omega\alpha _4\beta _5-\omega\alpha _5 \beta _4-\omega\alpha _6 \beta _6
\end{array}
\right)
\end{equation}

\begin{equation}
\nonumber\mathbf{3'_2}\sim\left(
\begin{array}{c}
\alpha _1 \beta _3 +\alpha _2 \beta _2 +\alpha _3 \beta _1+\omega\alpha _4 \beta _6+\omega\alpha _5\beta _5 +\omega\alpha _6 \beta _4 \\
\alpha _1 \beta _1+\alpha _2 \beta _3 +\alpha _3 \beta _2 +\omega\alpha _4 \beta _4+\omega\alpha _5 \beta _6+\omega\alpha _6\beta _5 \\
\alpha _1 \beta _2 +\alpha _2 \beta _1 +\alpha _3 \beta _3+\omega\alpha _4\beta _5 +\omega\alpha _5 \beta _4+\omega\alpha _6 \beta _6
\end{array}
\right)
\end{equation}

\begin{equation}
\nonumber\mathbf{6}_S\sim\left(
\begin{array}{c}
\alpha _4 \beta _6-2 \alpha _5 \beta _5 +\alpha _6 \beta _4 \\
-2 \alpha _4 \beta _4+\alpha _5 \beta _6+\alpha _6 \beta _5 \\
\alpha _4 \beta _5 +\alpha _5 \beta _4-2 \alpha _6 \beta _6 \\
\alpha _1 \beta _3-2 \alpha _2 \beta _2+\alpha _3 \beta _1 \\
-2\alpha _1 \beta _1+\alpha _2 \beta _3+\alpha _3 \beta _2 \\
\alpha _1 \beta _2+\alpha _2 \beta _1-2 \alpha _3 \beta _3
\end{array}
\right)
\end{equation}

\begin{equation}
\nonumber\mathbf{6}_A\sim\left(
\begin{array}{c}
\alpha _4 \beta _6-\alpha _6 \beta _4 \\
\alpha _6 \beta _5-\alpha _5 \beta _6 \\
\alpha _5 \beta _4-\alpha _4 \beta _5 \\
\alpha _3 \beta _1-\alpha _1 \beta _3 \\
\alpha _2 \beta _3-\alpha _3 \beta _2 \\
\alpha _1 \beta _2-\alpha _2 \beta _1
\end{array}
\right)
\end{equation}
\end{itemize}

\end{document}